\documentclass[12pt,a4paper]{article}
\usepackage[english]{babel}
\usepackage[utf8]{inputenc}
\usepackage[T1]{fontenc}
\usepackage{enumitem}
\usepackage{amsmath,amssymb,amsfonts,bbm,amsthm,mathrsfs,mathtools}
\usepackage[subnum]{cases}	% multi-case equations with separate number for each
\allowdisplaybreaks			% breaks "align" in multiple pages
\makeatletter
\newcommand*{\distas}[1]{\mathbin{\overset{#1}{\kern\z@\sim}}}	
\makeatother
\newcommand*{\norm}[1]{\big\lVert#1\big\rVert}		% norm
\newcommand*\abs[1]{\big|#1\big|}					% absolute value
\DeclareMathOperator*{\vect}{vec}		% define vectorization operator
\newcommand*\vecc[1]{\vect\big(#1\big)}
\DeclareMathOperator*{\diagg}{diag}		% define \diag
\newcommand*\diag[1]{\diagg\big(#1\big)}
\usepackage{dsfont}
\def \I{\mathds{1}}
\def \R{\mathds{R}}

\usepackage{xcolor,epsfig,subfigure,graphicx}  \graphicspath{{./Images/},{./Images/Plots_Lag1e3/}}
\usepackage[font=footnotesize,width=0.95\linewidth]{caption}
\usepackage{tikz,tkz-berge}
\usetikzlibrary{shapes,arrows,bayesnet}	% flow charts
\usepackage{float,array,rotating,booktabs}	
\usepackage[section]{placeins}	% floats not go to next section

\newtheoremstyle{custom}%    <name>
                {\topsep}%   <space above>
                {\topsep}%   <space below>
                {\itshape}%  <body font>
                {}%          <indent amount>
                {\bfseries}% <Theorem head font>
                {.}%         <punctuation after theorem head>
                {\newline}%  <space after theorem head> (default .5em)
                {}%          <Theorem head spec>
\theoremstyle{custom}
\usepackage{hyperref} % linktoc=all
\hypersetup{bookmarks=true, pagebackref=false, colorlinks=false, unicode=true, urlcolor=blue, linkcolor=blue, filecolor=blue, citecolor=red}
\begin{document}

%%%%%%%%%%%%%%%%%%%%%%%%%%%%%%%%%%%%%%%%%%%%%%%%%%%%%%%%%%%%%%%%%%%%%%%%%%%%%%
\def\spacingset#1{\renewcommand{\baselinestretch}%
{#1}\small\normalsize} \spacingset{1}

\title{\bf Bayesian Markov Switching Tensor Regression for Time-varying Networks}

\author{Monica Billio\thanks{
We are grateful to Federico Bassetti, Sylvia Fr\"uhwirth-Schnatter, Christian Gouriéroux, Alain Monfort and Christian Robert for their comments and suggestions on earlier versions of this paper. Also, we thank the seminar participants in CREST. Moreover, we thank the conference and workshop participants at: ``12th CFENetwork'' in Pisa, 2018, ``DYSES'' in Paris, 2018, ``46th SSC'' in Montréal, 2018, ``1st QFFE'' in Marseille, 2018, ``8th MAF'' in Madrid, 2018, ``11th CFENetwork'' in London, 2017, ``8th ESOBE'' in Maastricht, 2017, ``1st Italian-French Statistics Seminar'' in Venice, 2017, ``41st AMASES'' in Cagliari, 2017 and ``1st EcoSta'' in Hong Kong, 2017.
This work was supported by the Université Franco-Italienne under Grant ``Bando Vinci 2016''. This research used the SCSCF multiprocessor cluster system at Ca' Foscari University of Venice.}\hspace{.2cm}\\
Department of Economics, Ca' Foscari University of Venice\\
and \\
Roberto Casarin \\
Department of Economics, Ca' Foscari University of Venice\\
and\\
Matteo Iacopini \\
Scuola Normale Superiore of Pisa}

\maketitle

\bigskip
\begin{abstract}
We propose a new Bayesian Markov switching regression model for multidimensional arrays (tensors) of binary time series. We assume a zero-inflated logit regression with time-varying parameters and apply it to multilayer temporal networks. The original contribution is threefold. First, to avoid over-fitting we propose a parsimonious parametrization based on a low-rank decomposition of the tensor of regression coefficients. Second, we assume the parameters are driven by a hidden Markov chain, thus allowing for structural changes in the network topology. We follow a Bayesian approach to inference and provide an efficient Gibbs sampler for posterior approximation. We apply the methodology to a real dataset of financial networks to study the impact of several risk factors on the edge probability.
Supplementary materials for this article are available online.
\end{abstract}

\noindent%
{\it Keywords:} multidimensional data; sparsity; nonlinear time series; zero-inflated logit
% 3 to 5 keywords, that do not appear in the title
\vfill
%\textbf{AMS 2000 subject classifications:} Primary 62; secondary 91B84.
%\textbf{JEL Classification:} C13, C33, C51, C53

\newpage
\spacingset{1.5} % DON'T change the spacing!
%%%%%%%%%%%%%%%%%%%%%%%%%%%%%%%%%%%%%%%%%%%%%%%%%%%%%%%%%%%%%%%%%%%%%%%%%%%%%%

\section{Introduction}
The analysis of large sets of binary data is a central issue in many fields such as biostatistics (\cite{Schildcrout05Longitudinal_Binary},~\cite{Wilbur02multivariate_Binary_regression}), image processing (\cite{Yue12Binary_neuroimaging}), machine learning (\cite{Banerjee08sparse_ML_binary},~\cite{Koh07logistic_l1regularized}), medicine (\cite{Christakis08socialNetwork_medicine}), text analysis (\cite{Taddy13multinomial_text}) and statistics (\cite{Ravikumar10Ising_logistic_l1regularized},~\cite{Sherman06Binary_autologistic_spatial},~\cite{Visaya15multivariate_Binary_longitudinal}).

In this paper we consider binary series representing the edge activation in time-varying (\cite{Holme12TemporalNetworks}) and multilayer networks (\cite{Boccaletti14Multilayer_Networks}).
The application focuses on financial networks since, in spite of the wide literature on theoretical models (e.g.~\cite{Acemoglu12Network_AggregateFluctuations},~\cite{Chaney14aer_InternTradeNetwork},~\cite{Mele17ecta_FormationDenseNetwork},~\cite{Graham17ecta_FormationNetwork_degree_heterogeneity}), the statistical analysis of their dynamical properties is still at its infancy (e.g., \cite{Billioetal12GrangerNet} and \cite{Diebold14NetworkTopology}). The study of temporal networks is very interdisciplinary and we expect our statistical framework to be of interest for many disciplines.

The first issue in building a dynamic network model concerns the impact of covariates on the dynamic process of link formation. We propose a parsimonious model that can be successfully used to this aim, relying on tensors and their decompositions. See~\cite{KoldaBader09},~\cite{Cichocki15Tensor_Multiway_Analysis} and~\cite{Cichocki16Tensor_theory} for a review. The main advantage in using tensors is the possibility of dealing with the complexity of novel data structures which are becoming increasingly available, such as networks, multilayer networks, three-way tables, spatial panels with multiple series observed for each unit (e.g., municipalities, regions, countries).
The use of tensor algebra has the advantage of preventing data reshape and manipulation, and preserving data intrinsic structure. Another advantage of tensors stems from the decompositions and approximations, which provide representations in lower dimensional spaces (see ch.7-8 of\cite{Hackbusch12Tensor_book}). In this paper, we exploit the parallel factor (PARAFAC) decomposition for reducing the number of parameters to estimate, thus making inference on network models feasible.

Another issue in network modelling regards the time variation of the network topology.
For example, structural breaks have been detected by~\cite{Billioetal12GrangerNet},~\cite{Ahelegbey16BStuctVAR} and~\cite{Bianchi19GraphicalSUR} in contagion networks and \cite{Giraitis16DynamicNetwork_estimating_Financial} found evidence of link persistence in interbank networks.
Starting from these stylized facts, we propose a new Markov switching model for capturing structural changes in temporal networks.
After~\cite{Hamilton89MS}, the Markov switching dynamics has been used in several time series models, such as VARs (\cite{Sims08LargeMarkovSwitch}), factor models (\cite{Kim98MarkovSwitch_Factor}), dynamic panels (\cite{Kaufmann15MSwitch_timevarying_transitions}), stochastic volatility (\cite{Chib02MCMC_MarkovSwitch_StochVol}), ARCH and GARCH (\cite{Haas04MarkovSwitch_GARCH}) and stochastic correlation (\cite{Casarin18BayesMS_DynCorrelation}).
See \cite{Fruhwirth06FiniteMixtures_MarkovSwitch_book} for an introduction to Markov switching models. We contribute to this literature by applying Markov switching to tensor valued data.

Many real world temporal networks exhibit sparsity (\cite{Newman10Networks_book}) and sudden abrupt changes in the sparsity level across time. See also~\cite{Ahelegbey16SparseGVAR} for an empirical evidence on financial networks.
Motivated by this observation, we propose a zero-inflated logit regression for the edge activation and allow for Markov switching sparsity levels.
We contribute to the statistics literature on models for network data (\cite{DuranteDunson14LogitDynamicNetwork_GP},~\cite{WangDuranteDunson17BayesLogitNetwork},~\cite{Carvalho08Sparse_Factor_gene_network}, \cite{Chen18Bayes_Dynamic_Network},~\cite{Berry19Bayes_Count_Network},~\cite{Snijders10MLE_Network_Dynamic},~\cite{Kolar10Estimate_Temporal_Nets}) and matrix-valued data (\cite{Windle14StateSpace_matrices},~\cite{Carvalho07Dynamic_Matrix_Graphical}) by proposing a nonlinear model for sparse tensor-valued data.

The remainder of this paper is organized as follows. Section~\ref{sec:model} presents the model. Sections~\ref{sec:inference}-\ref{sec:posterior_approx} discuss the Bayesian inference procedure. Section~\ref{sec:application} provides an application to financial network data. Concluding remarks are given in Section~\ref{sec:conclusions}. Further details and results are provided in the supplementary material.

\section{A Markov Switching Model for Networks} \label{sec:model}
Relevant objects in our modelling framework are $D$-order tensors $\mathcal{X} \in \R^{d_1\times\ldots\times d_D}$ of size $(d_1\times\ldots\times d_D)$, that are $D$-dimensional arrays, elements of the tensor product of $D$ vector spaces, each one endowed with a coordinate system. See~\cite{Hackbusch12Tensor_book} for an introduction to tensor spaces. A tensor can be though of as the multidimensional extension of a matrix (i.e., a $2$-order tensor), where each dimension is called mode. Other objects of interest in this paper are tensor slices, i.e. matrices obtained by fixing all but two of the indices of the array, and tensor fibers, i.e. vectors resulting from keeping fixed all indices but one. See \autoref{sec:apdx_tensor} for some background material on tensors.

Tensors are particularly useful for representing multilayer temporal networks (\cite{Boccaletti14Multilayer_Networks} and~\cite{Kivela14Multilayer_Networks}). Let $G_t = (V_1,V_2,M,E_t)$ be a multilayer temporal network, where $V_1 = \lbrace 1,\ldots,I \rbrace$, $V_2 = \lbrace 1,\ldots,J \rbrace$ are two vertex sets, $M = \lbrace 1,\ldots,K \rbrace$ is the set of layers and $E_t \subset (V_1 \times V_2 \times M)$ is the edge set at time $t=1,\ldots,T$. The network connectivity can be encoded in a $4$-order tensor $\mathcal{X}$ of size $(I\times J\times K\times T)$, with entries
\begin{equation}
x_{ijk,t} = \Bigg\{ \begin{array}{cc}
1 & \text{if } \lbrace i,j,k \rbrace \in E_t \\
0 & \text{if } \lbrace i,j,k \rbrace \notin E_t.
\end{array}
\end{equation}
This definition is general enough to include undirected and directed networks, and undirected bipartite networks. It can be further extended to account for other types of networks (\cite{Kivela14Multilayer_Networks}).
One of the most recurrent features of observed networks is sparsity. In random graph theory sparsity is defined asymptotically as the feature of a network where the number of edges grows subquadratically with the number of nodes \cite[see][ch.7]{Diestel12GraphTheory}.
In finite graphs, sparsity occurs when there is an excess of zeros in the connectivity tensor, that is, when the degree distribution has a peak at $0$.

To describe network sparsity we assume that the probability of observing an edge in each layer of the network is a mixture of a Dirac mass at $0$ and a Bernoulli distribution. Since the sparsity pattern in many real networks is not time homogeneous, we assume that both the mixing and the Bernoulli probabilities are time-varying. Finally, a logistic regression is assumed to include covariates. In summary, for each entry $x_{ijk,t}$ of the tensor $\mathcal{X}_t$ (that is, each edge of the corresponding network) we assume a zero-inflated logit regression model
\begin{equation}
\begin{split}
x_{ijk,t}|\rho(t),\mathbf{g}_{ijk}(t) & \sim \rho(t) \delta_{\lbrace 0 \rbrace}(x_{ijk,t}) + (1-\rho(t))\delta_{\lbrace d_{ijk,t} \rbrace}(x_{ijk,t}) \\
d_{ijk,t} & = \I_{\R_+} (x_{ijk,t}^*) \\
x_{ijk,t}^* & = \mathbf{z}_{ijk,t}' \mathbf{g}_{ijk}(t) + \varepsilon_{ijk,t} \qquad \varepsilon_{ijk,t} \distas{iid} \text{Logistic}(0,1).
\label{eq:model_all_xijkt}
\end{split}
\end{equation}
where $\mathbf{z}_{ijk,t}\in\R^Q$ is a vector of edge-specific covariates and $\mathbf{g}_{ijk}(t) \in \R^Q$ is a time-varying edge-specific vector of parameters and $\rho(t)$ is the time-varying probability of excess of zeros in the network. Without loss of generality, we assume the set of covariates is common to all edges, i.e. $\mathbf{z}_{ijk,t} = \mathbf{z}_t$.
The specification of the model is completed with the assumption that the parameters $\rho(t)$ and $\mathbf{g}_{ijk}(t)$ are driven by a hidden Markov chain $\lbrace s_t \rbrace_{t=1}^T$ with finite state space $\lbrace 1,\ldots,L \rbrace$, that is $\rho(t) = \rho_{s_t}$ and $\mathbf{g}_{ijk}(t) = \mathbf{g}_{ijk,s_t}$. The transition matrix of the chain is assumed to be time-invariant and denoted by $\boldsymbol{\Xi} = (\boldsymbol{\xi}_{1}',\ldots,\boldsymbol{\xi}_{L}')'$, where $\boldsymbol{\xi}_{l} = (\xi_{l,1},\ldots,\xi_{l,L})$ is a probability vector and $\xi_{i,j} = p(s_t=j | s_{t-1}=i)$ is the transition probability from state $i$ to state $j$.

By integrating out $x_{ijk,t}^*$ in eq. \eqref{eq:model_all_xijkt}, we obtain the regime-specific probabilities of observing an edge from $i$ to $j$ in the layer $k$
\begin{align}
p(x_{ijk,t} = 1 | \rho_l,\mathbf{g}_{ijk,l}) & = (1-\rho_l) \frac{\exp( \mathbf{z}_t' \mathbf{g}_{ijk,l})}{1+\exp( \mathbf{z}_t' \mathbf{g}_{ijk,l})} \\
p(x_{ijk,t} = 0 | \rho_l,\mathbf{g}_{ijk,l}) & = \rho_l + (1-\rho_l) \bigg( 1-\frac{\exp( \mathbf{z}_t' \mathbf{g}_{ijk,l})}{1+\exp( \mathbf{z}_t' \mathbf{g}_{ijk,l})} \bigg).
\end{align}
For the ease of notation, we provide a compact representation of the general model. First, we define $\mathbb{X}^{d} = \lbrace \mathcal{X}\in \R^{i_1\times\ldots\times i_d} \rbrace$ the set of real valued $d$-order tensors of size $(i_1\times\ldots\times i_d)$, $\mathbb{X}_{0,1}^{d} = \lbrace  \mathcal{X}\in \R^{i_1\times\ldots\times i_d} : \mathcal{X}_{i_1,\ldots,i_d} \in \lbrace 0,1 \rbrace \rbrace \subset \mathbb{X}^{d}$ the set of adjacency tensors of size $(i_1\times\ldots\times i_d)$, and $\Psi : \mathbb{X}^{d} \rightarrow \mathbb{X}_{0,1}^{d}$ a linear operator such that $\mathcal{X}^* \mapsto \Psi(\mathcal{X}^*) \in \lbrace 0,1 \rbrace^{i_1\times\ldots\times i_d}$. 
For a tensor $\mathcal{X}_t^*$ with $k$-th slice $\mathbf{X}_{k,t}^* \in\mathcal{X}^{I,J}$ it is possible to write the model in tensor form by $\Psi(\mathbf{X}_{k,t}^*) = (\I_{\R_+}(x_{ijk,t}^*))_{i,j}$, where $\I_A (x)$ is the indicator function, which takes value $1$ if $x\in A$ and $0$ otherwise.

Second, we define the mode-$n$ product between a $D$-order tensor $\mathcal{X}\in\R^{d_1\times\ldots\times d_D}$ and a vector $\mathbf{v}\in\R^{d_n}$, as a $(D-1)$-order tensor $\mathcal{Y}\in\R^{d_1\times\ldots\times d_{n-1}\times d_{n+1}\times\ldots\times d_D}$ whose entries are
\begin{equation}
\mathcal{Y}_{(i_1,\ldots,i_{n-1},i_{n+1},\ldots,i_D)} = (\mathcal{X} \times_n \mathbf{v})_{(i_1,\ldots,i_{n-1},i_{n+1},\ldots,i_D)} = \sum_{i_n=1}^{d_n} \mathcal{X}_{i_1,\ldots,i_n,\ldots,i_D} \mathbf{v}_{i_n} \, .
\label{eq:tensor_moden_vector}
\end{equation}

By collecting the coefficients $\mathbf{g}_{ijk}(t)$ along the indices $i,j,k$ in a $4$-order tensor $\mathcal{G}(t) \in \R^{I\times J \times K\times Q}$, we can rewrite eq.~\eqref{eq:model_all_xijkt} in the compact form:
\begin{equation}
\Bigg\lbrace
\begin{array}{ll}
\mathcal{X}_t = \mathcal{B}(t) \odot \Psi(\mathcal{X}_t^*) & \qquad b_{ijk}(t) \distas{iid} \mathcal{B}ern(1-\rho(t))  \\
\mathcal{X}^*_{t} = \mathcal{G}(t) \times_4 \mathbf{z}_t + \mathcal{E}_t & \qquad \varepsilon_{ijk,t} \distas{iid} \text{Logistic}(0,1)
\end{array}
\label{eq:model_compact_first}
\end{equation}
where $\mathcal{B}(t) \in \lbrace 0,1 \rbrace^{I\times J\times K}$ and $\mathcal{E}_t \in \R^{I\times J\times K}$ are tensors of the same size of $\mathcal{X}_t$, with entries $b_{ijk}(t)$ and $\varepsilon_{ijk,t}$, respectively, and the symbol $\odot$ is the Hadamard product (\cite{KoldaBader09}). Matrix operations and results from linear algebra can be generalized to tensors (see~\cite{Hackbusch12Tensor_book}, \cite{Kroonenberg08AppliedMultiwayDataAnalysis}).
This model is closely related to a switching regression representation (see Ch. 8 of \cite{Fruhwirth06FiniteMixtures_MarkovSwitch_book}) which can be used to carry out inference simultaneously for all coefficient tensors.
By introducing a dummy coding for $s_t$ through $L$ binary variables $\zeta_{t,l} = \I_{\lbrace l \rbrace}(s_t)$, $l=1,\ldots,L$, model \eqref{eq:model_compact_first} is written as
\begin{equation}
\begin{cases}
\mathcal{X}_t = \mathcal{B}(t) \odot \Psi(\mathcal{X}_t^*) & \quad b_{ijk}(t) \distas{iid} \mathcal{B}ern(1-\rho(t)) \\[2pt]
\mathcal{X}_t^* = \mathcal{G} \times_4 (\boldsymbol{\zeta}_t \otimes \tilde{\mathbf{z}}_t)' + \mathcal{E}_t = \mathcal{G} \times_4 (\boldsymbol{\zeta}_t, \boldsymbol{\zeta}_t \otimes \mathbf{z}_t)' + \mathcal{E}_t  &  \quad \varepsilon_{ijk,t} \distas{iid} \text{Logistic}(0,1)  \\[2pt]
\boldsymbol{\zeta}_{t+1} = \boldsymbol{\Xi} \boldsymbol{\zeta}_t + \tilde{\mathbf{u}}_t   &  \quad \mathbb{E}[\tilde{\mathbf{u}}_t | \tilde{\mathbf{u}}_{t-1}] =0 
\end{cases}
\label{eq:model_factor_statespace_form}
\end{equation}
which is a switching SUR (\cite{Zellner62SUR}, \cite{Bianchi19GraphicalSUR}), where $\otimes$ denotes the Kronecker product, $\lbrace \tilde{\mathbf{u}}_t \rbrace_t$ is a martingale difference process, $\tilde{\mathbf{z}}_t = (1, \mathbf{z}_t)'$ and $\boldsymbol{\zeta}_t = (\zeta_{t,1}, \ldots, \zeta_{t,L})'$.

We propose a parsimonious parametrisation of the model by exploiting tensor representations (see~\cite{KoldaBader09} for a review). In particular we assume a PARAFAC decomposition with fixed rank $R$ for the tensor $\mathcal{G}(t) = \mathcal{G}_{s_t}$:
\begin{equation}
\mathcal{G}(t) = \sum_{r=1}^R \boldsymbol{\gamma}_{1}^{(r)}(t) \circ \boldsymbol{\gamma}_{2}^{(r)}(t) \circ \boldsymbol{\gamma}_{3}^{(r)}(t) \circ \boldsymbol{\gamma}_{4}^{(r)}(t) \, ,
\label{eq:CP_decomposition}
\end{equation}
where the vectors $\boldsymbol{\gamma}_{h}^{(r)}(t) = \boldsymbol{\gamma}_{h,s_t}^{(r)}$, $h=1,\ldots,4$, $r=1,\ldots,R$, are called the marginals of the PARAFAC decomposition and have length $I$, $J$, $K$ and $Q$, respectively. See \autoref{sec:apdx_tensor} and the supplement for further details. This specification permits us to: (i) achieve parsimony of the model, since for each value of the state $s_t$ the dimension of the parametric space is reduced from $IJKQ$ to $R(I+J+K+Q)$; (ii) introduce sparsity in the coefficient tensor, through a suitable choice of the prior distribution for the PARAFAC marginals.

\section{Bayesian Inference} \label{sec:inference}
As regards the prior distributions for the parameters of interest, we choose the following specifications. We assume a global-local shrinkage prior for on $\boldsymbol{\gamma}_{h,l}^{(r)}$
\begin{equation}
p(\boldsymbol{\gamma}_{h,l}^{(r)} | \bar{\boldsymbol{\zeta}}_{h,l}^r, \tau, \phi_r, w_{h,r,l}) \sim \mathcal{N}_{n_h}(\bar{\boldsymbol{\zeta}}_{h,l}^r, \: \tau \phi_r w_{h,r,l} \mathbf{I}_{n_h})
\label{eq:prior_gammas}
\end{equation}
for $r=1,\ldots,R$, each $h=1,\ldots,4$ and each $l=1,\ldots,L$, where $n_1= I$, $n_2=J$, $n_3=K$, $n_4=Q$. The parameter $\tau$ represents the global component of the variance, common to all marginals, $\phi_r$ is the level component and $w_{h,r}$ is the local component. 
The choice of a global-local shrinkage prior, as opposed to a spike-and-slab distribution, is motivated by the reduced computational complexity and the capacity to handle high-dimensional settings. In what follows we denote with $p(\mathcal{G}|\mathcal{W},\boldsymbol{\phi},\tau)$ the joint prior of the $\boldsymbol{\gamma}_{h,l}^{(r)}$, where $\mathcal{W} = \lbrace w_{h,r,l} \rbrace_{h,r,l}$.
We assume the following hyperpriors for the variance components\footnote{We use the shape-rate formulation for the gamma distribution, such that $\mathbb{E}(x) = \alpha/\beta$, $Var(x)=\alpha/\beta^2$.}:
\begin{align}
\label{eq:prior_tau}
p(\tau) & \sim \mathcal{G}a(\bar{a}^\tau, \bar{b}^\tau) \qquad \bar{a}^\tau = \bar{\alpha} R \\\label{eq:prior_phi}
p(\boldsymbol{\phi}) & \sim \mathcal{D}ir(\bar{\boldsymbol{\alpha}}) \quad \qquad \bar{\boldsymbol{\alpha}} = \bar{\alpha}\boldsymbol{\iota}_R \\
\label{eq:prior_w}
p(w_{h,r,l}|\lambda_l) & \sim \mathcal{E}xp(\lambda_l^2/2) \qquad \forall \, h,r,l \\
\label{eq:prior_lambda}
p(\lambda_l) & \sim \mathcal{G}a(\bar{a}_l^\lambda,\bar{b}_l^\lambda) \qquad \forall \, l \, ,
\end{align}
where $\boldsymbol{\iota}_n$ is the $n$-dimensional vector of ones. 
The further level of hierarchy for the local components $w_{h,r,l}$ is added with the aim of favouring information sharing across local components of the variance (indices $h$ and $r$) within a given regime $l$.
%This hierarchical prior induces the following marginal prior on the vector $\mathbf{w}_l = (w_{1,1,l},\ldots,w_{4,R,l})'$:
%\begin{align}
%\notag
%p(\mathbf{w}_l) & = \int_{\R_+} \prod_{r=1}^R \prod_{h=1}^4 p(w_{h,r,l}|\lambda_l) p(\lambda_l) \: \mathrm{d}\lambda_l \\
% & = \int_{\R_+} \frac{(\bar{b}_l^\lambda)^{\bar{a}_l^\lambda}}{2\Gamma(\bar{a}_l^\lambda)} \lambda_l^{\bar{a}_l^\lambda + 8R - 1} \exp\left( -\bar{b}_l^\lambda \lambda_l - \left( \sum_{r=1}^R \sum_{h=1}^4 w_{h,r,l} \right) \frac{\lambda_l^2}{2} \right) \: \mathrm{d}\lambda_l \, .
%\label{eq:prior_joint_marginal_w}
%\end{align}
%The marginal prior for a generic entry $w_{h,r,l}$ is a compound gamma distribution, that is $p(w_{h,r,l}) \sim \text{CoGa}(1,\bar{a}_l^\lambda,1,\bar{b}_l^\lambda)$, with $\bar{a}_l^\lambda > -1$. In the univariate case (i.e $H=1$, $R=1$ and $L=1$), we obtain a generalized Pareto distribution $p(w) = gP(0,a_\lambda, b_\lambda/a_\lambda)$.
The specification of an exponential distribution for the local component of the variance of the $\boldsymbol{\gamma}_{h,l}^{(r)}$ yields a Laplace (or Double Exponential) distribution for each component of the vectors once the $w_{h,r,l}$ is integrated out, that is $\boldsymbol{\gamma}_{h,l,i}^{(r)}|\lambda_l,\tau,\phi_r \sim \text{Laplace}(0,\lambda_l/\sqrt{\tau\phi_r})$ for all $i=1,\ldots,n_h$. The marginal distribution of each entry, integrating all remaining random components, is a generalized Pareto distribution, which favours sparsity.

In logit models it is not possible to identify the coefficients of the latent regression equation as well as the variance of the noise. As a consequence, we make the usual identifying restriction by imposing unitary variance for each $\varepsilon_{ijk,t}$.

\begin{figure}[t]
\centering
\includegraphics[trim= 0mm 0mm 0mm 0mm,clip,scale= 1.00]{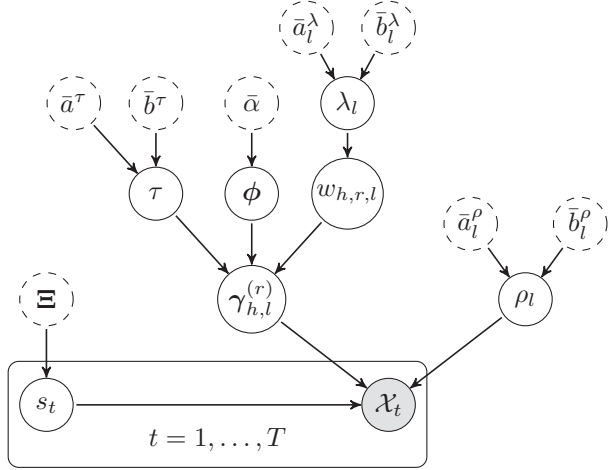}
\caption{Directed acyclic graph of the model in eq.~\eqref{eq:model_compact_first} and prior structure in eq.~\eqref{eq:prior_gammas}-\eqref{eq:prior_xi}. Gray circles denote observable variables, white solid circles indicate parameters, white dashed circles indicate fixed hyperparameters. Directed edges represent the conditional independence relationships.}
\label{fig:prior}
\end{figure}

The mixing probability of the observation model is assumed beta distributed:
\begin{equation}
p(\rho_l) \sim \mathcal{B}e(\bar{a}_l^\rho, \bar{b}_l^\rho) \qquad \forall \, l \, .
\label{eq:prior_rho}
\end{equation}
A well known identification issue for mixture models is the label switching problem (e.g., see \cite{Fruhwirth01LabelSwitch_PermutationSampler}). When the specific application provides meaningful restrictions on the value of some parameters (e.g., from theory, or interpretation), they can be used for identifying the regimes. Following this approach, we assume $\rho_1 > \rho_2 > \ldots > \rho_L$, meaning that regime $1$ represents the sparsest and regime $L$ the densest.
Finally, we assume each row of the transition matrix $\boldsymbol{\xi}_l$ follows a Dirichlet distribution
\begin{equation}
p(\boldsymbol{\xi}_{l}) \sim \mathcal{D}ir(\bar{\mathbf{c}}_{l}) \quad \forall \, l \, .
\label{eq:prior_xi}
\end{equation}
The overall structure of the hierarchical prior distribution is represented graphically by means of the directed acyclic graph in Fig.~\ref{fig:prior}.

%%%%%%%%%%%%%%%%%%%%%%%%%%%%%%%%%%%%%%%%%%%%%%%%%%%%%%%%%%%%%%%%%%%%

\section{Posterior Approximation} \label{sec:posterior_approx}
Since the joint posterior is not tractable, we apply Markov chain Monte Carlo (MCMC) combined with a data augmentation strategy (\cite{Tanner87DataAugmentation}). We introduce allocation variables for the mixture in eq.~\eqref{eq:model_all_xijkt} and the P\'olya-Gamma augmentation of~\cite{Polsonetal13PolyaGamma}, which allows for conjugate full conditional distributions and a better mixing of the MCMC chain.
See also~\cite{WangDuranteDunson17BayesLogitNetwork} and~\cite{Holsclaw17PolyaGamma_NotHomogMarkovmodel} for an application of the P\'olya-Gamma scheme to network-response regression and hidden Markov models, respectively.
Define $\boldsymbol{\mathcal{X}} = \lbrace \mathcal{X}_t \rbrace_{t=1}^T$, $\mathbf{s} = \lbrace s_t \rbrace_{t=0}^T$ and let $\boldsymbol{\theta}$ denote the set of parameters. For each $l=1,\ldots,L$, we define $\mathcal{T}_l = \lbrace t : \zeta_{t,l}=1 \rbrace$ and $T_l = \#\mathcal{T}_l$.
The data augmented likelihood of the model in eq. \eqref{eq:model_compact_first} is
\begin{align}
L(\boldsymbol{\mathcal{X}},\mathbf{s}|\boldsymbol{\theta}) & = L(\boldsymbol{\mathcal{X}}|\mathbf{s}, \boldsymbol{\theta}) L(\mathbf{s} | \boldsymbol{\theta}),
\label{eq:likelihood_X_y_state_1}
\end{align}
where
\begin{align}
L(\boldsymbol{\mathcal{X}} | \mathbf{s},\boldsymbol{\theta}) & = \prod_{l=1}^L \prod_{t\in \mathcal{T}_l} \prod_{i=1}^I \prod_{j=1}^J \prod_{k=1}^K \bigg( \frac{(1-\rho_l) \exp( \mathbf{z}_t' \mathbf{g}_{ijk,l})}{1+\exp( \mathbf{z}_t' \mathbf{g}_{ijk,l})} \bigg)^{x_{ijk,t}} \bigg( \rho_l + \frac{1-\rho_l}{1+\exp( \mathbf{z}_t' \mathbf{g}_{ijk,l})} \bigg)^{1-x_{ijk,t}}
\label{eq:likelihood_X_y_state_2}
\end{align}
and
\begin{align}
L(\mathbf{s}|\boldsymbol{\theta}) & = \prod_{g=1}^L \prod_{l=1}^L \xi_{g,l}^{N_{gl}(\mathbf{s})},
\label{eq:likelihood_X_y_state_3}
\end{align}
%\begin{align}
%\notag
%L(\boldsymbol{\mathcal{X}},\mathbf{s}|\boldsymbol{\theta}) & = \prod_{l=1}^L \prod_{t\in \mathcal{T}_l} \prod_{i=1}^I \prod_{j=1}^J \prod_{k=1}^K \bigg( \frac{(1-\rho_l) \exp( \mathbf{z}_t' \mathbf{g}_{ijk,l})}{1+\exp( \mathbf{z}_t' \mathbf{g}_{ijk,l})} \bigg)^{x_{ijk,t}} \bigg( \rho_l + \frac{1-\rho_l}{1+\exp( \mathbf{z}_t' \mathbf{g}_{ijk,l})} \bigg)^{1-x_{ijk,t}} \\
% & \quad \cdot \bigg( \prod_{g=1}^L \prod_{l=1}^L \xi_{g,l}^{N_{gl}(\mathbf{s})} \bigg),
%\label{eq:likelihood_X_y_state}
%\end{align}
with $N_{gl}(\mathbf{s}) = \# \lbrace \zeta_{t-1,g}=1, \zeta_{t,l}=1, \; t=1,\ldots,T \rbrace$, $g,l=1,\ldots,L$, with $\#$ the cardinality of a set.
To make the likelihood more tractable, we further augment the data in two steps. First, we introduce the latent allocation variable for the mixture in eq.~\eqref{eq:model_all_xijkt}, $d_{ijk,t} \in \lbrace 0,1 \rbrace$, and obtain the conditional distribution
\begin{align}
p(x_{ijk,t}|d_{ijk,t},s_t=l,\mathcal{G}_l) = \big( \delta_{\lbrace 0 \rbrace}(x_{ijk,t}) \big)^{d_{ijk,t}} \frac{ \big( \exp(\mathbf{z}_t' \mathbf{g}_{ijk,l}) \big)^{x_{ijk,t}(1-d_{ijk,t})}}{ \big( 1+\exp( \mathbf{z}_t' \mathbf{g}_{ijk,l}) \big)^{(1-d_{ijk,t})}} \, ,
\label{eq:likelihood_cond_X_d}
\end{align}
and the marginal distribution
\begin{equation}
p(d_{ijk,t}|s_t=l,\rho_l) = \rho_l^{d_{ijk,t}} (1-\rho_l)^{1-d_{ijk,t}} \, .
\label{eq:marginal_d}
\end{equation}
Second, we decompose the ratio in eq.~\eqref{eq:likelihood_cond_X_d} and obtain
\begin{align}
p(x_{ijk,t} & | d_{ijk,t},\omega_{ijk,t},s_t=l,\mathcal{G}_l) = \frac{\big( 2\delta_{\lbrace 0 \rbrace}(x_{ijk,t}) \big)^{d_{ijk,t}}}{2} \exp\Big( -\frac{\omega_{ijk,t}}{2}(\mathbf{z}_t' \mathbf{g}_{ijk,l})^2 +\kappa_{ijk,t}(\mathbf{z}_t' \mathbf{g}_{ijk,l}) \Big),
\label{eq:likelihood_conditional_X_d_omega}
\end{align}
where $\kappa_{ijk,t} = (1-d_{ijk,t}) (x_{ijk,t} - 1/2)$ and $\omega_{ijk,t} \sim PG(1,0)$, with $PG(b,c)$ the P\'olya-Gamma distribution with parameters $b >0$ and $c \in \R$ \cite[][Theorem 1]{Polsonetal13PolyaGamma}.
Defining $\mathcal{D} = \lbrace d_{ijk,t} \rbrace_{ijkt}$ and $\boldsymbol{\Omega} = \lbrace \omega_{ijk,t} \rbrace_{ijkt}$ and combining the previous steps one gets the complete data likelihood
\begin{align}
\notag
 & L(\boldsymbol{\mathcal{X}},\mathcal{D},\boldsymbol{\Omega},\mathbf{s}|\boldsymbol{\theta}) = \bigg( \prod_{t=1}^T \prod_{i=1}^I \prod_{j=1}^J \prod_{k=1}^K p(\omega_{ijk,t}) \bigg) \cdot \bigg( \prod_{g=1}^L \prod_{l=1}^L \xi_{g,l}^{N_{gl}(\mathbf{s})} \bigg) \\
 & \quad \cdot \prod_{l=1}^L \prod_{t\in \mathcal{T}_l} \prod_{i=1}^I \prod_{j=1}^J \prod_{k=1}^K \bigg( \frac{2\rho_l \delta_{\lbrace 0 \rbrace}(x_{ijk,t})}{1-\rho_l} \bigg)^{d_{ijk,t}} \frac{1-\rho_l}{2} \exp\Big( -\dfrac{\omega_{ijk,t}}{2}(\mathbf{z}_t' \mathbf{g}_{ijk,l})^2 +\kappa_{ijk,t}(\mathbf{z}_t' \mathbf{g}_{ijk,l}) \Big).
\label{eq:complete_likelihood_final}
\end{align}
In the following, we define $\mathcal{G} = \lbrace \mathcal{G}_l \rbrace_{l=1}^L$ and $\boldsymbol{\rho} = \lbrace \rho_l \rbrace_{l=1}^L$, and let $\mathbf{W}_l$ and $\mathbf{W}^{(r)}$ be the $(4\times R)$ and $(4\times L)$ matrices representing the $l$- and $r$-th slices of $\mathcal{W}$, along the third and second mode, respectively.
The complete data likelihood and the prior distributions yield a posterior sampling scheme consisting of four blocks (see the supplement for the derivation of the posterior full conditional distributions).

In block (I) the sampler draws the latent variables from the full conditional distribution:
\begin{align}
p(\mathbf{s},\mathcal{D},\boldsymbol{\Omega} | \boldsymbol{\mathcal{X}}, \mathcal{G}, \boldsymbol{\Xi}, \boldsymbol{\rho}) & = p( \mathbf{s} | \boldsymbol{\mathcal{X}}, \mathcal{G}, \boldsymbol{\Xi}, \boldsymbol{\rho}) p( \mathcal{D} | \boldsymbol{\mathcal{X}}, \mathcal{G}, \boldsymbol{\rho}, \mathbf{s}) p( \boldsymbol{\Omega} | \boldsymbol{\mathcal{X}}, \mathcal{G}, \boldsymbol{\rho}, \mathbf{s}).
%& \cdot \prod_{ijkt} p( \omega_{ijk,t} | x_{ijk,t}, s_t, \mathcal{G}_{s_t}) p( d_{ijk,t} | x_{ijk,t}, s_t, \mathcal{G}_{s_t}, \rho_{s_t}).
\end{align}
Samples of $\mathbf{s}$ are drawn via the Forward Filter Backward Sampler (see ch.13 of \cite{Fruhwirth06FiniteMixtures_MarkovSwitch_book}). The latent variables $\omega_{ijk,t}$ are sampled independently from
\begin{equation}
p( \omega_{ijk,t} | x_{ijk,t}, s_t, \mathcal{G}_{s_t}) \propto PG(1, \mathbf{z}_t' \mathbf{g}_{ijkq,s_t}).
\label{eq:posterior_omega}
\end{equation}
The latent variables $\omega_{ijk,t}$ are sampled in block for each $t$. The latent variables $d_{ijk,t}$ are sampled independently from
\begin{equation}
\begin{split}
p( d_{ijk,t}=1 | x_{ijk,t}, s_t, \mathcal{G}_{s_t}, \rho_{s_t}) & \propto \rho_{s_t} \delta_{\lbrace 0 \rbrace}(x_{ijk,t}) \\
p( d_{ijk,t}=0 | x_{ijk,t}, s_t, \mathcal{G}_{s_t}, \rho_{s_t}) & \propto (1-\rho_{s_t}) \frac{\exp( (\mathbf{z}_t' \mathbf{g}_{ijkq,s_t})x_{ijk,t})}{1+\exp( \mathbf{z}_t' \mathbf{g}_{ijkq,s_t} )} \, .
\end{split}
\label{eq:posterior_d}
\end{equation}
The hyperparameters which control the variance of the PARAFAC marginals are sampled in block (II) from the full conditional distribution
\begin{equation}
p(\tau, \boldsymbol{\phi}, \mathcal{W} | \lbrace \boldsymbol{\gamma}_{h,l}^{(r)} \rbrace_{h,l,r}) = p( \boldsymbol{\phi} | \lbrace \boldsymbol{\gamma}_{h,l}^{(r)} \rbrace_{h,l,r}, \mathcal{W}) p( \tau | \lbrace \boldsymbol{\gamma}_{h,l}^{(r)} \rbrace_{h,l,r}, \mathcal{W}, \boldsymbol{\phi}) p( \mathcal{W} | \lbrace \boldsymbol{\gamma}_{h,l}^{(r)} \rbrace_{h,l,r}, \boldsymbol{\phi}, \tau).
\end{equation}
We enable better mixing by blocking together the parameters $\boldsymbol{\phi}$. We set $\phi_r = \psi_r / (\psi_1 + \ldots + \psi_R)$, where the auxiliary variables $\psi_r$ are sampled independently for each $r$ from
\begin{equation}
p( \psi_r | \lbrace \boldsymbol{\gamma}_{h,1}^{(r)} \rbrace_{h,l}, \mathbf{W}^{(r)}) \propto \text{GiG} \Big( 2\bar{b}^\tau, \sum_{h=1}^4 \sum_{l=1}^L \frac{\boldsymbol{\gamma}_{h,l}^{(r)\prime} \boldsymbol{\gamma}_{h,l}^{(r)}}{w_{h,r,l}}, \bar{\alpha}-n \Big),
\label{eq:posterior_psir}
\end{equation}
where $\textnormal{GiG}(a,b,p)$ is Generalized Inverse Gaussian distribution with parameters $p \in \R$, $a >0$ and $b >0$, and $n=\sum_{h=1}^4 n_h$.
The global variance parameter $\tau$ is drawn from
\begin{equation}
p( \tau | \lbrace \boldsymbol{\gamma}_{h,l}^{(r)} \rbrace_{h,l,r}, \mathcal{W}, \boldsymbol{\phi}) \propto \text{GiG} \Big( 2\bar{b}^\tau, \sum_{r=1}^R \sum_{h=1}^4 \sum_{l=1}^L \frac{\boldsymbol{\gamma}_{h,l}^{(r)\prime} \boldsymbol{\gamma}_{h,l}^{(r)}}{\phi_r w_{h,r,l}}, (\bar{\alpha}-n)R \Big).
\label{eq:posterior_tau}
\end{equation}
The local variance parameters $w_{h,r,l}$ are independently drawn from
\begin{equation}
p( w_{h,r,l} | \boldsymbol{\gamma}_{h,l}^{(r)}, \phi_r, \tau, \lambda_l) \propto \text{GiG} \Big( \lambda_l^2, \frac{\boldsymbol{\gamma}_{h,l}^{(r)\prime} \boldsymbol{\gamma}_{h,l}^{(r)}}{\tau \phi_r}, 1-\frac{n_h}{2} \Big).
\label{eq:posterior_w}
\end{equation}
Finally, the hyperparameters $\lambda_l$ are independently drawn from
\begin{equation}
p( \lambda_l | \mathbf{W}_l ) \propto \lambda_l^{\bar{a}_l^\lambda +8R -1} \exp\Big( -\lambda_l \bar{b}_l^\lambda -\frac{\lambda_l^2}{2}\sum_{r=1}^R \sum_{h=1}^4 w_{h,r,l} \Big). 
%\exp\Bigg( -\lambda_l \Big( \bar{b}_l^\lambda + \sum_{r=1}^R \sum_{h=1}^4 \frac{\norm{\boldsymbol{\gamma}_{h,l}^{(r)}}_1}{\sqrt{\tau \phi_r}} \Big) \Bigg)
\label{eq:posterior_lambda}
\end{equation}
Block (III) concerns the marginals of the PARAFAC decomposition for the tensors $\mathcal{G}_l$. The vectors $\boldsymbol{\gamma}_{h,l}^{(r)}$ are sampled independently from
\begin{equation}
p(\boldsymbol{\gamma}_{h,l}^{(r)} | \mathcal{X}, \mathcal{W}, \boldsymbol{\phi}, \tau, \mathbf{s}, \mathcal{D}, \boldsymbol{\Omega}) \propto \mathcal{N}_{n_h}\Big( \tilde{\boldsymbol{\zeta}}_{h,l}^r, \tilde{\boldsymbol{\Lambda}}_{h,l}^r \Big).
\label{eq:posterior_gamma}
\end{equation}
Finally, in block (IV) are drawn the mixing probability $\rho_l$ and the row $\boldsymbol{\xi}_l$ of the transition matrix $\boldsymbol{\Xi}$ from
\begin{align}
\label{eq:posterior_rho}
p(\rho_l | \mathcal{D}, \mathbf{s}) & \propto \mathcal{B}e(\tilde{a}_l^\rho, \tilde{b}_l^\rho), \\
\label{eq:posterior_xi}
p(\boldsymbol{\xi}_l|\mathbf{s}) & \propto \mathcal{D}ir(\tilde{\mathbf{c}}).
\end{align}
Blocks (I) and (II) are Rao-Blackwellized Gibbs steps: in block (I) we have marginalised over both $(\mathcal{D},\boldsymbol{\Omega})$ in the full joint conditional distribution of the state $\mathbf{s}$ and $\mathcal{D}$ (together with $\boldsymbol{\rho}$) in the full conditional of $\boldsymbol{\Omega}$, while in (II) we have integrated out $\tau$ from the full conditional of $\boldsymbol{\phi}$. The derivation of the full conditional distributions is given in \autoref{sec:apdx_proofs}.
The supplement provides details on the Gibbs sampler and the results of a simulation study in which we show the efficiency of the proposed MCMC and its effectiveness in recovering the true value of the latent Markov chain and parameters.

\section{Empirical Application} \label{sec:application}
We apply the proposed methodology to temporal financial networks for European institutions obtained as in~\cite{Billioetal12GrangerNet}.
The application is appealing since there are few empirical studies on this dataset and, to the best of our knowledge, none of them considers a dynamic network model.
The dataset consists of $110$ binary, directed networks estimated at the monthly frequency, from December 2003 to January 2013, by Granger\footnote{See e.g. \cite{Swanson97IRF_Granger_causal}, \cite{Boudjellaba92Gramger_causal_test}.} causality\footnote{We define a binary adjacency matrix for each month by setting an entry to 1 only if the corresponding Granger-causality link existed for the whole month (i.e. for each trading day of the corresponding month), and setting the entry to 0 otherwise.}, where the nodes are $61$ European financial institutions ($25$ banks, $11$ insurance companies and $25$ investment companies, in this order).
$x_{ij,t} = 1$ represents a Granger-causal link from institution $i$ to institution $j$ at time $t$.
The most striking features of the data are time-varying sparsity (see Fig.~\ref{fig:data_graphs}) and temporal clustering of sparse and dense network topologies (see the supplement for a representation of the temporal network dataset).

\begin{figure}[H]
\setlength{\abovecaptionskip}{2pt}
\centering
\includegraphics[trim=12mm 17mm 12mm 17mm,clip,height= 4.0cm, width= 4.0cm]{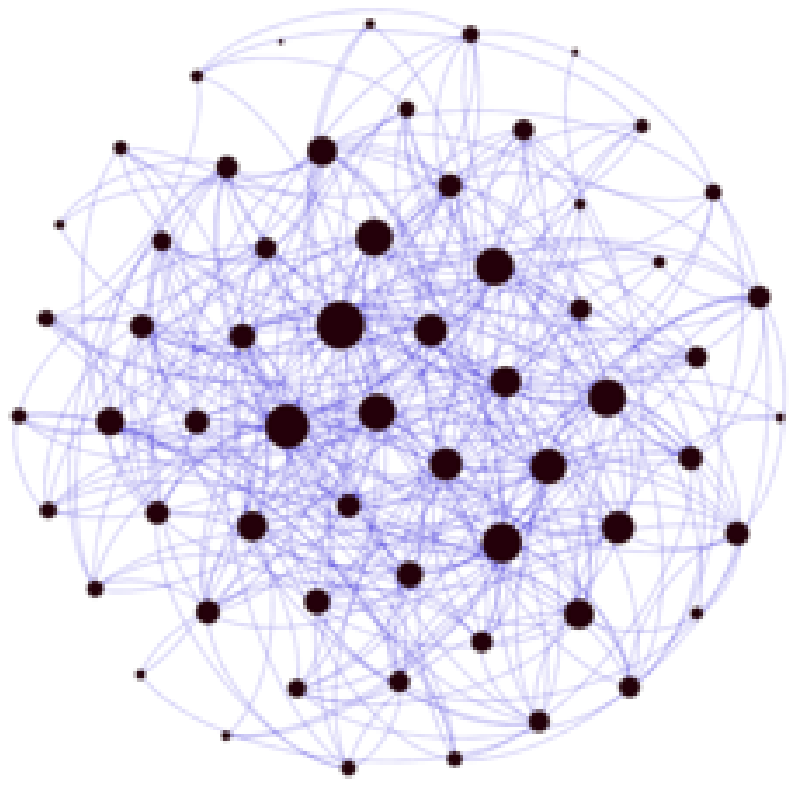} \qquad
\includegraphics[trim=12mm 17mm 12mm 17mm,clip,height= 4.0cm, width= 4.0cm]{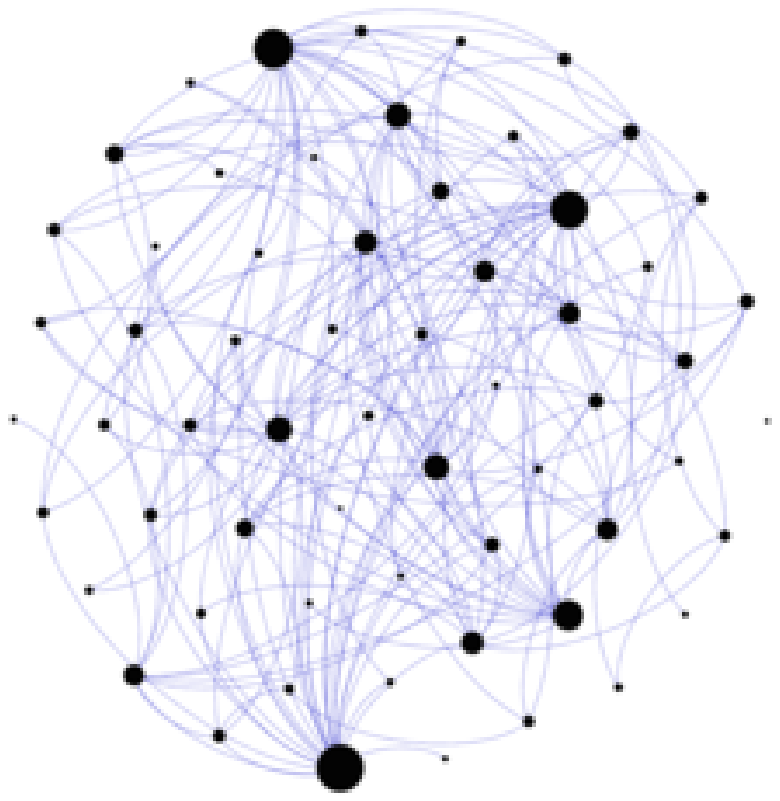} \qquad
\includegraphics[trim=12mm 17mm 12mm 17mm,clip,height= 4.0cm, width= 4.0cm]{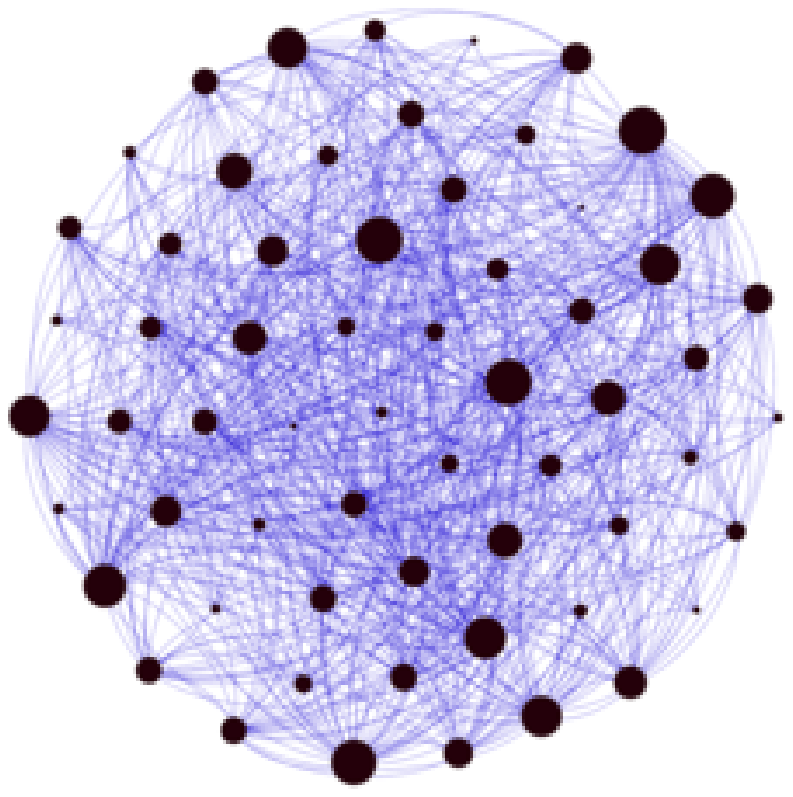}
\caption{Graphical representation of networks at time $t=25$ (Dec 2005), $t=43$ (Jul 2007) and $t=69$ (Aug 2009), respectively. Node size is proportional to its total degree. Edge $(i,j)$ is clockwise oriented when $i$ Granger causes $j$.}
\label{fig:data_graphs}
\end{figure}

The set of covariates $\mathbf{z}_t$ used to explain each edge's probability includes a constant term and some risk factors usually employed in empirical finance: the monthly change of the VSTOXX index (DVX), the monthly log-returns on the STOXX50 index (STX), the credit spread (CRS), the term spread (TRS) and the momentum factor (MOM). In addition, we include a connectedness risk measure to account for financial linkages persistence: the network total degree (DTD).
All covariates have been standardised and included with one lag, except DVX which is contemporaneous, following the standard practice (e.g., see \cite{Majewski15OptionPricing_VolatilityLeverage}).

We estimated the model in eq.~\eqref{eq:model_compact_first} with tensor rank $R=5$, $L=2$ regimes and use the Gibbs sampler of \autoref{sec:posterior_approx} to obtain 5,000 draws from the posterior, after thinning and burn-in.
%\footnote{After discarding 25,000 iterations as burn-in, we retained 10,000 iterations and thinned every 2, thus obtaining 5,000 draws from the posterior.}
See the supplement for details about the initialization.
%We initialised the latent state variables $\lbrace\mathbf{s}_t \rbrace_t$ according to suitable network statistics and the marginals of the tensor decomposition in both regimes (see the supplement for further details).

For comparison purposes, we estimate a restricted model which does not allow for heterogeneous effects of the covariates within each regime. The model is obtained by pooling parameters cross-edges for each covariate, $\mathbf{g}_{ijk,l} = \mathbf{g}_l \in \R^Q$, for each $i,j,k,l$, and by assuming the prior distributions (see \autoref{sec:apdx_pooled} and supplement for posteriors)
\begin{align*}
\mathbf{g}_l | \tau,w_l \sim \mathcal{N}_Q(\bar{\zeta}_l,\tau w_l \mathbf{I}_Q), \quad w_l|\lambda_l \sim \mathcal{E}xp(\lambda_l^2/2), \quad \lambda_l \sim \mathcal{G}a(\bar{a}_l^\lambda,\bar{b}_l^\lambda), \quad \tau \sim \mathcal{G}a(\bar{a}^\tau,\bar{b}^\tau).
\end{align*}
In both models the identification constraint $\rho_1 > \rho_2$ allows us to label state 1 and 2 as the sparse and dense regime, respectively.

\begin{figure}[t!h]
\setlength{\abovecaptionskip}{-1pt}
\centering % TotDegStates.eps
\includegraphics[trim= 10mm 0mm 17mm 5mm,clip,height= 4.0cm, width= 9.0cm]{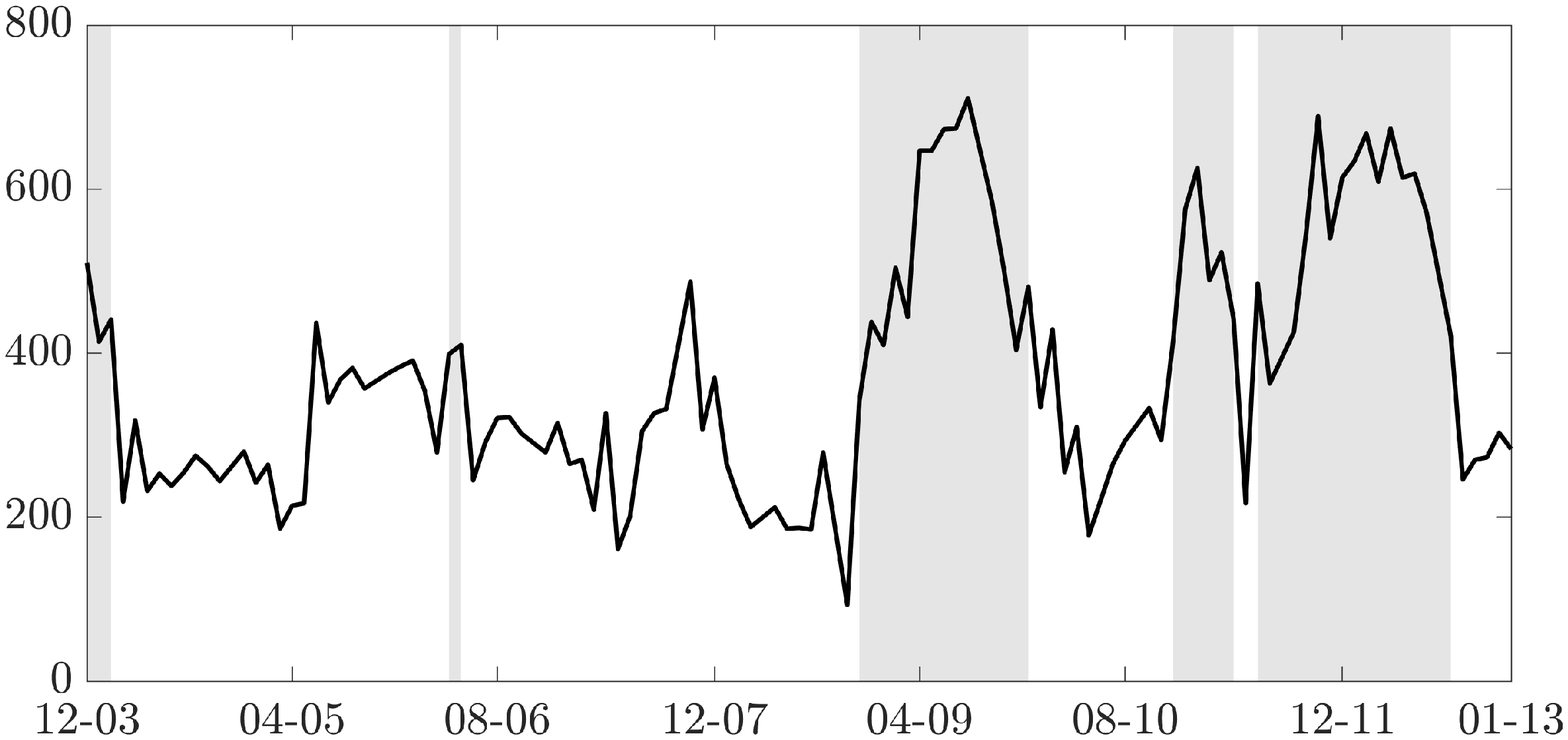}
\includegraphics[trim= 0mm 0mm 10mm 10mm,clip,height= 4.0cm, width= 3.5cm]{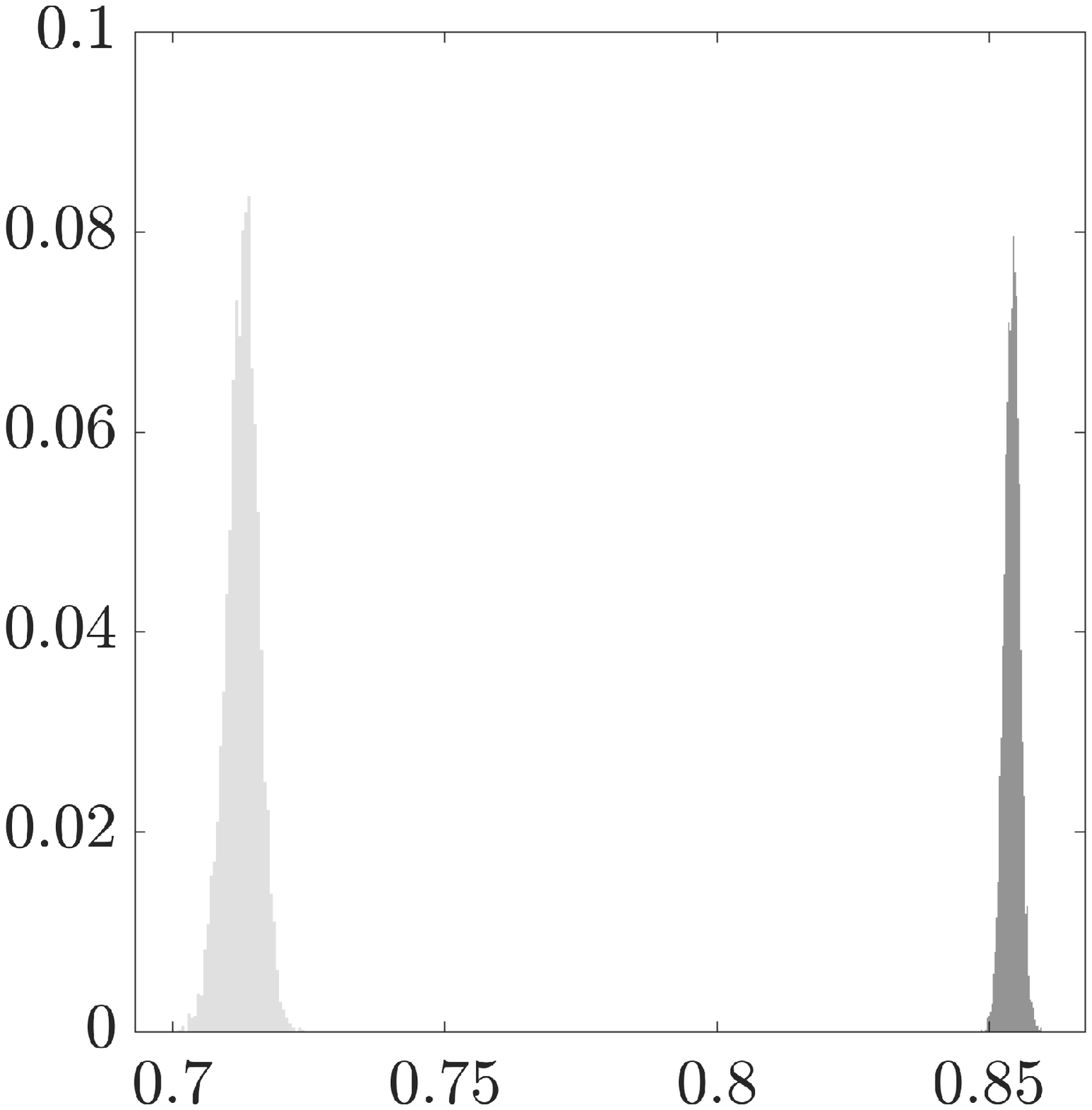}
\includegraphics[trim= 18mm 0mm 25mm 10mm,clip,height= 4.0cm, width= 3.5cm]{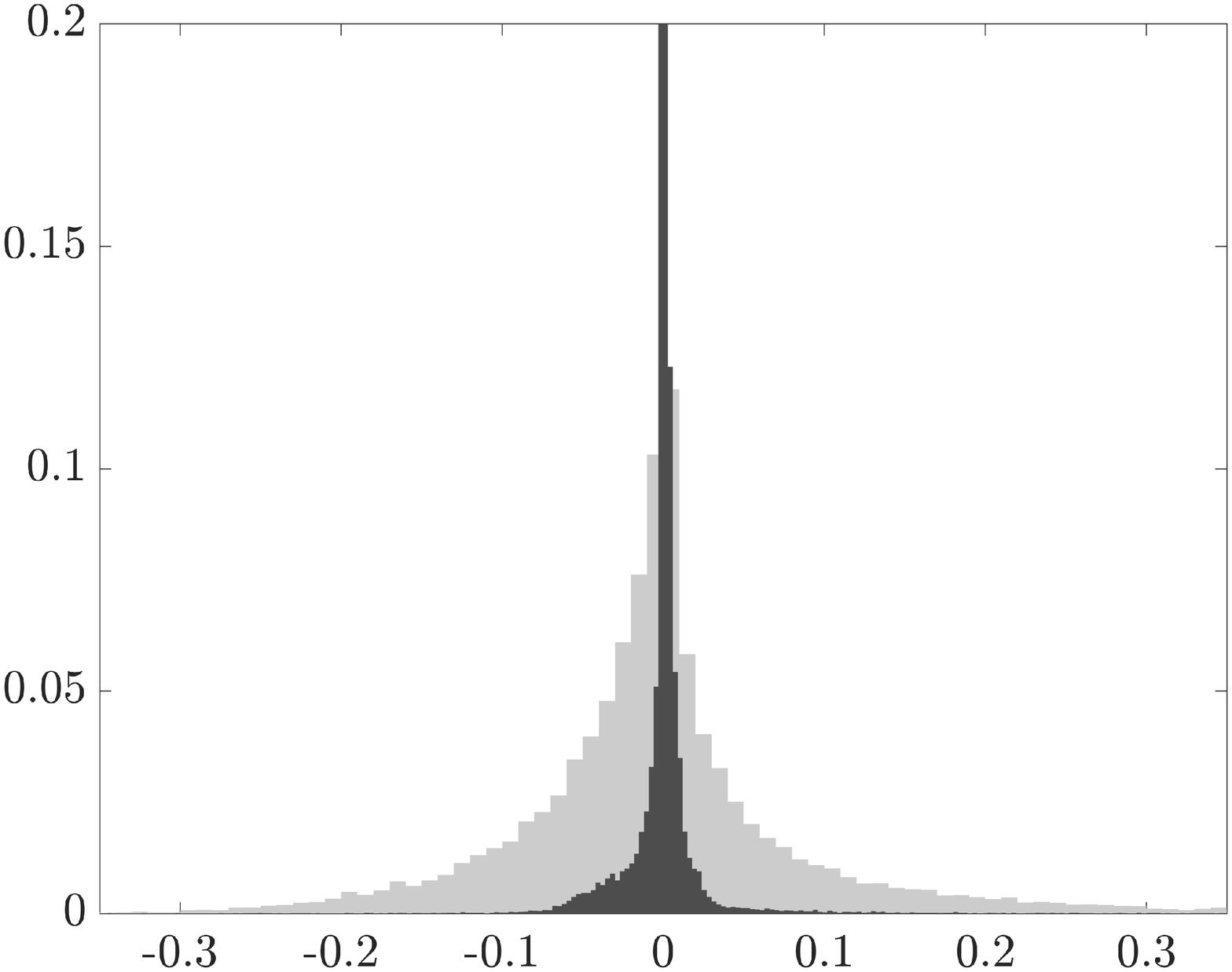}
\caption{In all plots light (dark) grey identifies the dense (sparse) regime. \textit{Left:} total degree of the temporal network (\textit{line}) and estimated regimes (\textit{vertical bars}) over time (format \textit{mm-yy}).  \textit{Middle:} posterior distribution of the sparsity parameters $\rho_1$ and $\rho_2$. \textit{Right:} distribution of the entries of the estimated coefficient tensor.}
\label{fig:degree_states}
\end{figure}

\begin{sidewaystable}[h!p]
\captionsetup{width=0.9\linewidth}
%\begin{tabular}{c c c c c c c c}
% & {\small CON} & {\small DTD} & {\small DVX} & {\small STX} & {\small CRS} & {\small TRS} & {\small MOM} \\
\begin{tabular}{c c c c c c c}
 & {\small DTD} & {\small DVX} & {\small STX} & {\small CRS} & {\small TRS} & {\small MOM} \\
\begin{rotate}{90} \hspace*{10pt} {\small sparse regime} \end{rotate} & %S1MatTensCov1
\includegraphics[trim= 0mm 0mm 20mm 0mm,clip,height= 3cm, width= 2.8cm]{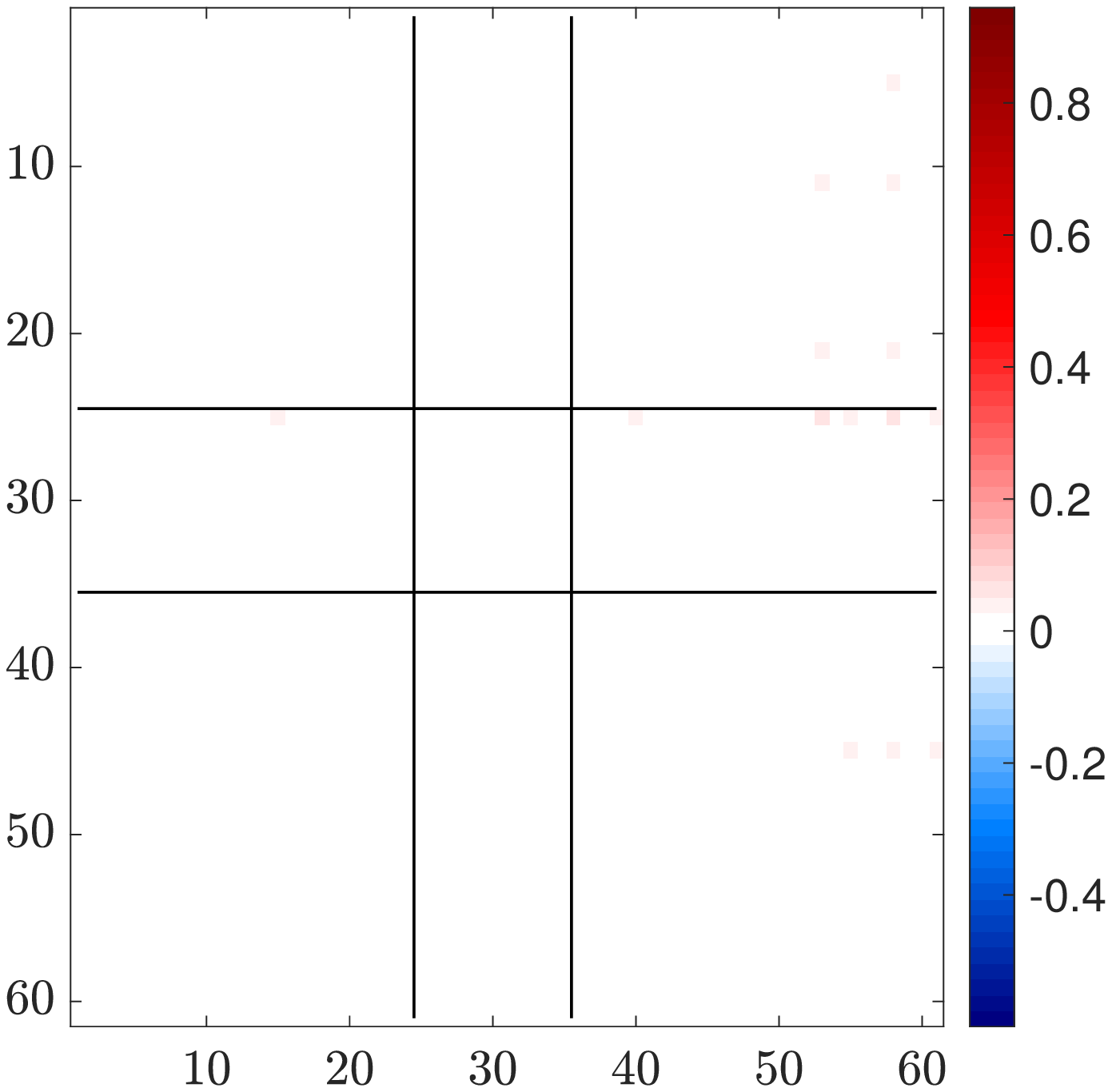} & \includegraphics[trim= 0mm 0mm 20mm 0mm,clip,height= 3cm, width= 2.8cm]{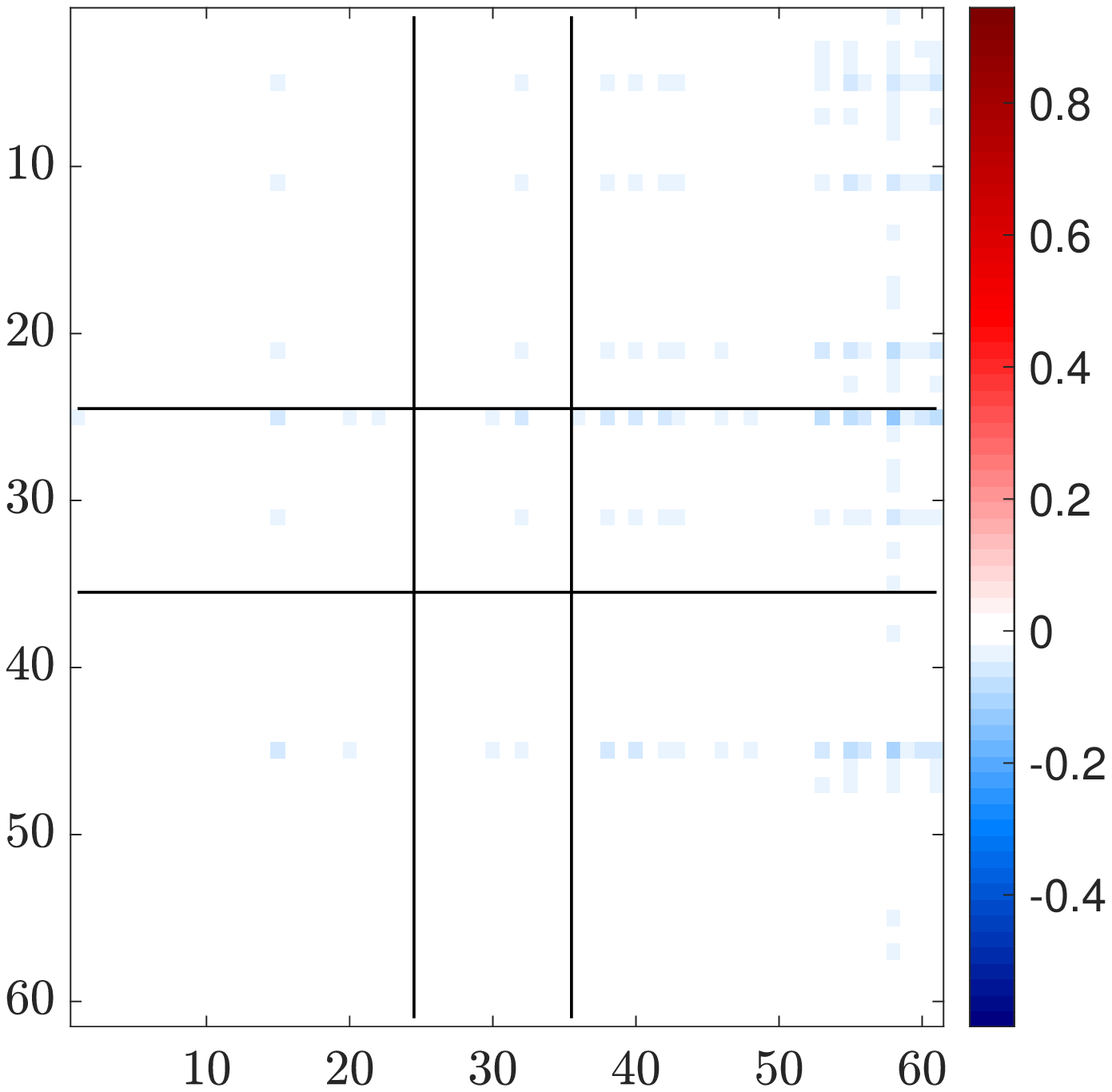} & \includegraphics[trim= 0mm 0mm 20mm 0mm,clip,height= 3cm, width= 2.8cm]{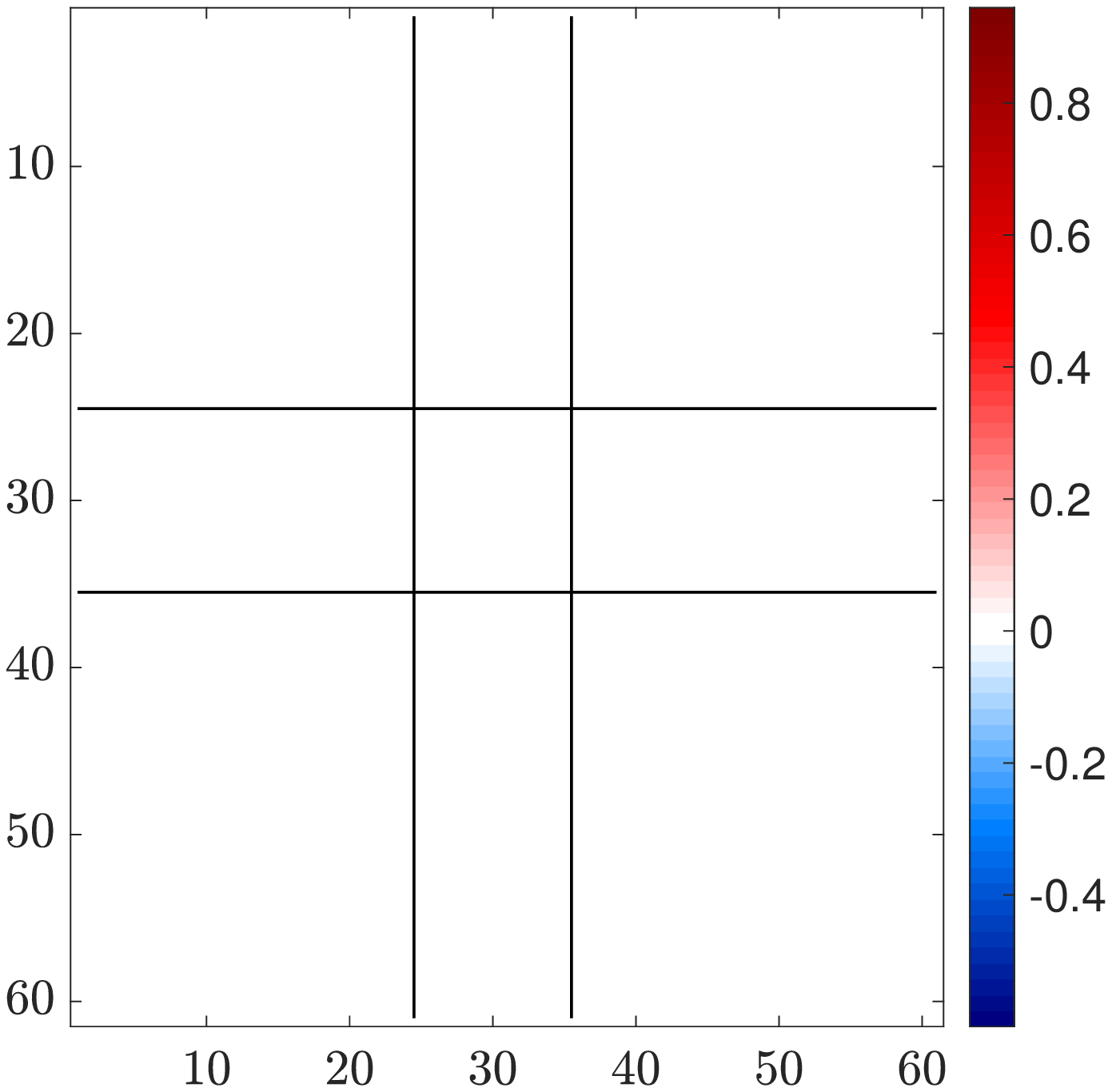} & \includegraphics[trim= 0mm 0mm 20mm 0mm,clip,height= 3cm, width= 2.8cm]{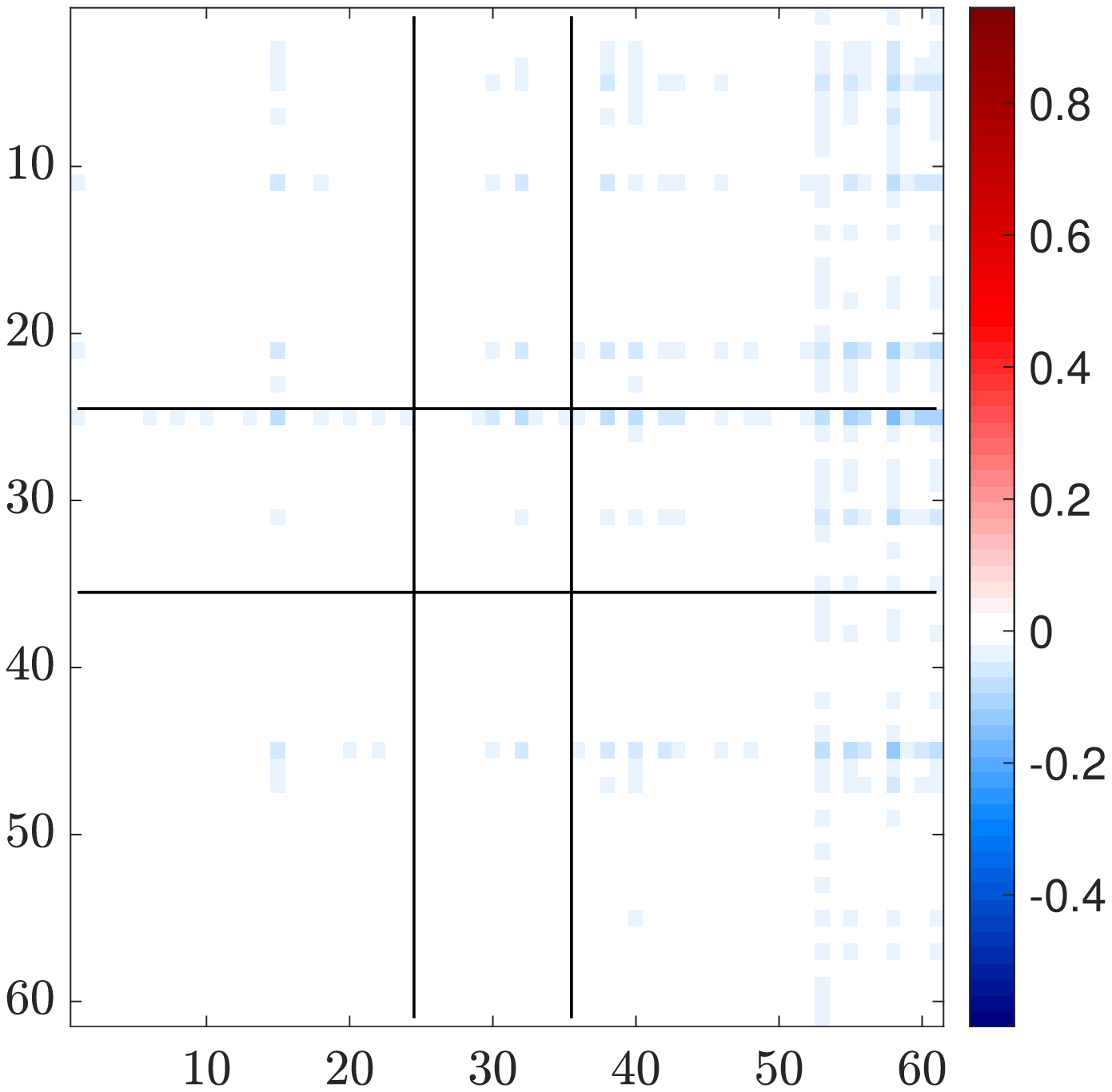} & \includegraphics[trim= 0mm 0mm 20mm 0mm,clip,height= 3cm, width= 2.8cm]{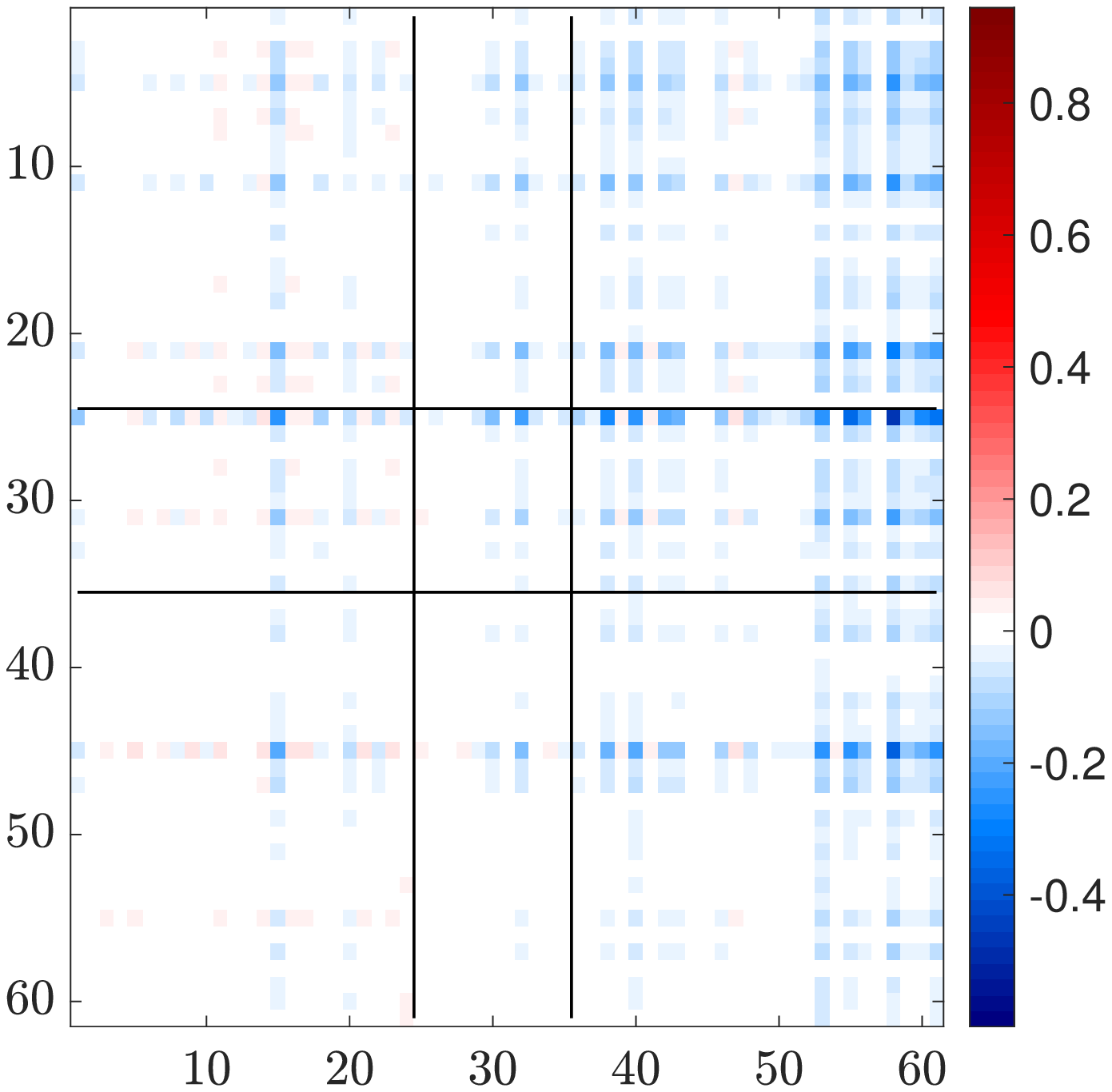} & \includegraphics[trim= 0mm 0mm 20mm 0mm,clip,height= 3cm, width= 2.8cm]{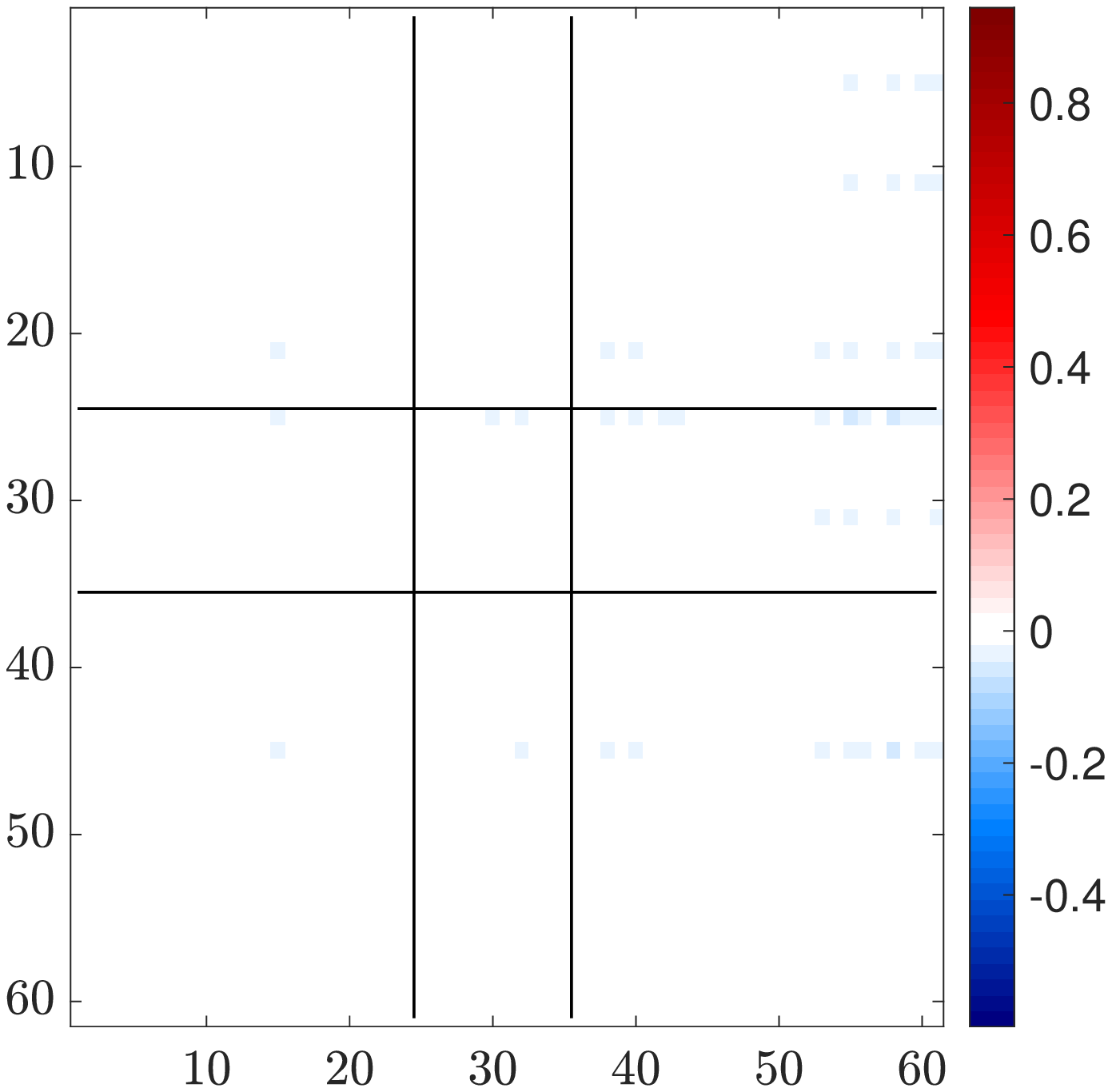} \\
\begin{rotate}{90} \hspace*{10pt} {\small dense regime} \end{rotate} & %S2MatTensCov1
\includegraphics[trim= 0mm 0mm 20mm 0mm,clip,height= 3cm, width= 2.8cm]{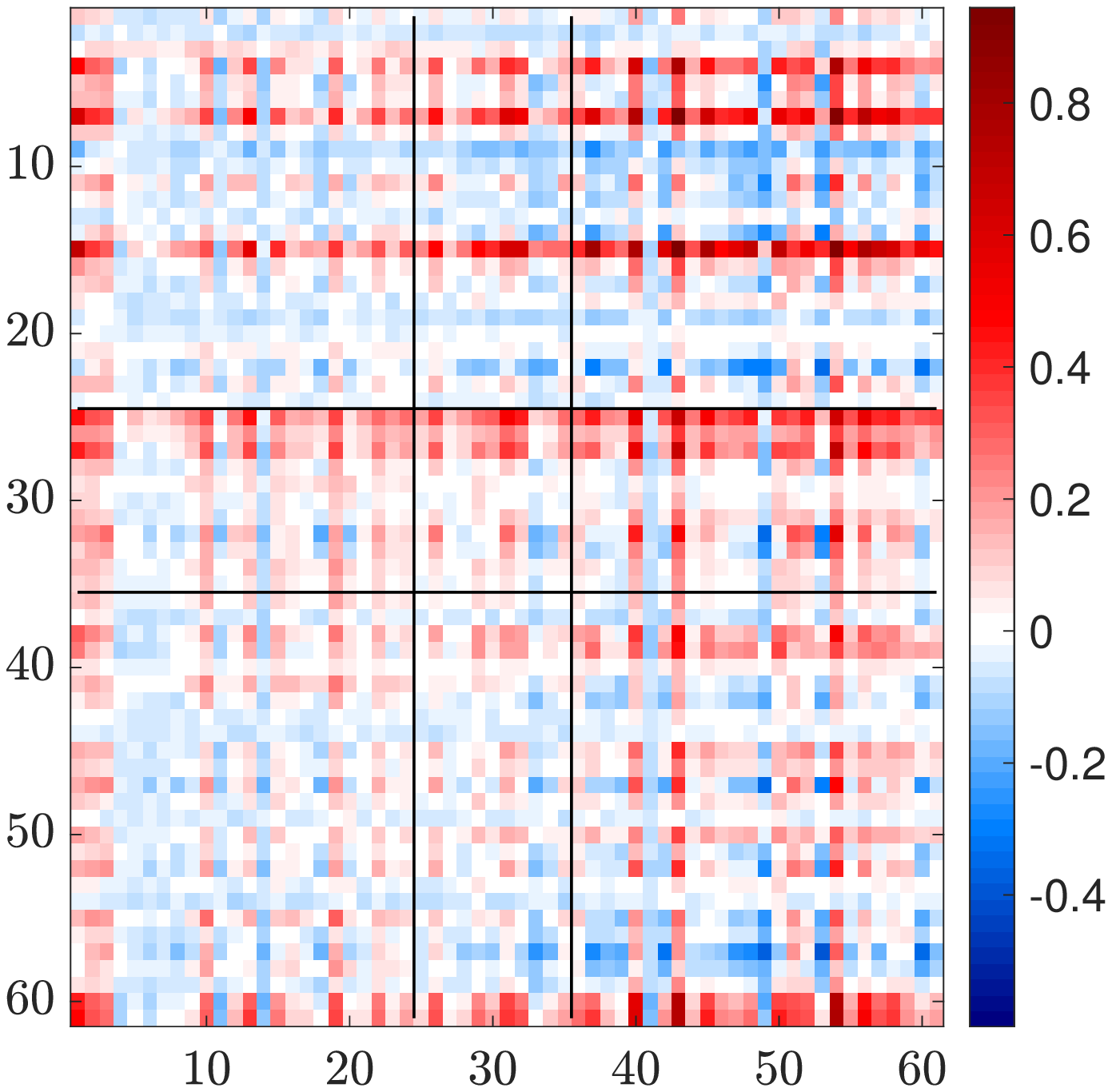} & \includegraphics[trim= 0mm 0mm 20mm 0mm,clip,height= 3cm, width= 2.8cm]{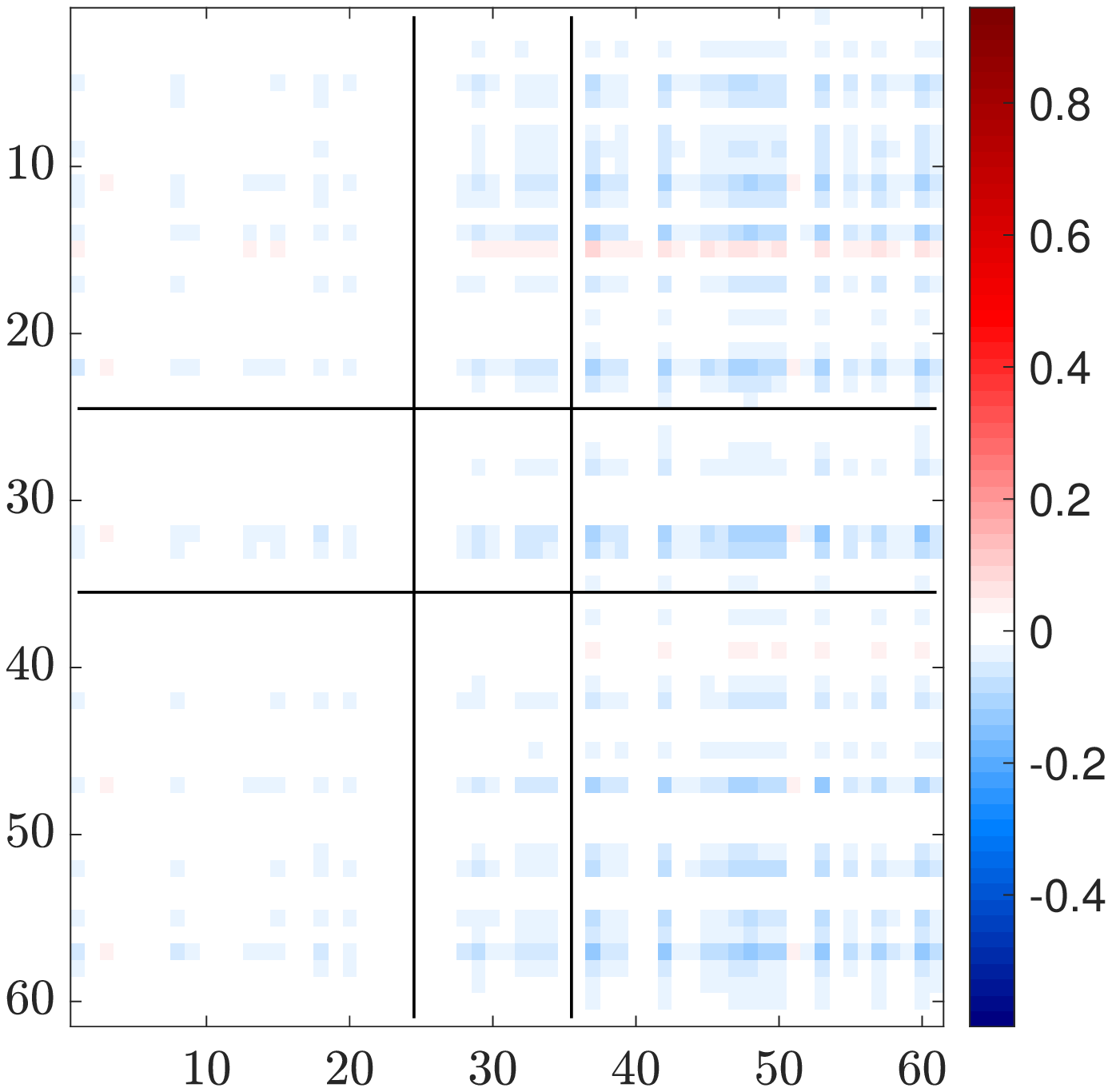} & \includegraphics[trim= 0mm 0mm 20mm 0mm,clip,height= 3cm, width= 2.8cm]{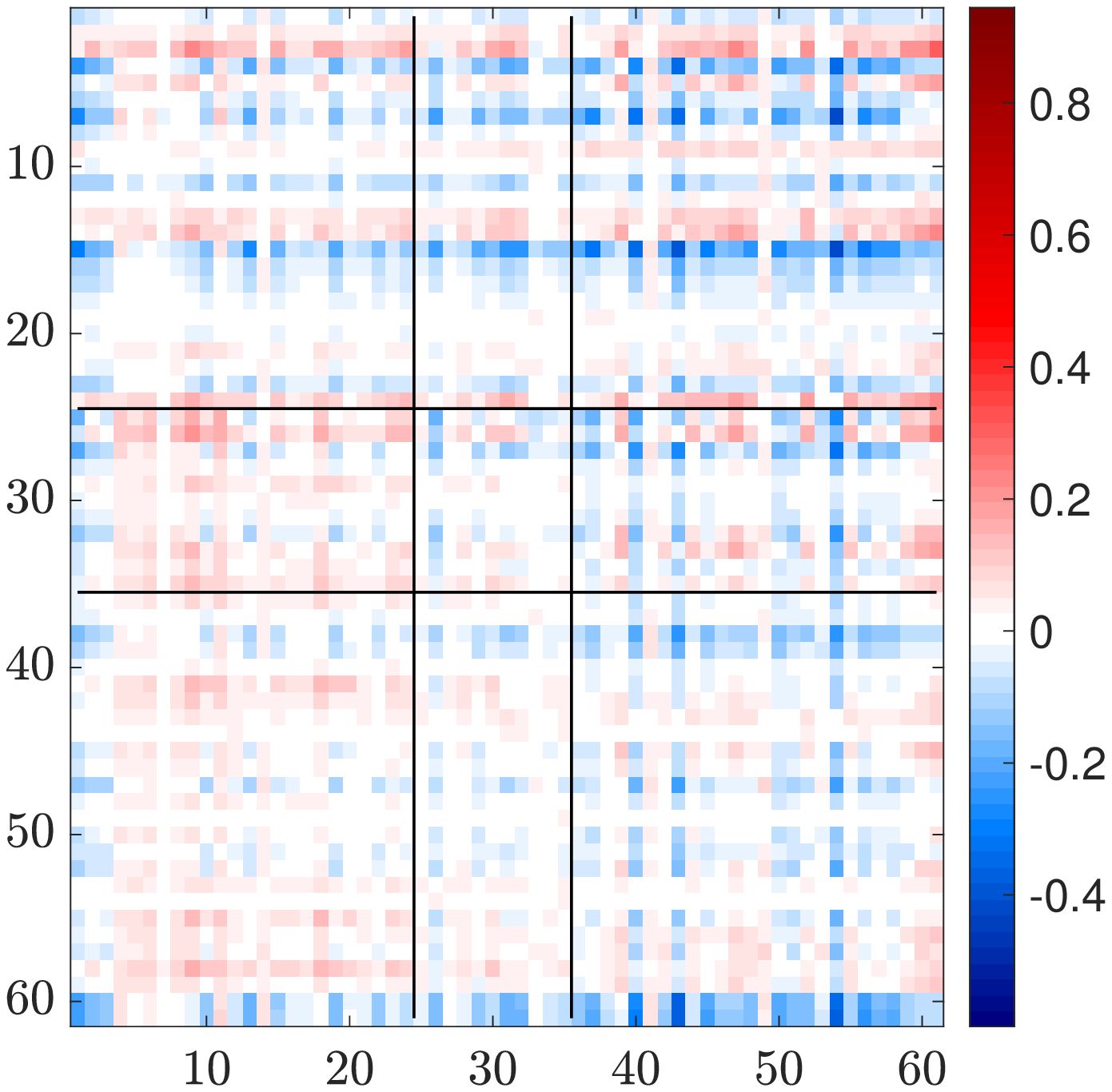} & \includegraphics[trim= 0mm 0mm 20mm 0mm,clip,height= 3cm, width= 2.8cm]{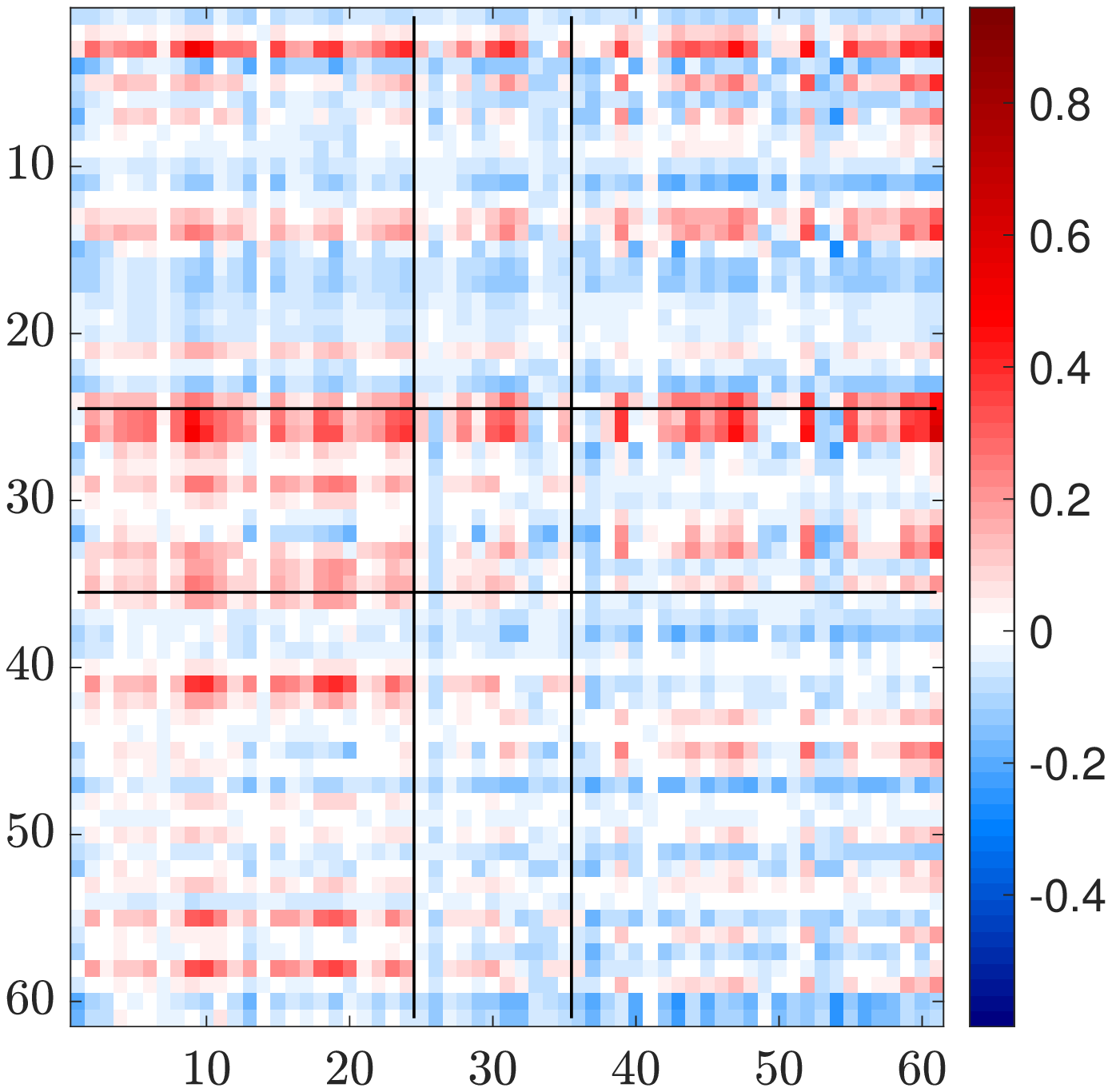} & \includegraphics[trim= 0mm 0mm 20mm 0mm,clip,height= 3cm, width= 2.8cm]{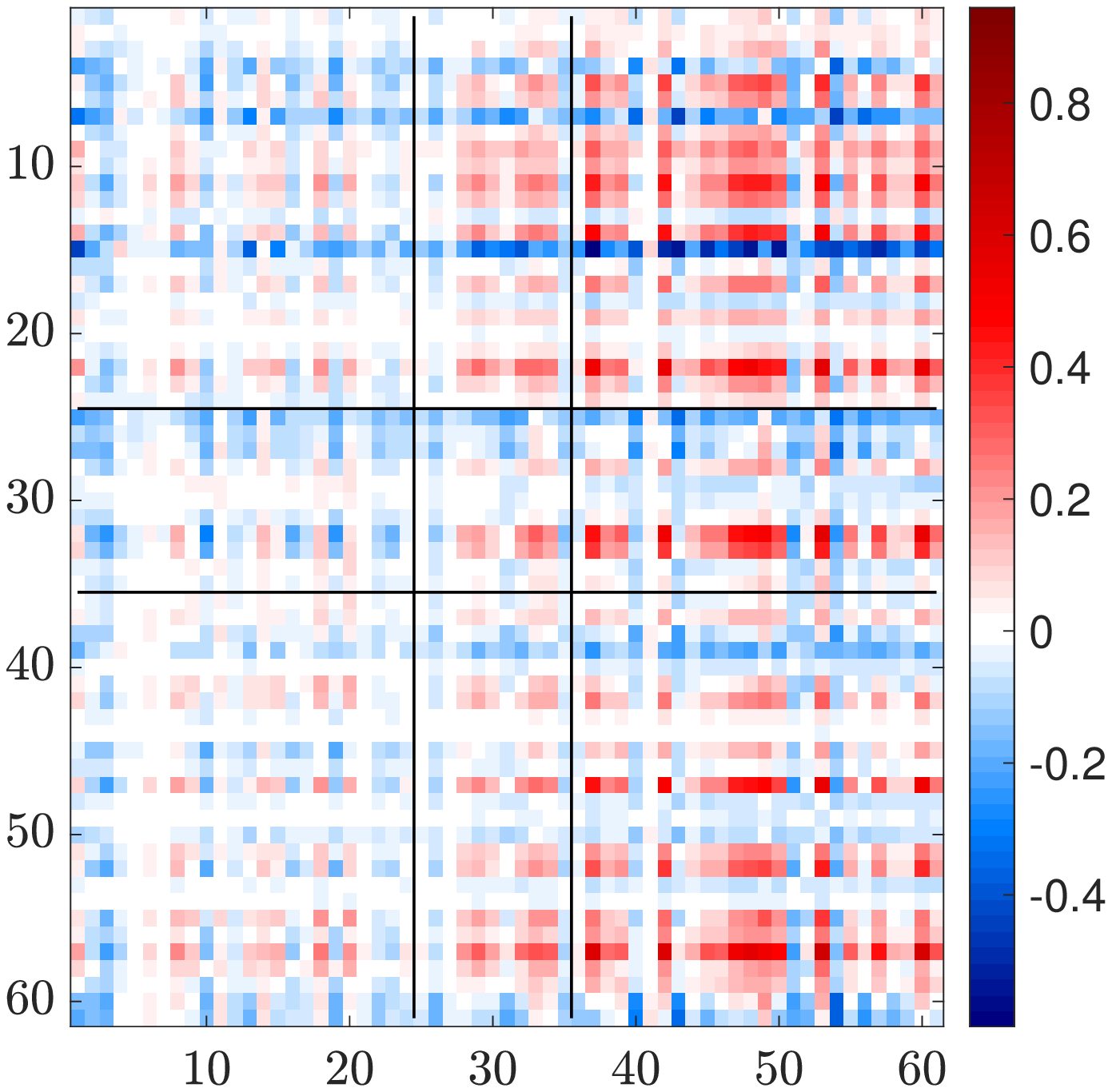} & \includegraphics[trim= 0mm 0mm 20mm 0mm,clip,height= 3cm, width= 2.8cm]{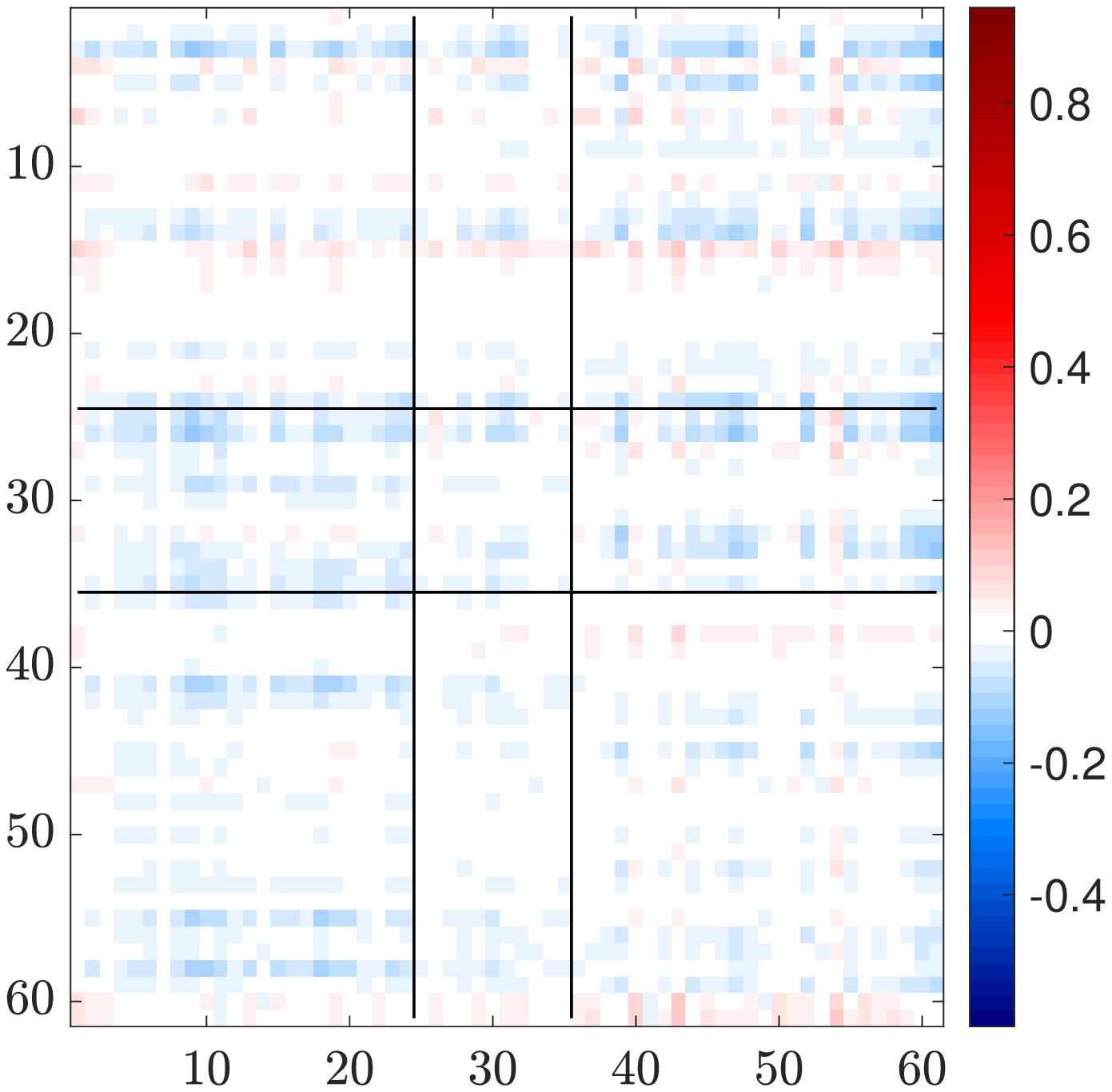} 
\end{tabular}
\captionof{figure}{Posterior mean of the coefficient tensor, in matricised form, in the sparse (\textit{top}) and dense (\textit{bottom}) state of the hidden Markov chain. In each plot, entry $(i,j)$ represents the effect of the covariate reported in column on the probability of observing the edge between institution $i$ and institution $j$. Black lines separate groups of institutions: banks ($i$ and $j$ in $\{1,\ldots,25\}$), insurance ($\{26,\ldots,36\}$) and investment companies ($\{37,\ldots,61\}$). Same color scale, with red, blue and white colors indicating positive, negative and zero valued coefficients, respectively.}
\label{fig:matTens}
\end{sidewaystable}

\begin{sidewaystable}[h!p]
\captionsetup{width=0.95\linewidth}
\setlength{\abovecaptionskip}{1pt}
%\begin{tabular}{c c c c c c c c}
% & {\small CON} & {\small DTD} & {\small DVX} & {\small STX} & {\small CRS} & {\small TRS} & {\small MOM} \\[-2pt]
\begin{tabular}{c c c c c c c}
  & {\small DTD} & {\small DVX} & {\small STX} & {\small CRS} & {\small TRS} & {\small MOM} \\[-2pt]
\begin{rotate}{90} \hspace*{10pt} {\small sparse regime} \end{rotate} &
\includegraphics[trim= 6mm 0mm 15mm 0mm,clip,height= 3.0cm, width= 2.8cm]{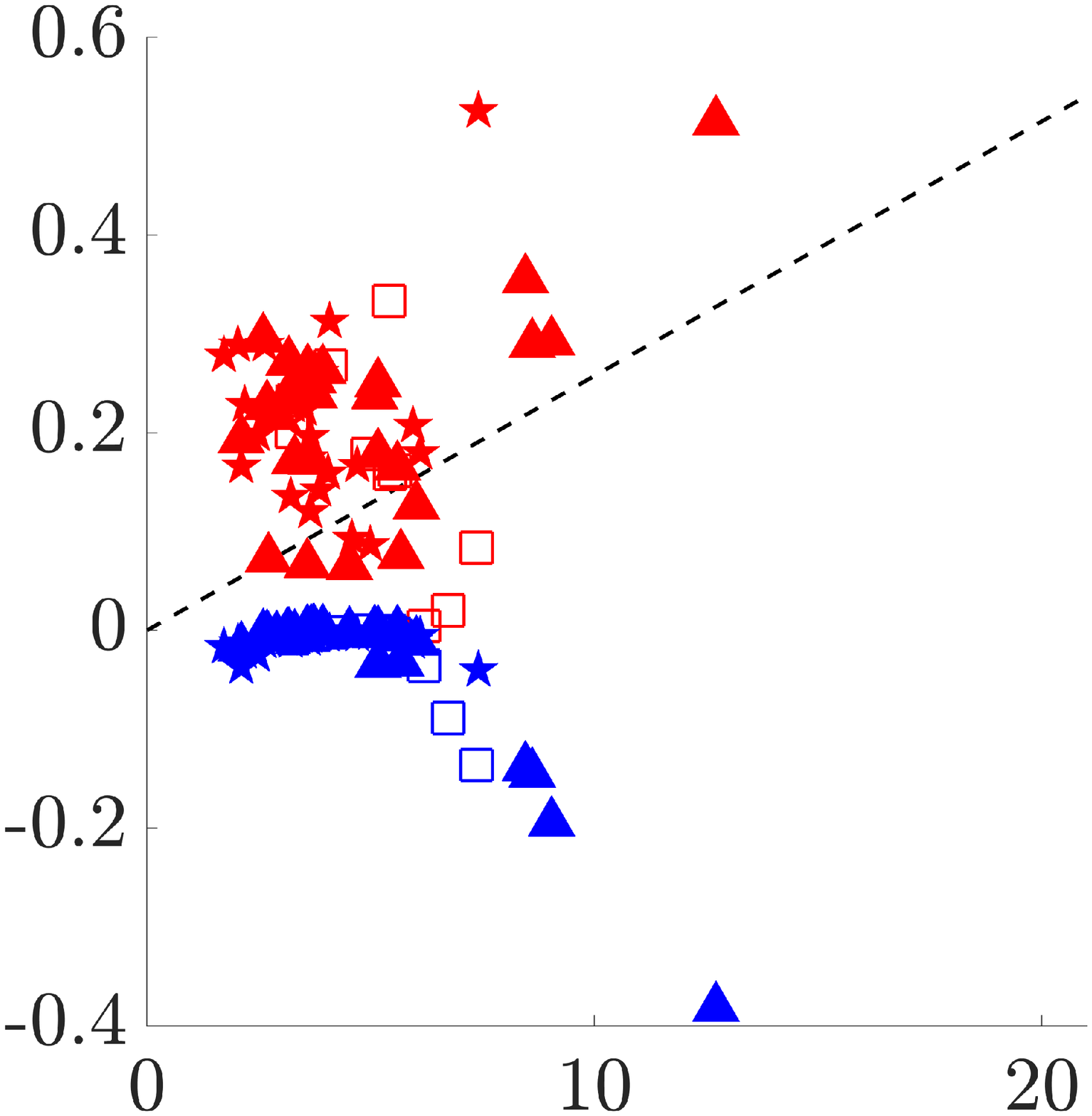} & \includegraphics[trim= 6mm 0mm 15mm 0mm,clip,height= 3.0cm, width= 2.8cm]{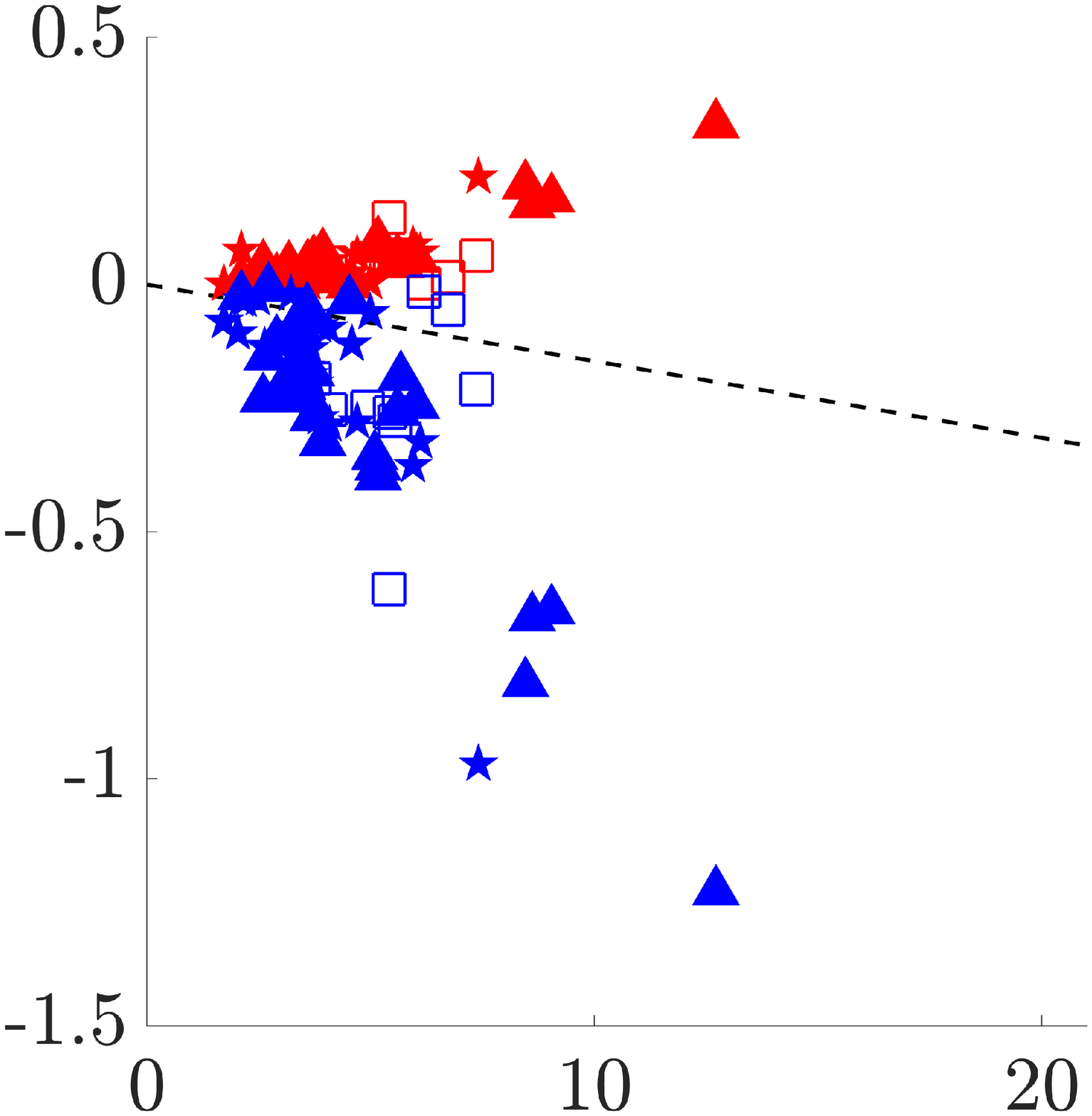} & \includegraphics[trim= 6mm 0mm 15mm 0mm,clip,height= 3.0cm, width= 2.8cm]{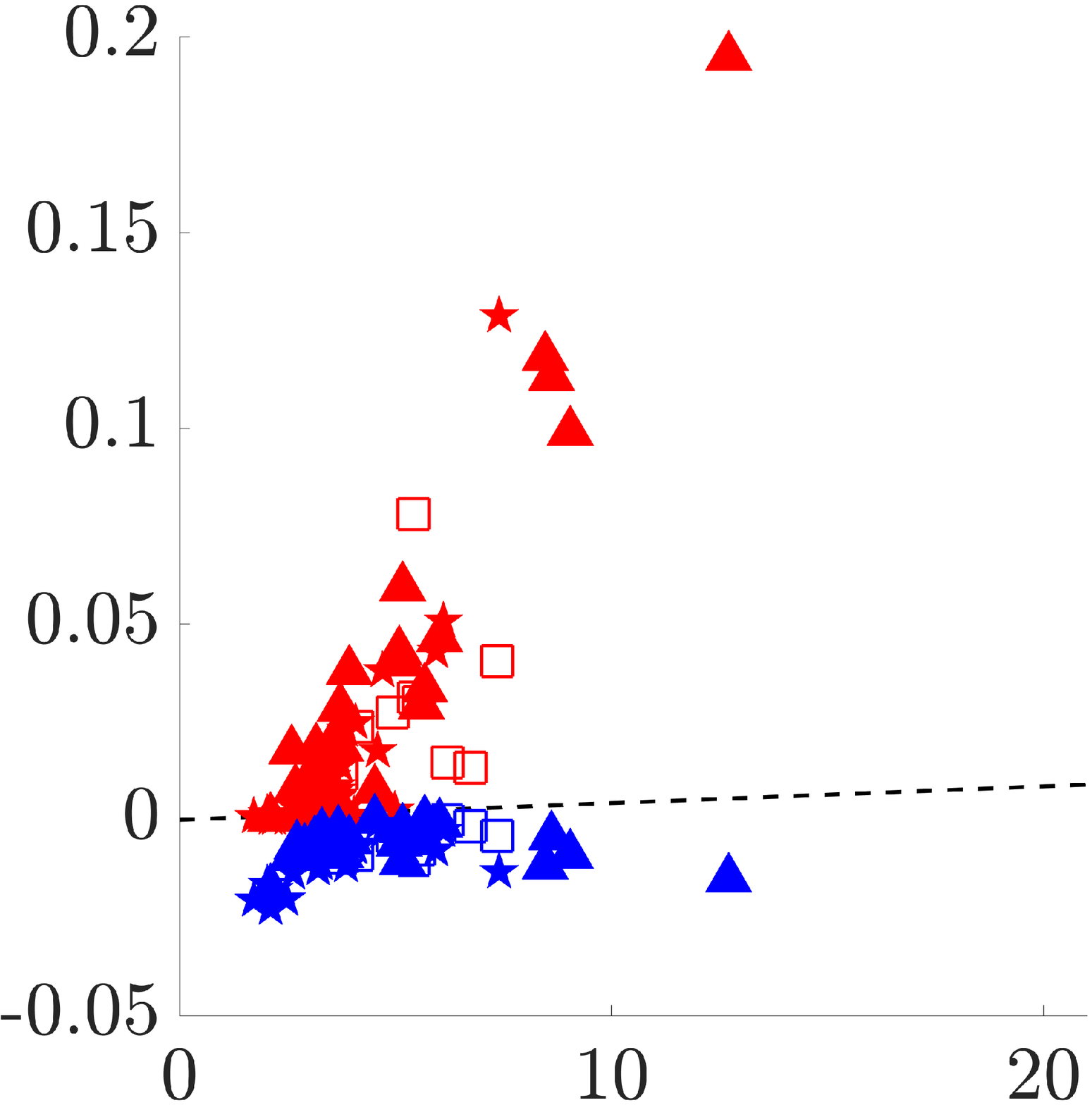} & \includegraphics[trim= 6mm 0mm 15mm 0mm,clip,height= 3.0cm, width= 2.8cm]{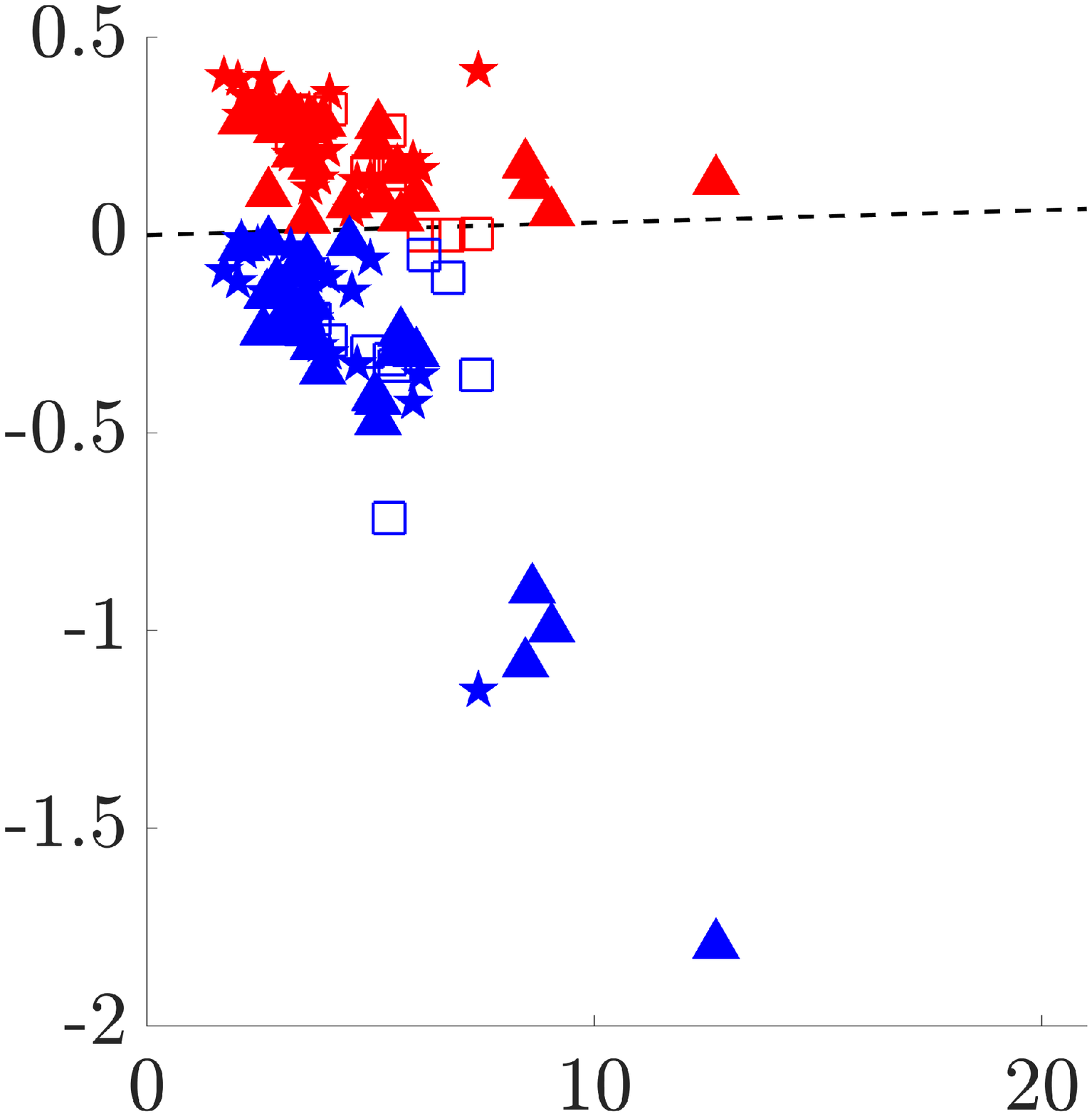} & \includegraphics[trim= 6mm 0mm 15mm 0mm,clip,height= 3.0cm, width= 2.8cm]{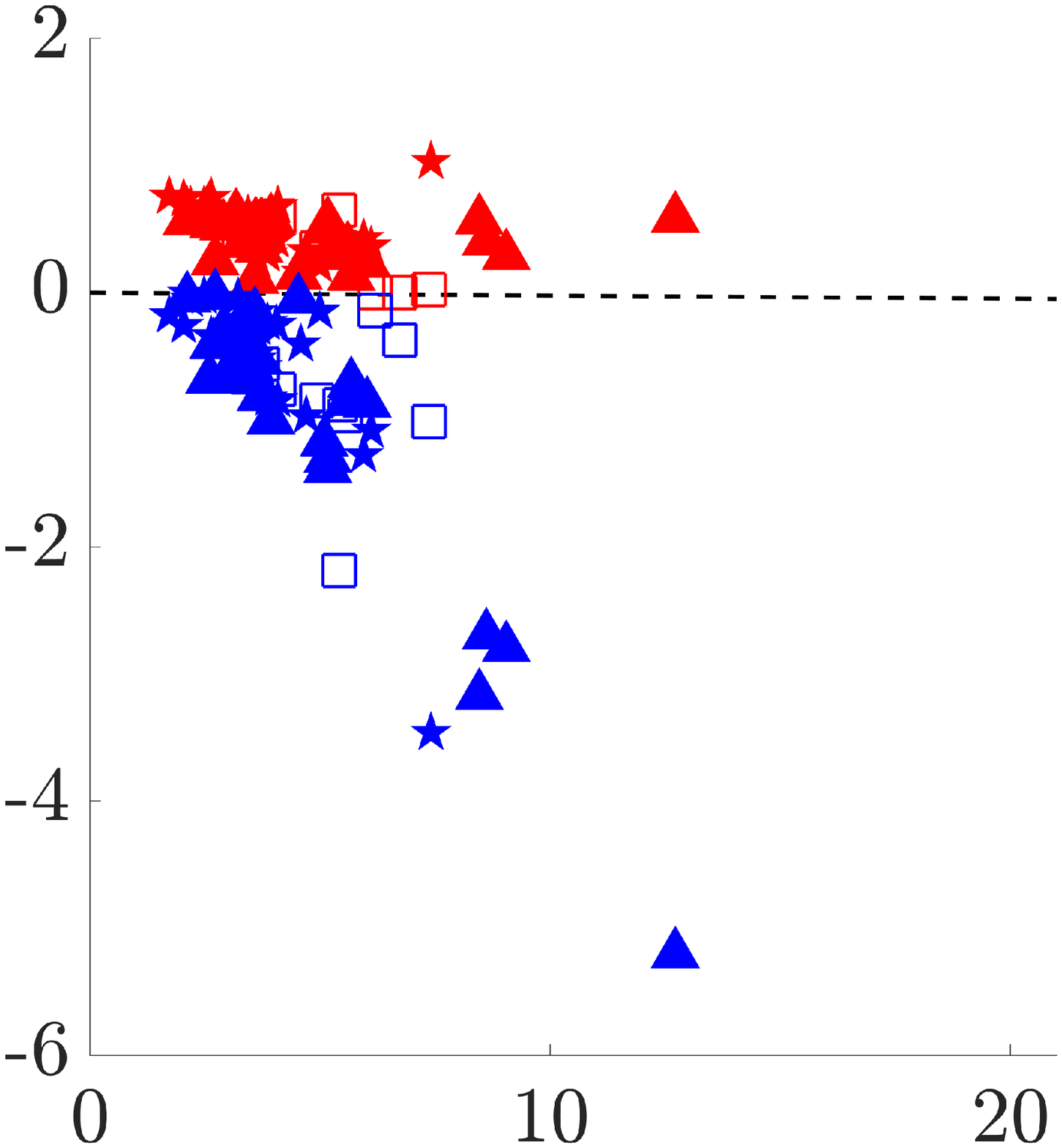} & \includegraphics[trim= 6mm 0mm 15mm 0mm,clip,height= 3.0cm, width= 2.8cm]{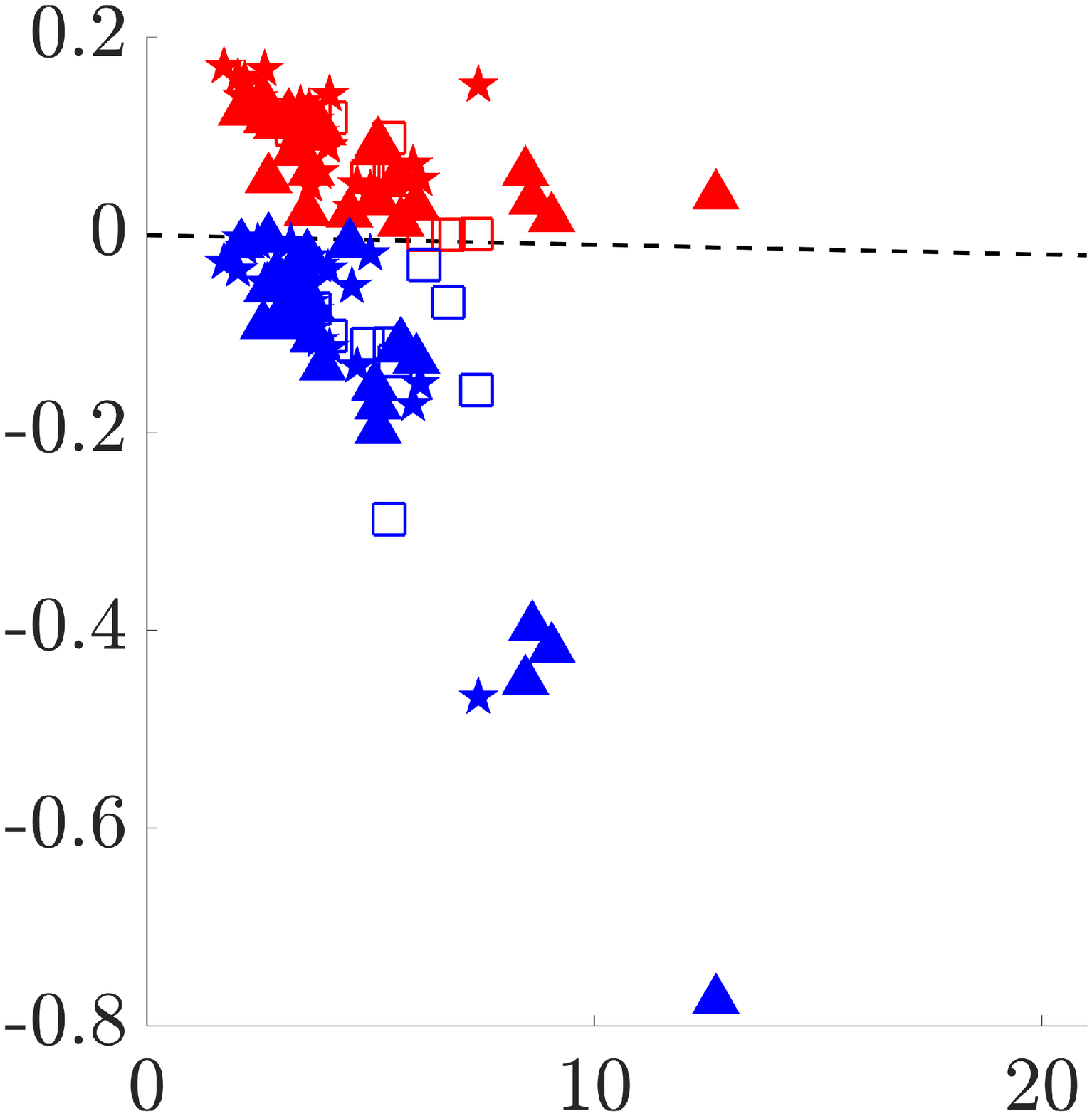} \\[-10pt]
\begin{rotate}{90} \hspace*{10pt} {\small dense regime} \end{rotate} &
\includegraphics[trim= 6mm 0mm 15mm 0mm,clip,height= 3.0cm, width= 2.8cm]{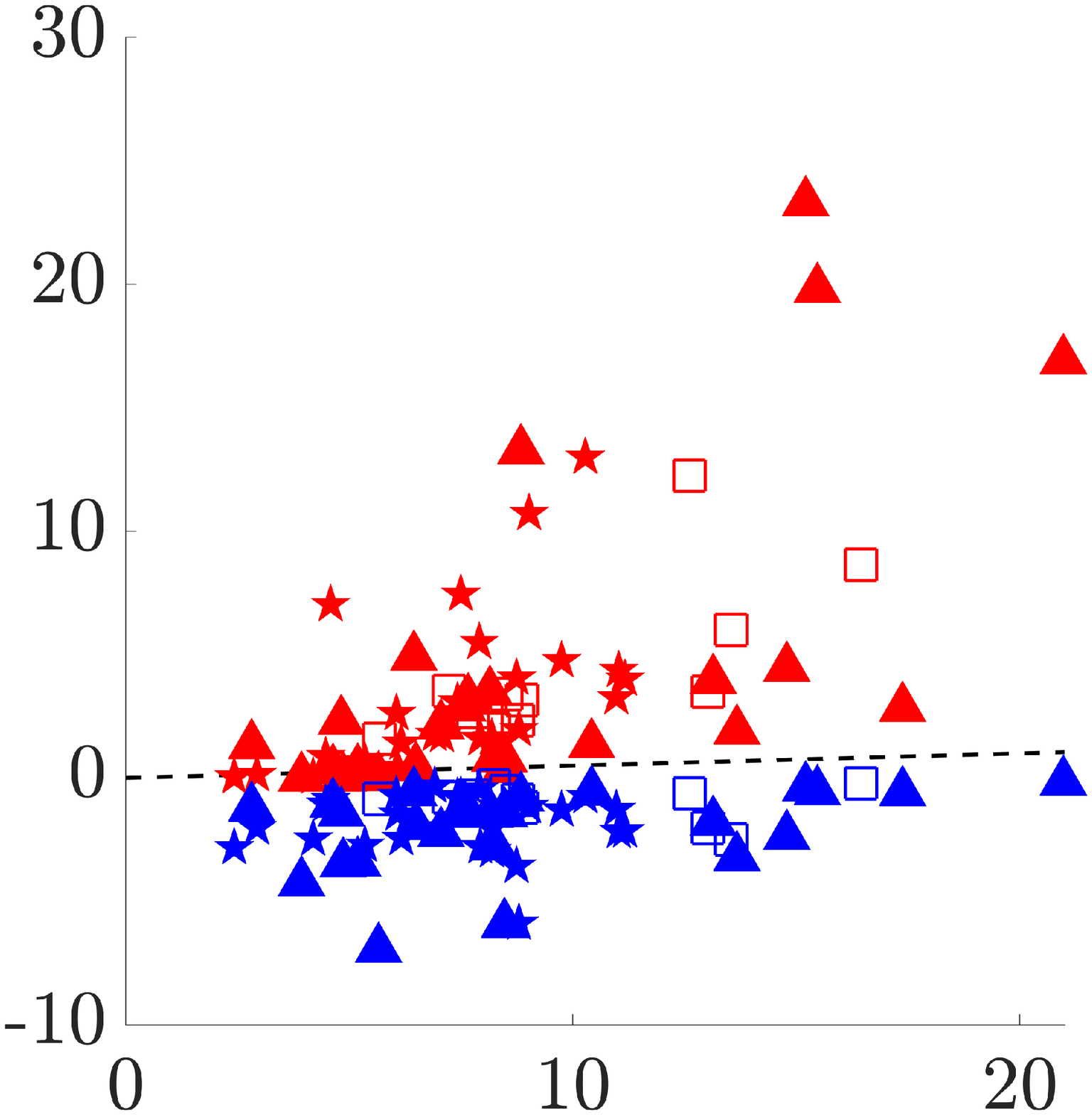} & \includegraphics[trim= 6mm 0mm 15mm 0mm,clip,height= 3.0cm, width= 2.8cm]{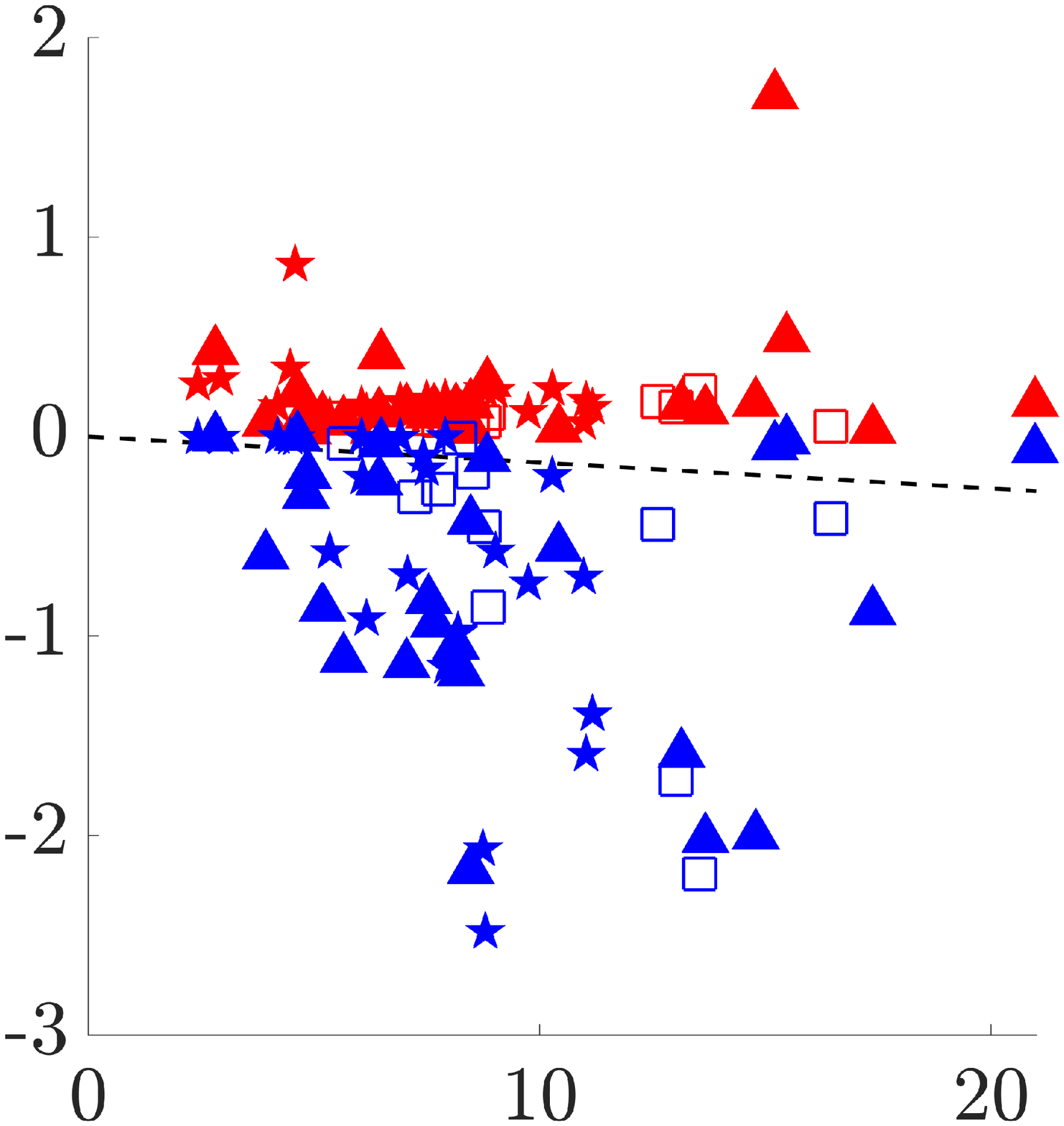} & \includegraphics[trim= 6mm 0mm 15mm 0mm,clip,height= 3.0cm, width= 2.8cm]{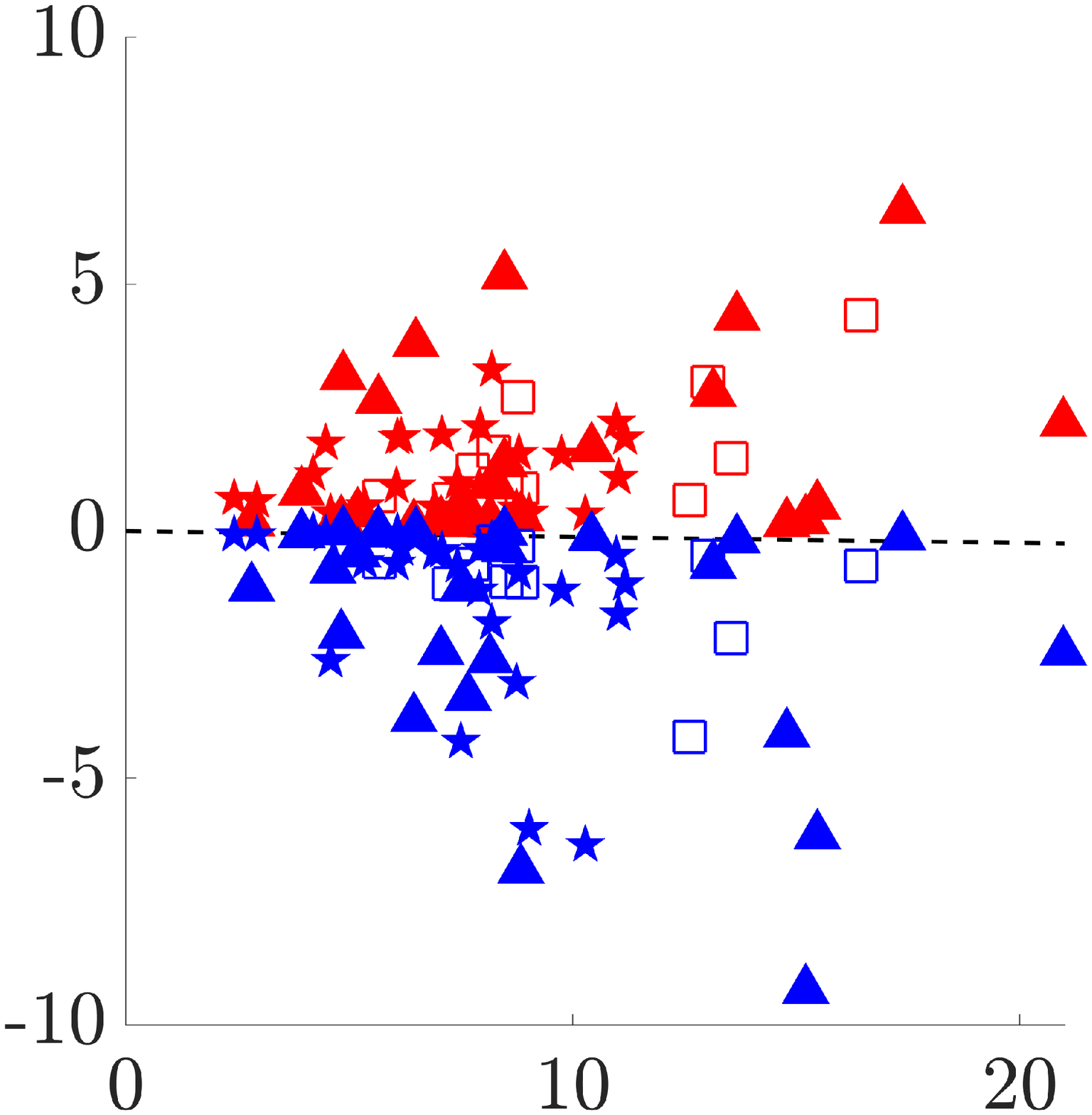} & \includegraphics[trim= 6mm 0mm 15mm 0mm,clip,height= 3.0cm, width= 2.8cm]{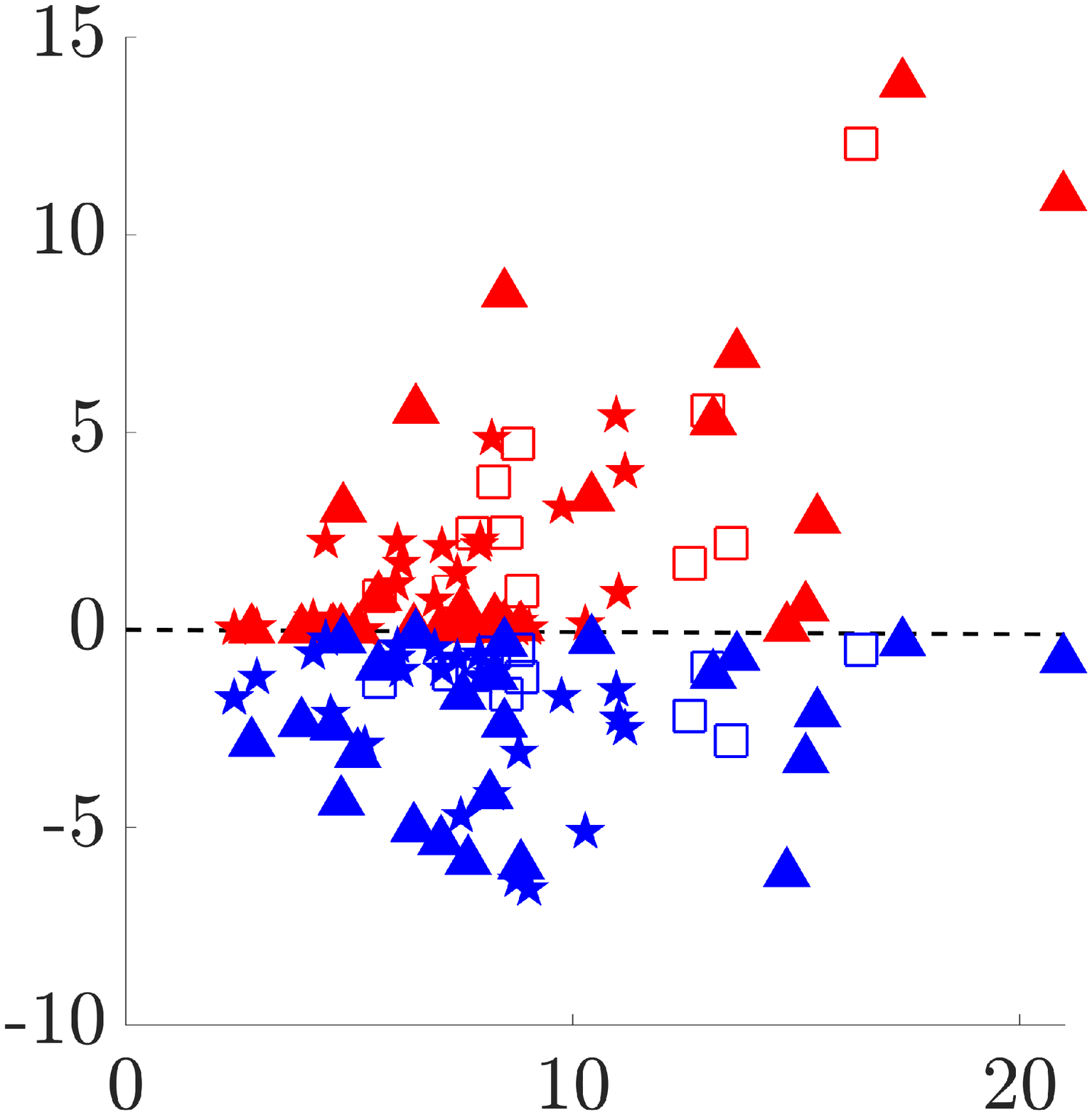} & \includegraphics[trim= 6mm 0mm 15mm 0mm,clip,height= 3.0cm, width= 2.8cm]{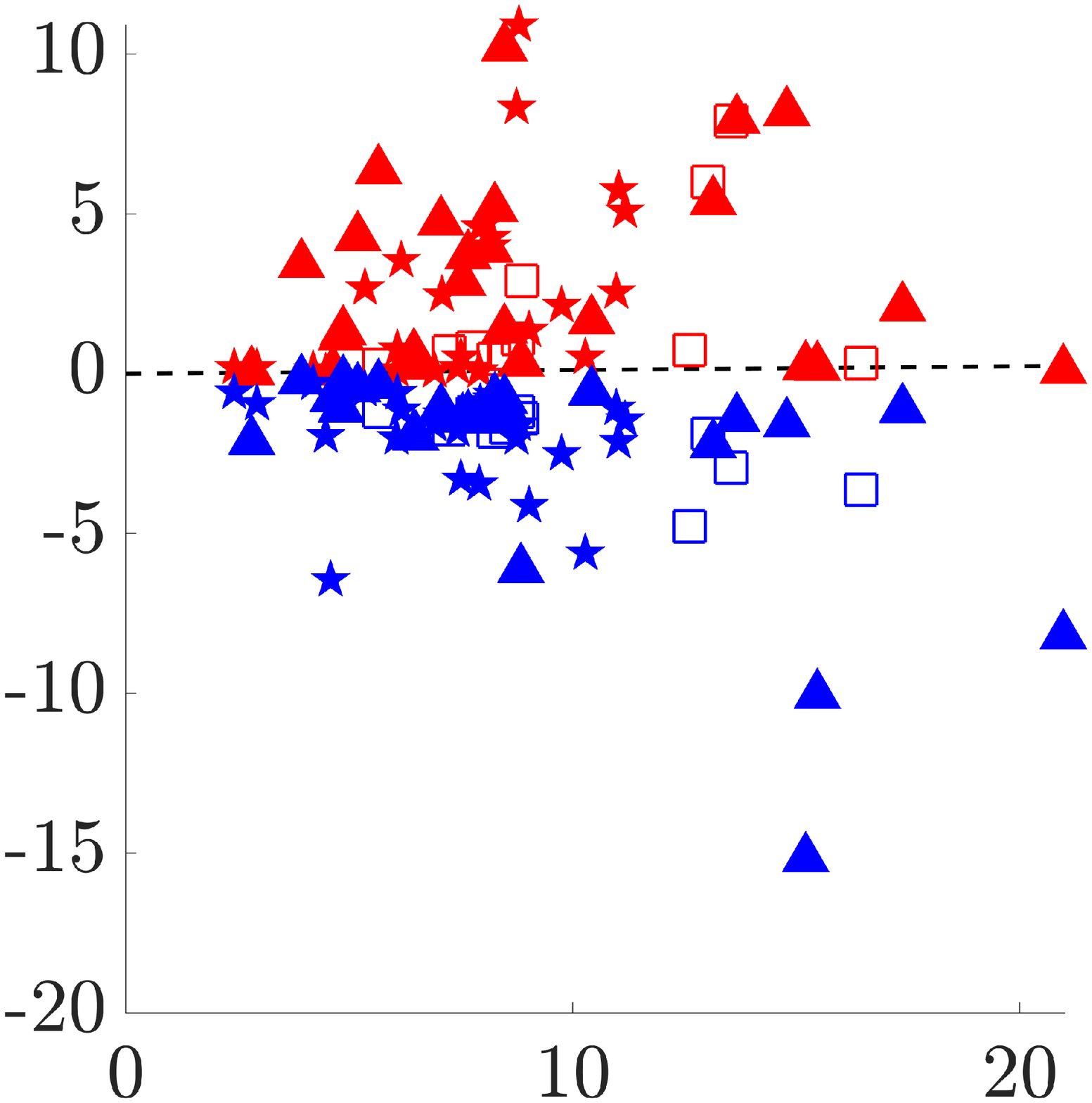} & \includegraphics[trim= 6mm 0mm 15mm 0mm,clip,height= 3.0cm, width= 2.8cm]{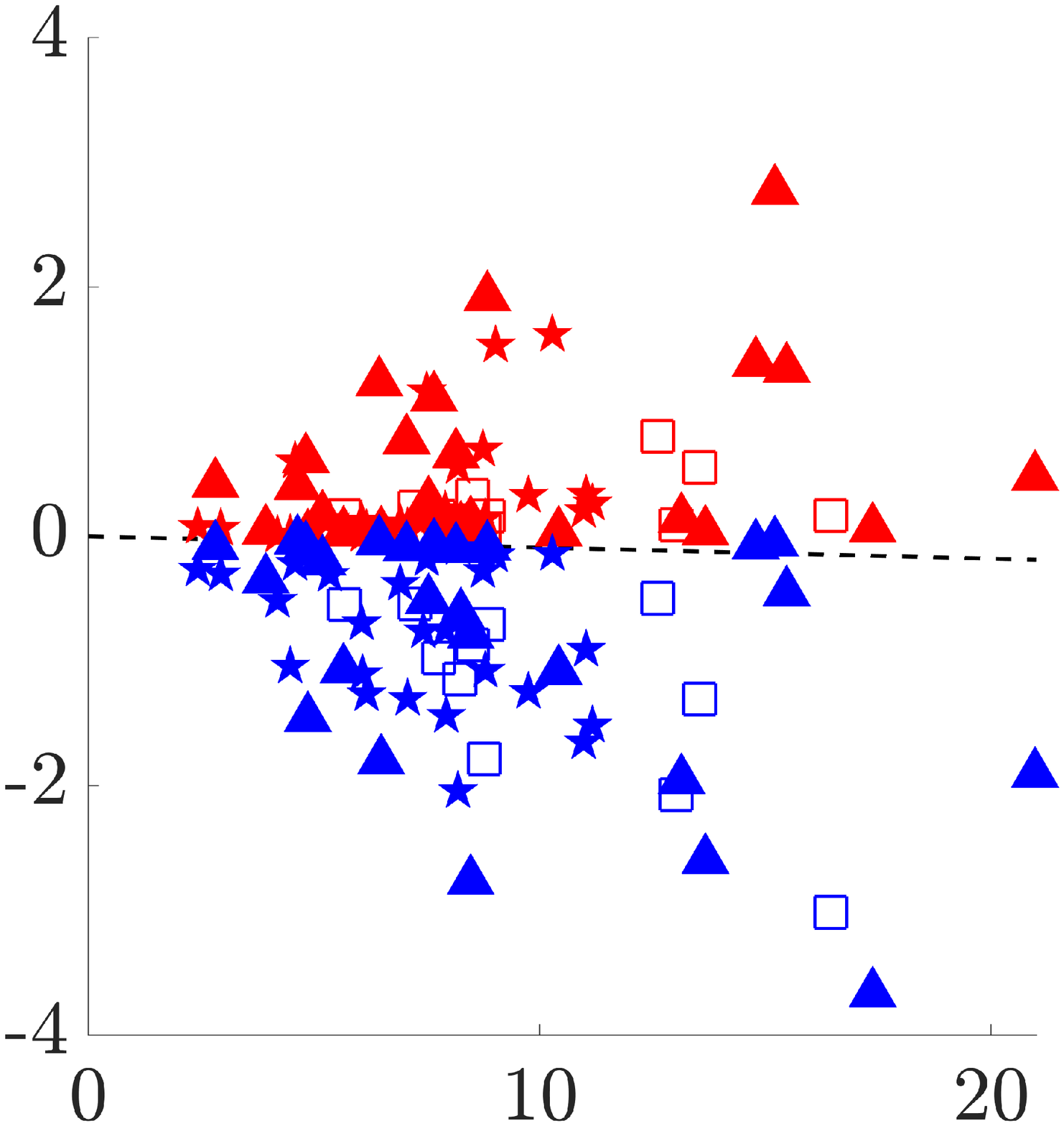}
\end{tabular}
\captionof{figure}{Covariate coefficients (columns) for the incoming edge probabilities in the two regimes (rows). In each scatterplot: total node degree averaged over time within each regime (horizontal axis) versus the sum of the negative (blue) and positive (red) node coefficients of a given variable (vertical axis). Nodes: banks ($\textcolor{red}{\blacktriangle},\textcolor{blue}{\blacktriangle}$), insurance companies ($\textcolor{red}{\square},\textcolor{blue}{\square}$) and investment companies ($\textcolor{red}{\star},\textcolor{blue}{\star}$). Dashed line: the sum of the coefficients for the pooled model.}
\label{fig:tens_IN}
\end{sidewaystable}

\begin{figure}[t!h]
%%% central nodes: banks= 15;  insurance= 32; investment= 43
\centering
\captionsetup{width=0.95\linewidth}
\setlength{\abovecaptionskip}{1pt}
\begin{tabular}{c c c c}
 & {\footnotesize In/Out banks connections} & {\footnotesize In/Out insurance connections} & {\footnotesize In/Out investment connections} \\[4pt]
\begin{rotate}{90} \hspace*{40pt} {\large $\substack{\text{between groups}\\ \text{(edge threshold)}}$} \end{rotate} & 
\includegraphics[trim= 10mm 20mm 10mm 28mm,clip,height= 5.0cm,width= 5.0cm]{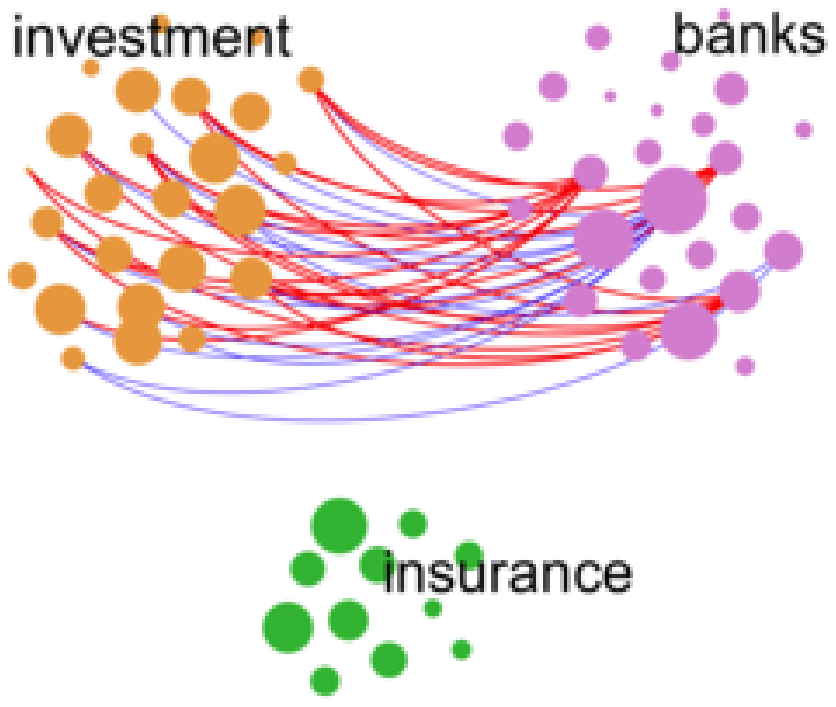} &
\includegraphics[trim= 10mm 25mm 10mm 28mm,clip,height= 5.0cm,width= 5.0cm]{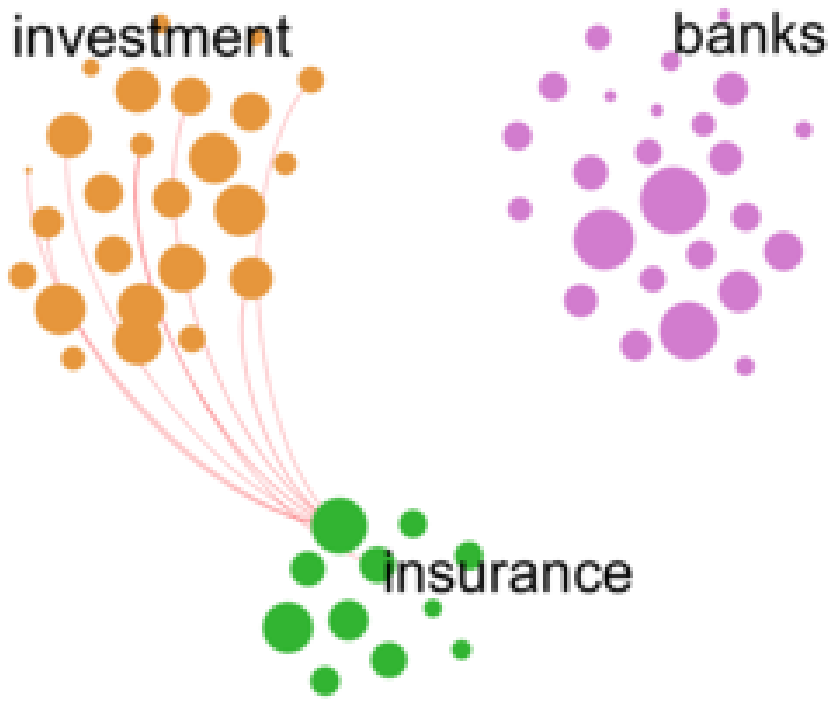} & 
\includegraphics[trim= 10mm 20mm 10mm 28mm,clip,height= 5.0cm,width= 5.0cm]{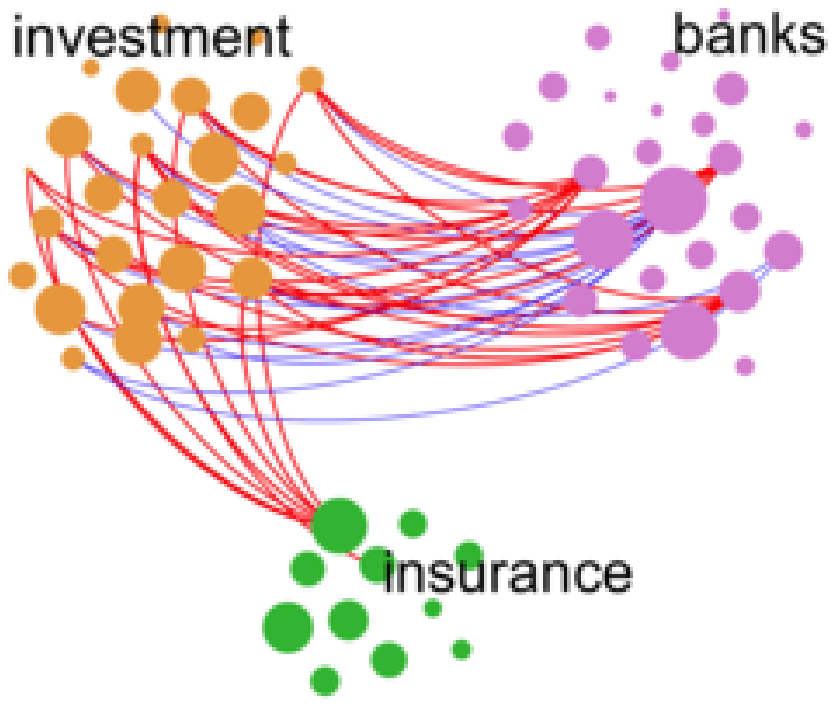} \\
\begin{rotate}{90} \hspace*{5pt} {\large $\substack{\text{within groups}\\ \text{(central institution)}}$} \end{rotate} & 
\includegraphics[trim= 170mm 140mm 10mm 28mm,clip,height= 3.5cm,width= 3.5cm]{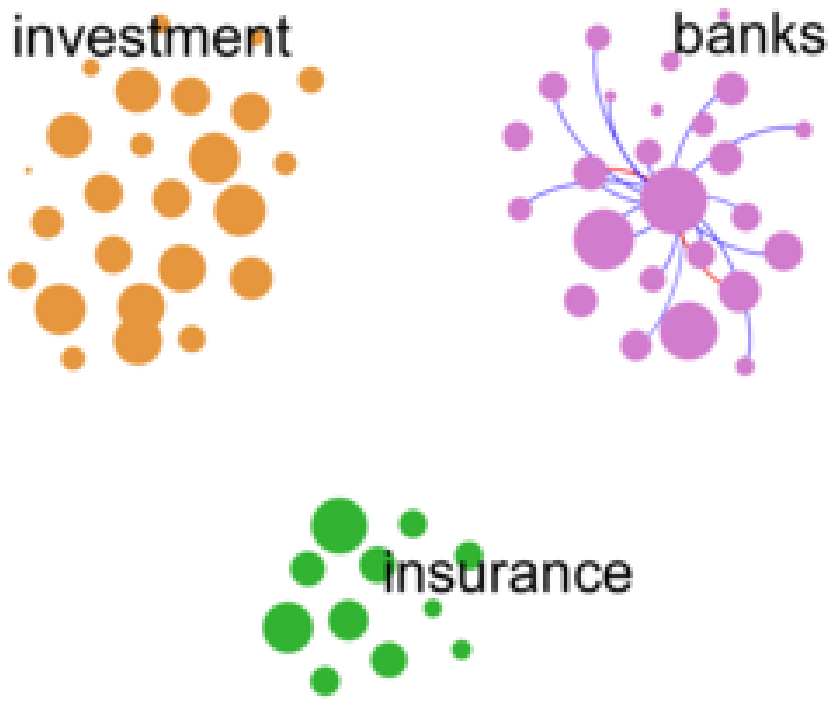} &
\includegraphics[trim= 110mm 5mm 90mm 240mm,clip,height= 2.3cm,width= 3.5cm]{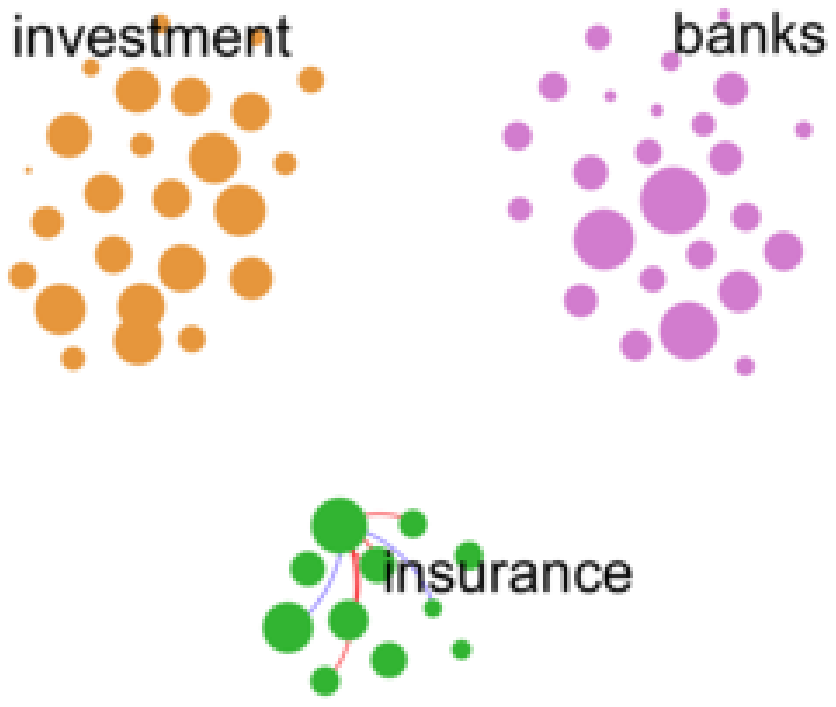} & 
\includegraphics[trim= 10mm 130mm 190mm 30mm,clip,height= 3.5cm,width= 3.5cm]{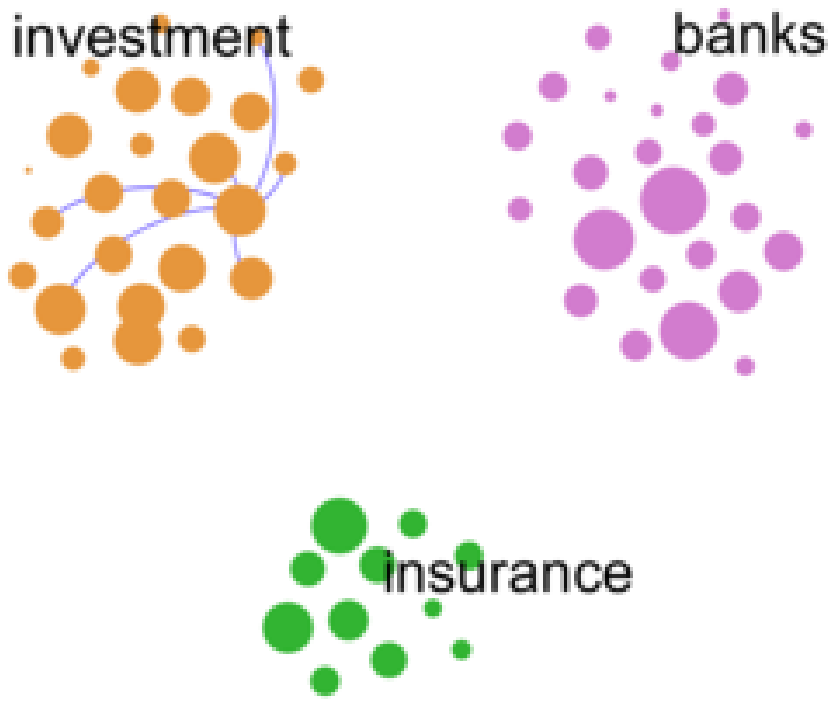} \\
\begin{rotate}{90} \hspace*{30pt} {\large $\substack{\text{between groups}\\ \text{(central institution)}}$} \end{rotate} &
\includegraphics[trim= 10mm 20mm 10mm 28mm,clip,height= 5.0cm,width= 5.0cm]{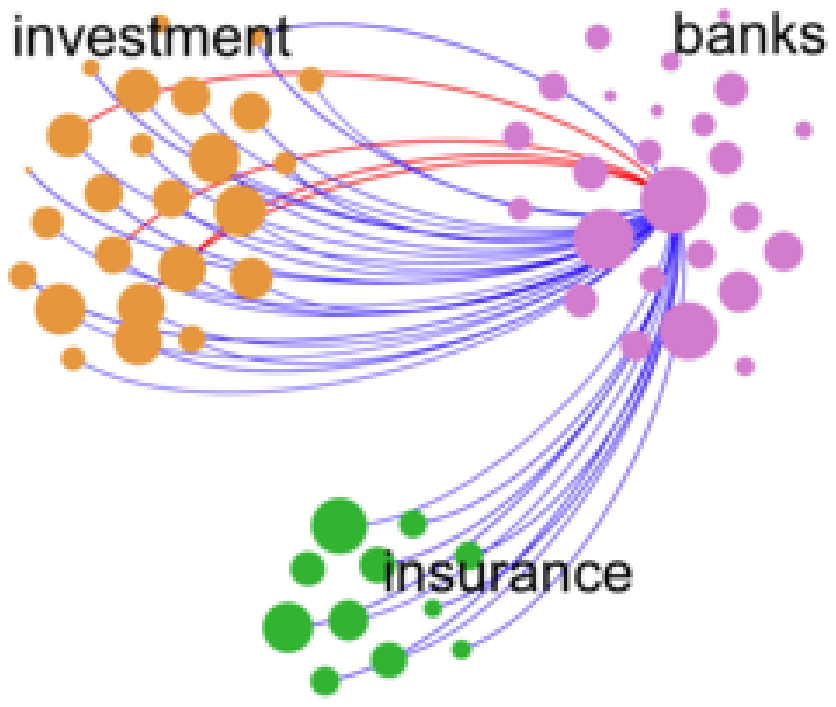} &
\includegraphics[trim= 10mm 25mm 10mm 28mm,clip,height= 5.0cm,width= 5.0cm]{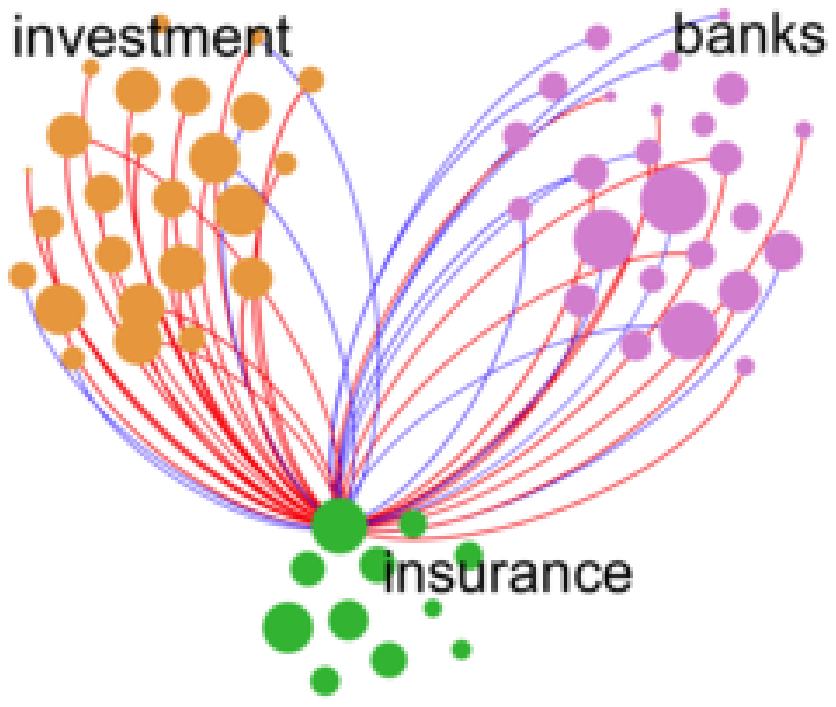} & 
\includegraphics[trim= 10mm 20mm 10mm 28mm,clip,height= 5.0cm,width= 5.0cm]{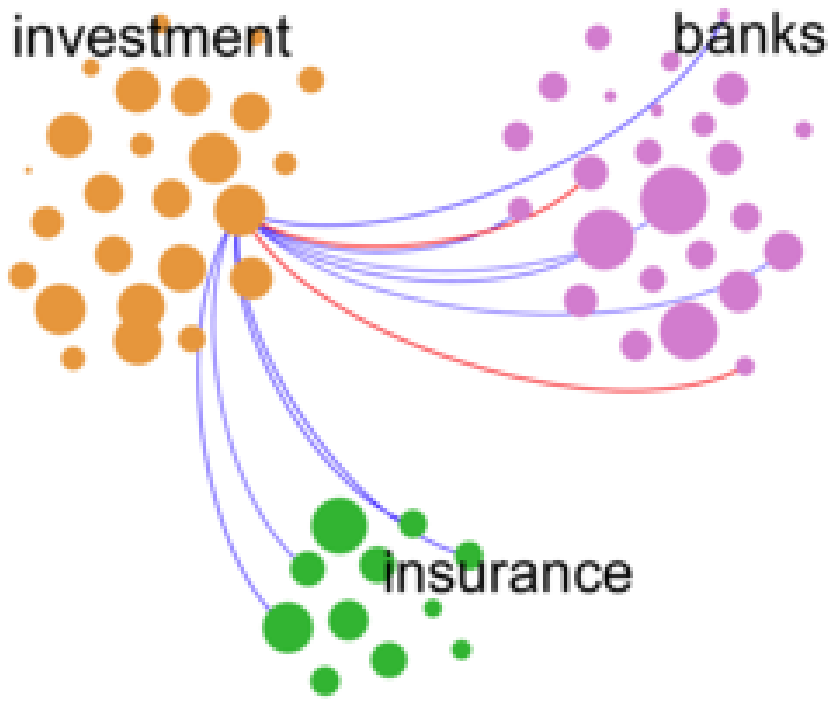} 
\end{tabular}
\captionof{figure}{TRS coefficients in the dense regime. In the columns the effect of TRS on the edges from and to a specific group of nodes: bank (purple), insurance (green), investment companies (orange). In the rows the effects of TRS on between and within groups connectivity, filtering relevant effects (first row) and central institutions (second and third row).
Node size: proportional to the total degree averaged over time within each regime. Edge color: blue for negative, red for positive. We show only edges with significant TRS coefficient.}
\label{fig:tens_IN_TRS_th10}
\end{figure}

The estimated regimes are given in the left plot of Fig.~\ref{fig:degree_states} together with the network total degree. The identification constraint permits to recognise low and high connectedness periods and is strongly supported by the data, since the posterior distributions are well separated (middle plot). The distribution of the estimated coefficients in the two regimes (right plot) highlights the higher heterogeneity across edges in the dense regime.
The unrestricted tensor model captures the edge-specific impact of each risk factor (different colors in each plot of Fig.~\ref{fig:matTens}) as opposed to the pooled model (see Fig.~\ref{S_fig:app_pool_tensor} in the supplement), and allows us to provide new insights on the dynamic relationship among financial institutions and risk factors.
In the dense regime, we find that the credit spread positively affects the probability to be connected to banks from all institutions, and a negative impact on the edge probabilities among investment companies. The term spread has a strong positive effect on connecting to insurance and investment companies, and from banks to insurances. Similarly, the stock index return positively affects the edge probability from insurance and investment companies to banks.
We find also that the autoregressive term has an average positive effect, which might account for either connectedness risk persistence or spurious autocorrelation due to the network estimation step.

In the sparse regime (first line of Fig.~\ref{fig:matTens}) there is no evidence of impact for almost all covariates. This is most striking for CRS, TRS and DTD, which are the most relevant predictors in the dense state. This finding supports the stylised fact that the risk factors have higher explanatory power in periods of higher connectivity of the financial network (\cite{Billioetal12GrangerNet}).

Fig.~\ref{fig:tens_IN} allows to detect potential relationships between covariate effects and node degree centrality. In particular, we find evidence of positive relationship (in absolute value) for DTD, CRS, TRS and MOM.
In the sparse regime, all institutions feature low average degree and there is evidence of a weaker relationship for CRS, TRS, and negative impact for MOM. In the dense regime, the most central institutions (banks and insurance) are the most affected both in terms of the number of connections and the risk factor impact (see the top- and bottom-right part of the scatterplots in Fig.~\ref{fig:tens_IN}). Furthermore, according to the estimated regimes, the most central institutions differ between regimes (see node size in Fig.~\ref{fig:tens_IN_TRS_th10}), with banks being the more connected in both states.

We focus on the term spread factor, since it is a key variable for monetary policy analysis. The unrestricted tensor model provides interesting results on the effect of term spread on the different types of institutions, especially in the dense regime.
We disentangle the relationship among institutions by highlighting the most affected linkages (first row of Fig.~\ref{fig:tens_IN_TRS_th10}) and the impact on all the linkages of the most central nodes (second and third row).
We find that the term spread mostly increases edge probability from banks and the most central insurance company to investment companies. There is no evidence of relevant impact on linkages between banks and insurances, which are strongly affected by the credit spread (see the supplement for further results).
Finally, the effect of the term spread is larger for between group connectivity than for within group connectivity. Most of the edges of the central investment company and bank are negatively affected by the term spread (left and right plots), whereas the connectivity of the central insurance company increases with the term spread (middle plot).

\section{Summary and Concluding Remarks} \label{sec:conclusions}
We present a new zero-inflated logit regression for time series of binary tensors, such as the connectivity tensors encoding the dependence structure of multilayer networks.
The mixing probability allows to capture the sparsity pattern in the data, and a set of coefficient tensors captures the effect of the covariates on each binary observation.
We propose a parsimonious parametrization based on the PARAFAC decomposition of the coefficient tensor and allow the regression parameters to switch between multiple regimes in order to capture the time-varying sparsity patterns. 

We consider the Bayesian paradigm in the inferential process and developed an efficient Gibbs sampler for posterior approximation.
We analyze a real dataset of time-varying networks among European financial institutions. There is strong evidence of heterogeneous effects of the covariates across edges and regimes, with the term spread and credit spread factors playing an important role in explaining the connectivity of central institutions.
Our new empirical results can give interesting insights to policy makers for financial stability and risk monitoring.

%Moreover, in each regime the most degree central institutions are more sensitive to the effect of covariates (either positive or negative). Finally, our model permits to uncover more stylised facts than the benchmark pooled model.

\section*{Supplementary Materials}
Background material on tensors, the derivation of the posterior, simulation experiments and the description of the data are given in an online supplement\footnote{\url{https://matteoiacopini.github.io/docs/BiCaIa_Supplement.pdf}}.

%%%%%%%%%%%%%%%%%%%%%%%%% REFERENCES %%%%%%%%%%%%%%%%%%%%%%%%%
\bibliographystyle{plain}  % plain, apalike (shorter)
%\bibliography{refsMSTens}

%%%%%%%%%%%%%%%%%%%%%%%%% APPENDIX %%%%%%%%%%%%%%%%%%%%%%%%%
\appendix
\renewcommand{\theequation}{\Alph{section}.\arabic{equation}} % equations as (A.1)...

\renewcommand{\thesection}{\Alph{section}}

\numberwithin{equation}{section}
\section{Background Material on Tensors} \label{sec:apdx_tensor}
This appendix provides the main definitions used in the paper. See the supplement for further results.
We introduce some notation for multilinear arrays (i.e. tensors), some basic operations defined on them and lower dimensional objects (such as matrices and vectors) and two tensor decompositions (see \cite{KoldaBader09}, \cite{Cichocki16Tensor_theory} for a noteworthy introduction to these topics).
The \textit{order} of a tensor is the number of dimensions, or modes (i.e., a matrix is a $2$-order tensor).
The mode-$k$ \textit{fiber} of $D$-order tensor $\mathcal{X}$ is the vector obtained along the dimension $k$ by fixing all the other dimensions, that is
\begin{equation}
\mathcal{X}_{(i_1,\ldots,i_{k-1},:,i_{k+2},\ldots,i_D)} \, .
\end{equation}
Similarly, \textit{slices} are matrices obtained by fixing all but two or more dimensions (or modes) of the tensor. By convention, we denote a whole dimension of a tensor by the symbol ``$:$''. 
The mode-$k$ \textit{matricization} operator, $\mathbf{X}_{(n)}$, transforms a $D$-array $\mathcal{X}$ into a matrix by rearranging all the mode-$n$ fibers to be the columns of a matrix, which will have size $\mathbf{X}_{(n)}\in\R^{d_n\times\bar{d}_{(-n)}}$ with $\bar{d}_{(-n)}=\Pi_{i\neq n} d_i$. The \textit{vectorization} of a tensor consists in stacking all the elements in a unique vector of dimension $\bar{d}=\Pi_{i} d_i$, following the inverse lexicographic order.
The \textit{outer product} $\circ$ of two tensors $\mathcal{Y}\in\R^{d_1^Y \times\ldots\times d_M^Y}$ and $\mathcal{X}\in\R^{d_1^X \times\ldots\times d_N^X}$ is the tensor $\mathcal{Z}\in\R^{d_1^Y \times\ldots\times d_M^Y \times d_1^X \times\ldots\times d_N^X}$ whose entries are obtained as
\begin{equation}
\mathcal{Z}_{i_1,\ldots,i_M,j_1,\ldots,j_N} = (\mathcal{Y} \circ \mathcal{X})_{i_1,\ldots,i_M,j_1,\ldots,j_N} = \mathcal{Y}_{i_1,\ldots,i_M} \mathcal{X}_{j_1,\ldots,j_N}.
\label{eq:apdx_tensor_outer_product}
\end{equation}
The \textit{PARAFAC($R$) decomposition} is a low rank decomposition that represents a $D$-order tensor $\mathcal{X}\in\R^{d_1\times\ldots\times d_D}$ as the sum of $R$ rank one tensors, that is, of outer products of vectors (also called marginals):
\begin{equation}
\mathcal{X} = \sum_{r=1}^{R} \mathcal{X}_r = \sum_{r=1}^{R} \mathbf{x}_1^{(r)} \circ \ldots \circ \mathbf{x}_D^{(r)}, \qquad \mathbf{x}_j^{(r)} \in \R^{d_j}.
\label{eq:PARAFAC_demposition}
\end{equation}

\section{Proofs of the Results in the Paper} \label{sec:apdx_proofs}
This appendix provides the derivation of the results. See the supplement for further details.

\subsection{Full conditional distribution of $\phi_r$} \label{sec:posterior_phi_r}
Let $n= n_1+n_2+n_3+n_4 = I+J+K+Q$, then the posterior full conditional is
\begin{align*}
p(\boldsymbol{\phi}| \mathcal{G}, \mathcal{W}) & \propto p(\boldsymbol{\phi}) \int_0^\infty p(\mathcal{G}|\mathcal{W},\boldsymbol{\phi}, \tau) p(\tau) \: \mathrm{d}\tau \\
 & \propto \int_0^\infty \Big( \prod_{r=1}^R (\tau \phi_r)^{\bar{\alpha}-n-1} \Big) \exp \bigg( -\frac{1}{2} \sum_{r=1}^R \Big( 2\bar{b}^\tau \tau\phi_r + \frac{1}{\tau\phi_r} \sum_{h=1}^4 \sum_{l=1}^L \frac{\boldsymbol{\gamma}_{h,l}^{(r)\prime} \boldsymbol{\gamma}_{h,l}^{(r)}}{w_{h,r,l}} \Big) \bigg) \: \mathrm{d}\tau.
\end{align*}
The integrand is the kernel of the GiG distribution given in eq. \eqref{eq:posterior_psir}. Following~\cite{GuhaniyogiDunson17BayesTensorReg} and~\cite{Kruijer10AdaptiveBayesDensityEstimation_LocationScaleMixtures}, it is possible to sample from the posterior of $\phi_r$, for each $r=1,\ldots,R$ by first sampling $\psi_r$, then setting $\phi_r = \psi_r / (\psi_1 + \ldots + \psi_R)$.

\subsection{Full conditional distribution of $\tau$} \label{sec:posterior_tau}
The posterior full conditional distribution is
\begin{align*}
p(\tau| \mathcal{G}, \mathcal{W}, \boldsymbol{\phi}) & \propto \tau^{\bar{a}^\tau -1} \exp(-\bar{b}^\tau \tau) \prod_{r=1}^R \prod_{h=1}^4 \prod_{l=1}^L \abs{\tau \phi_r w_{h,r,l} \mathbf{I}_{n_h}}^{-1/2} \exp \Big( -\frac{1}{2} \frac{\boldsymbol{\gamma}_{h,l}^{(r)\prime} \boldsymbol{\gamma}_{h,l}^{(r)}}{\phi_r w_{h,r,l}} \Big) \\
 & \propto \tau^{\bar{a}^\tau -4R-1} \exp\Bigg( -\frac{1}{2} \bigg( 2 \bar{b}^\tau \tau + \frac{1}{\tau} \sum_{r=1}^R \sum_{h=1}^4 \sum_{l=1}^L \frac{\boldsymbol{\gamma}_{h,l}^{(r)\prime} \boldsymbol{\gamma}_{h,l}^{(r)}}{\phi_r w_{h,r,l}} \bigg) \Bigg) ,
\end{align*}
which is the kernel of the GiG in eq.~\eqref{eq:posterior_tau}.

\subsection{Full conditional distribution of $w_{h,r,l}$} \label{sec:posterior_w_hrl_hierarchical}
The posterior full conditional distribution is
\begin{align*}
p(w_{h,r,l}|\boldsymbol{\gamma}_{h,l}^{(r)},\phi_r,\tau,\lambda_l) & \propto p(w_{h,r,l}|\lambda_l) p(\boldsymbol{\gamma}_{h,l}^{(r)}|w_{h,r,l},\phi_r,\tau) \\
 & \propto \exp\Big( -\frac{\lambda_l^2}{2} w_{h,r,l} \Big) w_{h,r,l}^{-n_h/2} \exp\Big( -\frac{1}{2} \frac{\boldsymbol{\gamma}_{h,l}^{(r)\prime} \boldsymbol{\gamma}_{h,l}^{(r)}}{\tau \phi_r w_{h,r,l}} \Big) ,
\end{align*}
which is the kernel of the GiG in eq.~\eqref{eq:posterior_w}.

\subsection{Full conditional distribution of $\lambda_l$} \label{sec:posterior_lambda_l}
The posterior full conditional distribution of $\lambda_l$ is
\begin{align*}
p( \lambda_l | \mathbf{W}_l ) & \propto \lambda_l^{a_l^\lambda-1} \exp(-b_l^\lambda \lambda_l) \prod_{r=1}^R \prod_{h=1}^4 \frac{\lambda_l^2}{2} \exp\Big( -\frac{\lambda_l^2}{2} w_{h,r,l} \Big) \\
 & \propto \lambda_l^{\bar{a}_l^\lambda +8R -1} \exp\Big( -\lambda_l \bar{b}_l^\lambda -\frac{\lambda_l^2}{2}\sum_{r=1}^R \sum_{h=1}^4 w_{h,r,l} \Big).
\end{align*}
%\begin{align*}
%p(\lambda_l|\lbrace \boldsymbol{\gamma}_{1,l}^{(r)},\ldots,\boldsymbol{\gamma}_{4,l}^{(r)} \rbrace_r,\tau,\phi_r)
%%& \propto p(\lambda_l) \prod_{r=1}^R \prod_{h=1}^4 \prod_{p=1}^{n_h} p(\gamma_{h,l,p}^{(r)}|\tau,\phi_r,\lambda_l) \\
% & \propto \lambda_l^{a_\lambda^l-1} \exp(-b_\lambda^l \lambda_l) \prod_{r=1}^R \prod_{h=1}^4 \prod_{p=1}^{n_h} \frac{\lambda_l}{2\sqrt{\tau \phi_r}} \exp\Biggl( -\frac{\abs{\gamma_{h,l,p}^{(r)}}}{\sqrt{\tau \phi_r}} \lambda_l \Biggr) \\
% & \propto \lambda_l^{\bar{a}_l^\lambda +IJKQR -1} \exp\Bigg( -\lambda_l \Big( \bar{b}_l^\lambda + \sum_{r=1}^R \sum_{h=1}^4 \frac{\norm{\boldsymbol{\gamma}_{h,l}^{(r)}}_1}{\sqrt{\tau \phi_r}} \Big) \Bigg).
%\end{align*}
We sample from this distribution using a Hamiltonian Monte Carlo step (\cite{Neal11HamiltonianMC}).

\subsection{Full conditional distribution of $\boldsymbol{\gamma}_{h,l}^{(r)}$} \label{sec:posterior_gamma_hlr}
For deriving the full conditional distribution of PARAFAC marginals, $\boldsymbol{\gamma}_{h,l}^{(r)}$, start by defining $u_{ijk,t} = \kappa_{ijk,t}/\omega_{ijk,t}$, $\mathcal{U}_t = (u_{ijk,t})_{ijk}$, $\boldsymbol{\Omega}_t = (\omega_{ijk,t})_{ijk}$, $\mathbf{u}_t = \vecc{\mathcal{U}_t}$ and $\bar{\bar{\boldsymbol{\Omega}}}_t = \diag{\vecc{\boldsymbol{\Omega}_t}}$. From eq.~\eqref{eq:likelihood_conditional_X_d_omega}, denoting with $p(\mathcal{G}_l)$ the joint prior distribution on $\lbrace \boldsymbol{\gamma}_{h,l}^{(r)} \rbrace_{h,r}$, one gets
\begin{align}
\notag
p(\mathcal{G}_l | \mathcal{X}_t, & \mathcal{D}_t,\boldsymbol{\Omega}_t, s_t=l, \rho_l) \propto p(\mathcal{G}_l) \prod_{t\in\mathcal{T}_l} \prod_{i=1}^I \prod_{j=1}^J \prod_{k=1}^K p(\omega_{ijk,t}) \\
 & \cdot \prod_{t\in\mathcal{T}_l} \exp\Big( -\frac{1}{2} \big( \vecc{\mathcal{G}_l \times_4 \mathbf{z}_t} -\mathbf{u}_t \big)' \bar{\bar{\boldsymbol{\Omega}}}_t \big( \vecc{\mathcal{G}_l \times_4 \mathbf{z}_t} -\mathbf{u}_t \big) \Big).
% & p(\mathcal{G}_l | \mathcal{X}_t,\mathcal{D}_t,\boldsymbol{\Omega}_t, s_t=l, \rho_l) \propto p(\mathcal{G}_l) \prod_{t\in\mathcal{T}_l} \prod_{i=1}^I \prod_{j=1}^J \prod_{k=1}^K p(\omega_{ijk,t}) \\
% & \cdot \prod_{t\in\mathcal{T}_l} \exp\Big( -\frac{1}{2} \Big( \vecc{\mathcal{G}_l \times_4 \mathbf{z}_t} - \vecc{\mathcal{U}_t} \Big)' \diag{\vecc{\boldsymbol{\Omega}_t}} \Big( \vecc{\mathcal{G}_l \times_4 \mathbf{z}_t} - \vecc{\mathcal{U}_t} \Big) \Big).
\label{eq:apdx_complete_likelihood_part_I_5}
\end{align}
%From eq.~\eqref{eq:complete_likelihood_final} and~\eqref{eq:apdx_complete_likelihood_part_I_5} we obtain
%\begin{equation}
%L(\boldsymbol{\mathcal{X}},\mathcal{D},\boldsymbol{\Omega},\mathbf{s}|\boldsymbol{\theta}) = \prod_{l=1}^L \prod_{t\in\mathcal{T}_l} p(\mathcal{X}_t,\mathcal{D}_t,\boldsymbol{\Omega}_t,s_t|\boldsymbol{\theta}) \propto \prod_{l=1}^L \prod_{t\in\mathcal{T}_l} f\left( \mathcal{G}_{l}, \mathbf{z}_t, \mathcal{U}_t, \boldsymbol{\Omega}_t \right) \, .
%\label{eq:apdx_likelihood_prop_tensor}
%\end{equation}
%Note that:
%\begin{equation}
%\mathcal{G}_l = \sum_{r=1}^R \boldsymbol{\gamma}_{1,l}^{(r)} \circ \boldsymbol{\gamma}_{2,l}^{(r)} \circ \boldsymbol{\gamma}_{3,l}^{(r)} \circ \boldsymbol{\gamma}_{4,l}^{(r)} = \mathcal{G}_l^{(r)} + \mathcal{G}_l^{(-r)} \, ,
%\label{eq:apdx_decomposition_Gl}
%\end{equation}
%where we have defined:
%\begin{align*}
%\mathcal{G}_l^{(r)} & = \boldsymbol{\gamma}_{1,l}^{(r)} \circ \boldsymbol{\gamma}_{2,l}^{(r)} \circ \boldsymbol{\gamma}_{3,l}^{(r)} \circ \boldsymbol{\gamma}_{4,l}^{(r)}, \qquad \mathcal{G}_l^{(-r)} = \sum_{\substack{v=1\\ v\neq r}}^R \mathcal{G}_l^{(v)} \, .
%\end{align*}
By the definitions of mode-$n$ product and PARAFAC decomposition, denoting by $\langle \cdot, \cdot \rangle$ the standard inner product in the Euclidean space $\R^n$, we obtain
\begin{equation}
\vecc{\mathcal{G}_l \times_4 \mathbf{z}_t} = \sum_{r=1}^R \big( \boldsymbol{\gamma}_{1,l}^{(r)} \circ \boldsymbol{\gamma}_{2,l}^{(r)} \circ \boldsymbol{\gamma}_{3,l}^{(r)} \big) \langle \boldsymbol{\gamma}_{4,l}^{(r)}, \mathbf{z}_t \rangle = \sum_{r=1}^R \bar{\mathbf{g}}_{l,t}^{(r)} \, .
\label{eq:apdx_vec_Glt}
\end{equation}
From eq.~\eqref{eq:apdx_vec_Glt} we have:
\begin{align}
\label{eq:apdx_vect_tensor_G4}
\bar{\mathbf{g}}_{l,t}^{(r)} & = \langle \boldsymbol{\gamma}_{4,l}^{(r)}, \mathbf{z}_t \rangle \vecc{\boldsymbol{\gamma}_{1,l}^{(r)} \circ \boldsymbol{\gamma}_{2,l}^{(r)} \circ \boldsymbol{\gamma}_{3,l}^{(r)}} = \vecc{\boldsymbol{\gamma}_{1,l}^{(r)} \circ \boldsymbol{\gamma}_{2,l}^{(r)} \circ \boldsymbol{\gamma}_{3,l}^{(r)}} \mathbf{z}_t' \boldsymbol{\gamma}_{4,l}^{(r)} = \mathbf{A}_4 \boldsymbol{\gamma}_{4,l}^{(r)}, \\
\label{eq:apdx_vect_tensor_G1}
 & = \langle \boldsymbol{\gamma}_{4,l}^{(r)}, \mathbf{z}_t \rangle \big( \boldsymbol{\gamma}_{3,l}^{(r)} \otimes \boldsymbol{\gamma}_{2,l}^{(r)} \otimes \mathbf{I}_I \big) \boldsymbol{\gamma}_{1,l}^{(r)} = \mathbf{A}_1 \boldsymbol{\gamma}_{1,l}^{(r)} \\
\label{eq:apdx_vect_tensor_G2}
 & = \langle \boldsymbol{\gamma}_{4,l}^{(r)}, \mathbf{z}_t \rangle \big( \boldsymbol{\gamma}_{3,l}^{(r)} \otimes \mathbf{I}_J \otimes \boldsymbol{\gamma}_{1,l}^{(r)} \big) \boldsymbol{\gamma}_{2,l}^{(r)} = \mathbf{A}_2 \boldsymbol{\gamma}_{2,l}^{(r)}, \\
\label{eq:apdx_vect_tensor_G3}
 & = \langle \boldsymbol{\gamma}_{4,l}^{(r)}, \mathbf{z}_t \rangle \big( \mathbf{I}_K \otimes \boldsymbol{\gamma}_{2,l}^{(r)} \otimes \boldsymbol{\gamma}_{1,l}^{(r)} \big) \boldsymbol{\gamma}_{3,l}^{(r)} = \mathbf{A}_3 \boldsymbol{\gamma}_{3,l}^{(r)}.
\end{align}
Setting $\bar{\mathbf{g}}_{l,t}^{(-r)} = \sum_{\substack{v=1 \\ v\neq r}}^R \bar{\mathbf{g}}_{l,t}^{(v)}$ we get $\vecc{\mathcal{G}_l \times_4 \mathbf{z}_t} = \bar{\mathbf{g}}_{l,t}^{(r)} + \bar{\mathbf{g}}_{l,t}^{(-r)}$.
Thus, for each $l$ we get
\begin{align}
L(\boldsymbol{\mathcal{X}},\mathcal{D},\boldsymbol{\Omega},\mathbf{s}|\boldsymbol{\theta}) \propto \prod_{t\in\mathcal{T}_l} \exp\Big( -\frac{1}{2} \big( \bar{\mathbf{g}}_{l,t}^{(r)} + \bar{\mathbf{g}}_{l,t}^{(-r)} - \mathbf{u}_t \big)' \; \bar{\bar{\boldsymbol{\Omega}}}_t \big( \bar{\mathbf{g}}_{l,t}^{(r)} + \bar{\mathbf{g}}_{l,t}^{(-r)} - \mathbf{u}_t \big) \Big).
\label{eq:apdx_likelihood_gltr}
\end{align}
We can now single out a specific component $\mathcal{G}_l^{(r)}$ of the PARAFAC decomposition of the tensor $\mathcal{G}$, which is incorporated in $\bar{\mathbf{g}}_{l,t}^{(r)}$. For each $l=1,\ldots,L$ we obtain
\begin{align}
L(\boldsymbol{\mathcal{X}},\mathcal{D},\boldsymbol{\Omega},\mathbf{s}|\boldsymbol{\theta}) & \propto \prod_{t\in\mathcal{T}_l} \exp\Big( -\frac{1}{2} \Big( \bar{\mathbf{g}}_{l,t}^{(r)\prime} \: \bar{\bar{\boldsymbol{\Omega}}}_t \bar{\mathbf{g}}_{l,t}^{(r)} -2 ( \mathbf{u}_t - \bar{\mathbf{g}}_{l,t}^{(-r)})' \: \bar{\bar{\boldsymbol{\Omega}}}_t \bar{\mathbf{g}}_{l,t}^{(r)} \Big) \Big).
\label{eq:apdx_likelihood_all_gammas}
\end{align}

\subsubsection{Full conditional distribution of $\boldsymbol{\gamma}_{1,l}^{(r)}$}
The full conditional distribution of $\boldsymbol{\gamma}_{1,l}^{(r)}$ is obtained from eqs.~\eqref{eq:apdx_vect_tensor_G1} and~\eqref{eq:apdx_likelihood_all_gammas}, where
\begin{align}
\label{eq:apdx_likelihood_gamma_1a}
\bar{\mathbf{g}}_{l,t}^{(r)\prime} \bar{\bar{\boldsymbol{\Omega}}}_t \bar{\mathbf{g}}_{l,t}^{(r)} & = \boldsymbol{\gamma}_{1,l}^{(r)\prime} \Big( \bar{\bar{\boldsymbol{\Sigma}}}_{1,l,t}^{(r)} \Big)^{-1} \boldsymbol{\gamma}_{1,l}^{(r)} \\
-2 (\mathbf{u}_t - \bar{\mathbf{g}}_{l,t}^{(-r)})' \bar{\bar{\boldsymbol{\Omega}}}_t \bar{\mathbf{g}}_{l,t}^{(r)} & = -2 \bar{\boldsymbol{\mu}}_{1,l,t}^{(r)\prime} \Big( \bar{\bar{\boldsymbol{\Sigma}}}_{1,l,t}^{(r)} \Big)^{-1} \boldsymbol{\gamma}_{1,l}^{(r)} \, .
\label{eq:apdx_likelihood_gamma_1b}
\end{align}
By Bayes' theorem and plugging eq.~\eqref{eq:apdx_likelihood_gamma_1a} and eq.~\eqref{eq:apdx_likelihood_gamma_1b} into eq.~\eqref{eq:apdx_likelihood_all_gammas} we get
\begin{align*}
p(\boldsymbol{\gamma}_{1,l}^{(r)} | \boldsymbol{\theta}) 
% & \propto L(\boldsymbol{\mathcal{X}}, \mathcal{D}, \boldsymbol{\Omega}, \mathbf{s}| \boldsymbol{\theta}) p(\boldsymbol{\gamma}_{1,l}^{(r)}|\mathbf{w}_{1,:},\boldsymbol{\phi},\tau) \\
 & = \exp \Bigg( -\frac{1}{2} \Bigg( \boldsymbol{\gamma}_{1,l}^{(r)\prime} \Big( \Big( \bar{\boldsymbol{\Lambda}}_{1,l}^{r} \Big)^{-1} + \sum_{t\in\mathcal{T}_l} \Big(  \bar{\bar{\boldsymbol{\Sigma}}}_{1,l,t}^{(r)} \Big)^{-1} \Big) \boldsymbol{\gamma}_{1,l}^{(r)} \\ 
 & \quad -2 \Big( \bar{\boldsymbol{\zeta}}_{1,l}^{r'} \Big( \bar{\boldsymbol{\Lambda}}_{1,l}^{r} \Big)^{-1} + \sum_{t\in\mathcal{T}_l} \bar{\boldsymbol{\mu}}_{1,l,t}^{(r)\prime} \Big( \bar{\bar{\boldsymbol{\Sigma}}}_{1,l,t}^{(r)} \Big)^{-1} \Big) \boldsymbol{\gamma}_{1,l}^{(r)} \Bigg) \Bigg),
\end{align*}
that is the kernel of the desired multivariate normal distribution.

\subsubsection{Full conditional distribution of $\boldsymbol{\gamma}_{2,l}^{(r)}$}
The full conditional distribution of $\boldsymbol{\gamma}_{2,l}^{(r)}$ is obtained from eqs.~\eqref{eq:apdx_vect_tensor_G2} and~\eqref{eq:apdx_likelihood_all_gammas}, where
\begin{align}
\label{eq:apdx_likelihood_gamma_2a}
\bar{\mathbf{g}}_{l,t}^{(r)\prime} \bar{\bar{\boldsymbol{\Omega}}}_t \bar{\mathbf{g}}_{l,t}^{(r)} & = \boldsymbol{\gamma}_{2,l}^{(r)\prime} \Big( \bar{\bar{\boldsymbol{\Sigma}}}_{2,l,t}^{(r)} \Big)^{-1} \boldsymbol{\gamma}_{2,l}^{(r)} \\
-2 (\mathbf{u}_t - \bar{\mathbf{g}}_{l,t}^{(-r)})' \bar{\bar{\boldsymbol{\Omega}}}_t \bar{\mathbf{g}}_{l,t}^{(r)} & = -2 \bar{\boldsymbol{\mu}}_{2,l,t}^{(r)\prime} \Big( \bar{\bar{\boldsymbol{\Sigma}}}_{2,l,t}^{(r)} \Big)^{-1} \boldsymbol{\gamma}_{2,l}^{(r)} \, .
\label{eq:apdx_likelihood_gamma_2b}
\end{align}
By Bayes' theorem and plugging eq.~\eqref{eq:apdx_likelihood_gamma_2a} and eq.~\eqref{eq:apdx_likelihood_gamma_2b} into eq.~\eqref{eq:apdx_likelihood_all_gammas} we get
\begin{align*}
p(\boldsymbol{\gamma}_{2,l}^{(r)} | \boldsymbol{\theta}) 
% & \propto L(\boldsymbol{\mathcal{X}}, \mathcal{D}, \boldsymbol{\Omega}, \mathbf{s}| \boldsymbol{\theta}) p(\boldsymbol{\gamma}_{2,l}^{(r)}|\mathbf{w}_{2,:},\boldsymbol{\phi},\tau) \\
 & = \exp \Bigg( -\frac{1}{2} \Bigg( \boldsymbol{\gamma}_{2,l}^{(r)\prime} \Big( \Big( \bar{\boldsymbol{\Lambda}}_{2,l}^{r} \Big)^{-1} + \sum_{t\in\mathcal{T}_l} \Big(  \bar{\bar{\boldsymbol{\Sigma}}}_{2,l,t}^{(r)} \Big)^{-1} \Big) \boldsymbol{\gamma}_{2,l}^{(r)} \\ 
 & \quad -2 \Big( \bar{\boldsymbol{\zeta}}_{2,l}^{r'} \Big( \bar{\boldsymbol{\Lambda}}_{2,l}^{r} \Big)^{-1} + \sum_{t\in\mathcal{T}_l} \bar{\boldsymbol{\mu}}_{2,l,t}^{(r)\prime} \Big( \bar{\bar{\boldsymbol{\Sigma}}}_{2,l,t}^{(r)} \Big)^{-1} \Big) \boldsymbol{\gamma}_{2,l}^{(r)} \Bigg) \Bigg),
\end{align*}
that is the kernel of the desired multivariate normal distribution.

\subsubsection{Full conditional distribution of $\boldsymbol{\gamma}_{3,l}^{(r)}$}
The full conditional distribution of $\boldsymbol{\gamma}_{3,l}^{(r)}$ is obtained from eqs.~\eqref{eq:apdx_vect_tensor_G3} and~\eqref{eq:apdx_likelihood_all_gammas}, where
\begin{align}
\label{eq:apdx_likelihood_gamma_3a}
\bar{\mathbf{g}}_{l,t}^{(r)\prime} \bar{\bar{\boldsymbol{\Omega}}}_t \bar{\mathbf{g}}_{l,t}^{(r)} & = \boldsymbol{\gamma}_{3,l}^{(r)\prime} \Big( \bar{\bar{\boldsymbol{\Sigma}}}_{3,l,t}^{(r)} \Big)^{-1} \boldsymbol{\gamma}_{3,l}^{(r)} \\
-2 (\mathbf{u}_t - \bar{\mathbf{g}}_{l,t}^{(-r)})' \bar{\bar{\boldsymbol{\Omega}}}_t \bar{\mathbf{g}}_{l,t}^{(r)} & = -2 \bar{\boldsymbol{\mu}}_{3,l,t}^{(r)\prime} \Big( \bar{\bar{\boldsymbol{\Sigma}}}_{3,l,t}^{(r)} \Big)^{-1} \boldsymbol{\gamma}_{3,l}^{(r)} \, .
\label{eq:apdx_likelihood_gamma_3b}
\end{align}
By Bayes' theorem and plugging eq.~\eqref{eq:apdx_likelihood_gamma_3a} and eq.~\eqref{eq:apdx_likelihood_gamma_3b} into eq.~\eqref{eq:apdx_likelihood_all_gammas} we get
\begin{align*}
p(\boldsymbol{\gamma}_{3,l}^{(r)} | \boldsymbol{\theta}) 
% & \propto L(\boldsymbol{\mathcal{X}}, \mathcal{D}, \boldsymbol{\Omega}, \mathbf{s}| \boldsymbol{\theta}) p(\boldsymbol{\gamma}_{3,l}^{(r)}|\mathbf{w}_{3,:},\boldsymbol{\phi},\tau) \\
 & = \exp \Bigg( -\frac{1}{2} \Bigg( \boldsymbol{\gamma}_{3,l}^{(r)\prime} \Big( \Big( \bar{\boldsymbol{\Lambda}}_{3,l}^{r} \Big)^{-1} + \sum_{t\in\mathcal{T}_l} \Big(  \bar{\bar{\boldsymbol{\Sigma}}}_{3,l,t}^{(r)} \Big)^{-1} \Big) \boldsymbol{\gamma}_{3,l}^{(r)} \\ 
 & \quad -2 \Big( \bar{\boldsymbol{\zeta}}_{3,l}^{r'} \Big( \bar{\boldsymbol{\Lambda}}_{3,l}^{r} \Big)^{-1} + \sum_{t\in\mathcal{T}_l} \bar{\boldsymbol{\mu}}_{3,l,t}^{(r)\prime} \Big( \bar{\bar{\boldsymbol{\Sigma}}}_{3,l,t}^{(r)} \Big)^{-1} \Big) \boldsymbol{\gamma}_{3,l}^{(r)} \Bigg) \Bigg) \, .
\end{align*}
that is the kernel of the desired multivariate normal distribution.

\subsubsection{Full conditional distribution of $\boldsymbol{\gamma}_{4,l}^{(r)}$}
The full conditional distribution of $\boldsymbol{\gamma}_{4,l}^{(r)}$ is obtained from eqs.~\eqref{eq:apdx_vect_tensor_G4} and~\eqref{eq:apdx_likelihood_all_gammas}, where
\begin{align}
\label{eq:apdx_likelihood_gamma_4a}
\bar{\mathbf{g}}_{l,t}^{(r)\prime} \bar{\bar{\boldsymbol{\Omega}}}_t \bar{\mathbf{g}}_{l,t}^{(r)} & = \boldsymbol{\gamma}_{4,l}^{(r)\prime} \Big( \bar{\bar{\boldsymbol{\Sigma}}}_{4,l,t}^{(r)} \Big)^{-1} \boldsymbol{\gamma}_{4,l}^{(r)} \\
-2 (\mathbf{u}_t - \bar{\mathbf{g}}_{l,t}^{(-r)})' \bar{\bar{\boldsymbol{\Omega}}}_t \bar{\mathbf{g}}_{l,t}^{(r)} & = -2 \bar{\boldsymbol{\mu}}_{4,l,t}^{(r)\prime} \Big( \bar{\bar{\boldsymbol{\Sigma}}}_{4,l,t}^{(r)} \Big)^{-1} \boldsymbol{\gamma}_{4,l}^{(r)} \, .
\label{eq:apdx_likelihood_gamma_4b}
\end{align}
By Bayes' theorem and plugging eq.~\eqref{eq:apdx_likelihood_gamma_4a} and eq.~\eqref{eq:apdx_likelihood_gamma_4b} into eq.~\eqref{eq:apdx_likelihood_all_gammas} we get
\begin{align*}
p(\boldsymbol{\gamma}_{4,l}^{(r)} | \boldsymbol{\theta}) 
% & \propto L(\boldsymbol{\mathcal{X}}, \mathcal{D}, \boldsymbol{\Omega}, \mathbf{s}| \boldsymbol{\theta}) p(\boldsymbol{\gamma}_{4,l}^{(r)}|\mathbf{w}_{4,:},\boldsymbol{\phi},\tau) \\
 & = \exp \Bigg( -\frac{1}{2} \Bigg( \boldsymbol{\gamma}_{4,l}^{(r)\prime} \Big( \Big( \bar{\boldsymbol{\Lambda}}_{4,l}^{r} \Big)^{-1} + \sum_{t\in\mathcal{T}_l} \Big(  \bar{\bar{\boldsymbol{\Sigma}}}_{4,l,t}^{(r)} \Big)^{-1} \Big) \boldsymbol{\gamma}_{4,l}^{(r)} \\ 
 & \quad -2 \Big( \bar{\boldsymbol{\zeta}}_{4,l}^{r'} \Big( \bar{\boldsymbol{\Lambda}}_{4,l}^{r} \Big)^{-1} + \sum_{t\in\mathcal{T}_l} \bar{\boldsymbol{\mu}}_{4,l,t}^{(r)\prime} \Big( \bar{\bar{\boldsymbol{\Sigma}}}_{4,l,t}^{(r)} \Big)^{-1} \Big) \boldsymbol{\gamma}_{4,l}^{(r)} \Bigg) \Bigg),
\end{align*}
that is the kernel of the desired multivariate normal distribution.

\subsection{Full conditional distribution of $\omega_{ijk,t}$} \label{sec:posterior_omega_ijkt}
Define $\psi_{ijk,t} = \mathbf{z}_t' \mathbf{g}_{ijk,s_t}$. The posterior full conditional distribution is
\begin{align*}
& p(\omega_{ijk,t}|x_{ijk,t},s_t,\mathcal{G}_{s_t}) = \sum_{d_{ijk,t} \in \lbrace 0,1 \rbrace} \int p(\omega_{ijk,t},d_{ijk,t}|x_{ijk,t},s_t,\mathcal{G}_{s_t},\rho_{s_t}) p(\rho_{s_t}) \: \mathrm{d}\rho_{s_t} \\
% & = \sum_{d_{ijk,t} \in \lbrace 0,1 \rbrace} \int \frac{p(x_{ijk,t},d_{ij,t}|\omega_{ijk,t},s_t,\mathcal{G}_{s_t},\rho_{s_t}) p(\omega_{ijk,t}) p(\rho_{s_t})}{\int_\Omega p(x_{ijk,t},\omega_{ijk,t},d_{ijk,t}|s_t,\mathcal{G}_{s_t},\rho_{s_t}) \: \mathrm{d}\omega_{ijk,t}} \: \mathrm{d}\rho_{s_t} \\
% & = \sum_{d_{ijk,t} \in \lbrace 0,1 \rbrace} \int \frac{p(x_{ijk,t},\omega_{ijk,t},d_{ijk,t}|s_t,\mathcal{G}_{s_t},\rho_{s_t})}{p(x_{ijk,t},d_{ijk,t}|s_t,\mathcal{G}_{s_t},\rho_{s_t})} p(\rho_{s_t}) \: \mathrm{d}\rho_{s_t} \\
% & = \sum_{d_{ijk,t} \in \lbrace 0,1 \rbrace} \int \frac{\big( \rho_{s_t} \delta_{\lbrace 0 \rbrace}(x_{ijk,t}) \right)^{d_{ijk,t}} \left( \frac{1-\rho_{s_t}}{2}\big)^{d_{ijk,t}} \exp\big( -\frac{\omega_{ijk,t}}{2}\psi_{ijk,t}^2 + \kappa_{ijk,t} \psi_{ijk,t} \big) p(\omega_{ijk,t}) p(\rho_{s_t})}{\big( \rho_{s_t} \delta_{\lbrace 0 \rbrace}(x_{ijk,t}) \big)^{d_{ijk,t}} \left( \frac{1-\rho_{s_t}}{2}\right)^{d_{ijk,t}} (\exp( \psi_{ijk,t} x_{ijk,t}) / (1+\exp( \psi_{ijk,t} )))^{1-d_{ijk,t}}} \: \mathrm{d}\rho_{s_t} \\ \notag
 & = \sum_{d_{ijk,t} \in \lbrace 0,1 \rbrace} \int \exp( \kappa_{ijk,t}^{(s_t)} \psi_{ijk,t} ) \frac{\exp( \psi_{ijk,t} x_{ijk,t}(1-d_{ijk,t}) )}{(1+\exp( \psi_{ijk,t} ))^{1-d_{ijk,t}}} \exp\bigg( -\frac{\omega_{ijk,t}}{2}\psi_{ijk,t}^2 \bigg) p(\omega_{ijk,t}) p(\rho_{s_t}) \: \mathrm{d}\rho_{s_t} \\ \notag
% & = \sum_{d_{ijk,t} \in \lbrace 0,1 \rbrace} \int \big( \frac{1+\exp(\psi_{ijk,t})}{\exp(\psi_{ijk,t}x_{ijk,t})} \cdot \frac{\exp( \psi_{ijk,t}x_{ijk,t})}{\exp(\psi_{ijk,t}/2)} \right]^{1-d_{ijk,t}} \big( \exp(-\psi_{ijk,t}^2 \omega_{ijk,t}/2) p(\omega_{ijk,t}) \big) p(\rho_{s_t}) \: \mathrm{d}\rho_{s_t} \\ \notag
 & = \bigg( 1+ \frac{1+\exp( \psi_{ijk,t} )}{\exp( \psi_{ijk,t}/2 )} \bigg) \Big( \exp( -\psi_{ijk,t}^2\omega_{ijk,t}/2) p(\omega_{ijk,t}) \Big) \\
 & \propto \exp( -\psi_{ijk,t}^2\omega_{ijk,t}/2) p(\omega_{ijk,t}) \, .
\end{align*}
Since $p(\omega_{ijk,t}) \sim PG(1,0)$, by \cite[Theorem 1]{Polsonetal13PolyaGamma} the result in eq. \eqref{eq:posterior_omega}.

\subsection{Full conditional distribution of $d_{ijk,t}$} \label{sec:posterior_d_ijkt}
The posterior full conditional posterior distribution is
\begin{align*}
p(d_{ijk,t} = 1|\boldsymbol{\mathcal{X}},\mathbf{s},\mathcal{G}_{s_t},\boldsymbol{\rho}_{s_t}) & \propto \rho_{s_t} \delta_{\lbrace 0 \rbrace}(x_{ijk,t}) \\
p(d_{ijk,t} = 0|\boldsymbol{\mathcal{X}},\mathbf{s},\mathcal{G}_{s_t},\boldsymbol{\rho}_{s_t}) & \propto (1-\rho_{s_t}) \frac{\exp \big( (\mathbf{z}_t' \mathbf{g}_{ijk,s_t})x_{ijk,t} \big)}{1+\exp( \mathbf{z}_t' \mathbf{g}_{ijk,s_t} )},
\end{align*}
which is the discrete distribution in eq. \eqref{eq:posterior_d}, and follows from
\begin{align*}
p(d_{ijk,t}|\boldsymbol{\mathcal{X}},\mathbf{s},\mathcal{G}_{s_t},\boldsymbol{\rho}_{s_t}) & \propto \big(\rho_{s_t} \delta_{\lbrace 0 \rbrace}(x_{ijk,t}) \big)^{d_{ijk,t}} \Bigg( (1-\rho_{s_t}) \frac{(\exp( \mathbf{z}_t' \mathbf{g}_{ijk,s_t} ))^{x_{ijk,t}}}{1+\exp( \mathbf{z}_t' \mathbf{g}_{ijk,s_t} )} \Bigg)^{1-d_{ijk,t}} \, .
% & = \big( \rho_l^{T_l} \prod_{t\in\mathcal{T}_l} \delta_{\lbrace 0 \rbrace}(x_{ijk,t}) \big)^{d_{ijk,t}} \big( (1-\rho_l)^{T_l} \prod_{t\in\mathcal{T}_l} \frac{(\exp( \mathbf{z}_t' \mathbf{g}_{ijk,l} ))^{x_{ijk,t}}}{1+\exp( \mathbf{z}_t' \mathbf{g}_{ijk,l} )} \big)^{1-d_{ijk,t}} \, .
\end{align*}

\subsection{Full conditional distribution of $\rho_l$} \label{sec:posterior_rho_lGB}
The posterior full conditional distribution is
\begin{align*}
p(\rho_l|\boldsymbol{\mathcal{X}},\mathcal{D},\mathbf{s}) 
% & \propto p(\rho_l) \int_G L(\boldsymbol{\mathcal{X}},\mathcal{D},\mathbf{s} | \rho_l, \mathcal{G}_l) p(\mathcal{G}_l) \: \mathrm{d}\mathcal{G}_l \\
% & \propto \Biggl[ \int_G \prod_{t\in\mathcal{T}_l} \prod_{i=1}^I \prod_{j=1}^J \prod_{k=1}^K \rho_l^{d_{ijk,t}} \cdot \big( \delta_{\lbrace 0 \rbrace}(x_{ijk,t}) \big)^{d_{ijk,t}} \cdot (1-\rho_l)^{1-d_{ijk,t}} \\
% & \quad \cdot \frac{ \big( \exp( \mathbf{z}_{ijk,t}' \mathbf{g}_{ijk,l}) \big)^{x_{ijk,t}(1-d_{ijk,t})}}{\big( 1+\exp( \mathbf{z}_{ijk,t}' \mathbf{g}_{ijk,l}) \big)^{(1-d_{ijk,t})}} \: \mathrm{d}\mathcal{G}_l \Biggr] \cdot \rho_l^{\bar{a}_l^\rho -1} (1-\rho_l)^{\bar{b}_l^\rho -1} \\
 & \propto \Bigg( \prod_{t\in\mathcal{T}_l} \prod_{i=1}^I \prod_{j=1}^J \prod_{k=1}^K \rho_l^{d_{ijk,t}} (1-\rho_l)^{1-d_{ijk,t}} \Bigg) \rho_l^{\bar{a}_l^\rho -1} (1-\rho_l)^{\bar{b}_l^\rho -1} \\
% & = \rho_l^{N_1^l} (1-\rho_l)^{N_0^l} \rho_l^{\bar{a}_l^\rho -1} (1-\rho_l)^{\bar{b}_l^\rho -1} \\
 & = \rho_l^{N_1^l + \bar{a}_l^\rho -1} (1-\rho_l)^{N_0^l + \bar{b}_l^\rho -1} \, ,
\end{align*}
which is the kernel of the Beta in eq. \eqref{eq:posterior_rho}, and we have defined
\begin{align*}
N_1^l = \sum_{t\in\mathcal{T}_l} \sum_{i=1}^I \sum_{j=1}^J \sum_{k=1}^K \I_{\lbrace 1 \rbrace}(d_{ijk,t}), \qquad N_0^l = \sum_{t\in\mathcal{T}_l} \sum_{i=1}^I \sum_{j=1}^J \sum_{k=1}^K \I_{\lbrace 0 \rbrace}(d_{ijk,t}).
\end{align*}

\subsection{Full conditional distribution of $\boldsymbol{\xi}_{l}$} \label{sec:posterior_xi_l}
The posterior full conditional distribution of each row $l$ is
\begin{align*}
p(\boldsymbol{\xi}_{l}|\mathbf{s}) & \propto \Big( \prod_{k=1}^{L} \xi_{l,k}^{\bar{c}_k-1} \Big) \Big( \prod_{g=1}^L \prod_{k=1}^L \xi_{g,k}^{N_{g,k}(\mathbf{s})} \Big) \\
% & \propto \prod_{k=1}^{L} \xi_{l,k}^{\bar{c}_k-1} \prod_{k=1}^L \xi_{l,k}^{N_{l,k}(\mathbf{s})} \\
 & \propto \prod_{k=1}^{L} \xi_{l,k}^{\bar{c}_k + N_{l,k}(\mathbf{s}) -1},
\end{align*}
which is the kernel of the Dirichlet in eq. \eqref{eq:posterior_xi}, and $N_{i,j}(\mathbf{s}) = \sum_t \zeta_{t-1,i} \zeta_{t,j}$.
%\I(s_{t-1}=i) \I(s_t=j)$.

\subsection{Full conditional distribution of $s_t$}\label{sec:posterior_s_t}
We update the whole path $\mathbf{s}$ from the posterior full joint conditional distribution via the Forward Filtering Backward Sampling algorithm (FFBS, see~\cite{Fruhwirth06FiniteMixtures_MarkovSwitch_book}). It is based on the factorisation of the full joint conditional distribution as the product of the entries of the transition matrix $\boldsymbol{\Xi}$ and the filtered probabilities $p(\mathbf{s}_t | \mathcal{X}_{1},\ldots,\mathcal{X}_t,\mathcal{G},\boldsymbol{\rho},\boldsymbol{\Xi})$. The full joint conditional distribution:
\begin{align*}
p(\mathbf{s} | \boldsymbol{\mathcal{X}},\mathcal{G},\boldsymbol{\rho},\boldsymbol{\Xi}) & \propto p(\mathbf{s}|\boldsymbol{\Xi}) \prod_{i=1}^I \prod_{j=i}^J \prod_{k=1}^K  p(x_{ijk,t} | s_t=l, \rho_l, \mathcal{G}_l).
\end{align*}

\section{Computational Details for the Pooled Model} \label{sec:apdx_pooled}
Let $\mathcal{H}$ be a tensor of size $(I\times J\times K\times Q\times Q)$ with entries defined by
\begin{equation*}
\mathcal{H}_{ijkqp} = \Bigg\lbrace \begin{array}{cc}
1 & \text{if } q = p \\
0 & \text{if } q \neq p
\end{array}
\end{equation*}
for each $ijk$. If $\mathbf{g}_{ijk,l} = \mathbf{g}_l \in \R^Q$ for each $ijk$ and regime $l$, then we have the representation $\mathcal{G}_l = \mathcal{H} \times_5 \mathbf{g}_l$.
In the pooled model we are assuming that the coefficient tensor in each regime $l$ satisfies $\mathcal{G}_{ijkq,l} = \mathbf{g}_{q,l}$. We assume the following prior distributions
\begin{align*}
\mathbf{g}_l | \tau,w_l \sim \mathcal{N}_Q(\bar{\boldsymbol{\zeta}}_l,\tau w_l \mathbf{I}_Q), \quad w_l|\lambda_l \sim \mathcal{E}xp(\lambda_l^2/2), \quad \lambda_l \sim \mathcal{G}a(\bar{a}_l^\lambda,\bar{b}_l^\lambda), \quad \tau \sim \mathcal{G}a(\bar{a}^\tau,\bar{b}^\tau) \, .
\end{align*}
The complete data likelihood thus becomes
\begin{align*}
\notag
L(\boldsymbol{\mathcal{X}}|\boldsymbol{\theta}) & \propto \prod_{t\in\mathcal{T}_l} \prod_{i=1}^I \prod_{j=1}^J \prod_{k=1}^K \exp\Big( -\frac{\omega_{ijk,t}}{2}(\mathbf{z}_t' \mathbf{g}_l)^2 +\kappa_{ijk,t}(\mathbf{z}_t' \mathbf{g}_l) \Big).
\end{align*}
This yields the posterior distribution for $\mathbf{g}_l$
%\begin{align*}
%p(\mathbf{g}_l | \boldsymbol{\Omega}_t, \tau, w_l) & \propto \exp \left( -\frac{1}{2} \sum_{t\in\mathcal{T}_l} \sum_{i,j,k} \mathbf{g}_l' \mathbf{z}_t \omega_{ijk,t} \mathbf{z}_t' \mathbf{g}_l - 2 \mathbf{z}_t' \mathbf{g}_l \kappa_{ijk,t} \right) \exp \left( -\frac{1}{2} \frac{\mathbf{g}_l' \mathbf{g}_l}{\tau w_l} \right) \\
% & = \exp \left( -\frac{1}{2} \left[ \mathbf{g}_l' \left( \frac{1}{\tau w_l} + \sum_{t\in\mathcal{T}_l} \sum_{i,j,k} \mathbf{z}_t \omega_{ijk,t} \mathbf{z}_t' \right) \mathbf{g}_l' - 2 \left( \sum_{t\in\mathcal{T}_l} \sum_{i,j,k} \kappa_{ijk,t} \mathbf{z}_t' \right) \mathbf{g}_l \right] \right) \, ,
%\end{align*}
\begin{align*}
p(\mathbf{g}_l & | \boldsymbol{\Omega}_t, \tau, w_l) \propto \exp\Big( -\frac{1}{2} \sum_{t\in\mathcal{T}_l} \sum_{i,j,k} \mathbf{g}_l' \mathbf{z}_t \omega_{ijk,t} \mathbf{z}_t' \mathbf{g}_l - 2 \mathbf{z}_t' \mathbf{g}_l \kappa_{ijk,t} \Big) \exp\Big( -\frac{1}{2} \frac{(\mathbf{g}_l-\bar{\boldsymbol{\zeta}}_l)' (\mathbf{g}_l-\bar{\boldsymbol{\zeta}}_l)}{\tau w_l} \Big) \\
 & = \exp\Bigg( -\frac{1}{2} \Big( \mathbf{g}_l' \bigg( \frac{1}{\tau w_l} + \sum_{t\in\mathcal{T}_l} \sum_{i,j,k} \mathbf{z}_t \omega_{ijk,t} \mathbf{z}_t' \Big) \mathbf{g}_l' - 2 \bigg( \frac{\bar{\boldsymbol{\zeta}_l'}}{\tau w_l} + \sum_{t\in\mathcal{T}_l} \sum_{i,j,k} \kappa_{ijk,t} \mathbf{z}_t' \bigg) \mathbf{g}_l \bigg) \Bigg),
\end{align*}
which is the kernel of a Normal distribution. The posterior distribution of $\tau$ is
\begin{align*}
p(\tau | \mathbf{g}, \mathbf{w}) & \propto \tau^{\bar{a}^\tau-1} \exp(-\bar{b}^\tau \tau) \prod_{l=1}^L \exp\Big( -\frac{(\mathbf{g}_l-\bar{\boldsymbol{\zeta}}_l)' (\mathbf{g}_l-\bar{\boldsymbol{\zeta}}_l)}{2\tau w_l} \Big) \\
 & = \tau^{\bar{a}^\tau-1} \exp\Big( -\frac{1}{2} \Big( 2\bar{b}^\tau \tau + \sum_{l=1}^L \frac{(\mathbf{g}_l-\bar{\boldsymbol{\zeta}}_l)' (\mathbf{g}_l-\bar{\boldsymbol{\zeta}}_l)}{w_l} \frac{1}{\tau} \Big) \Big),
\end{align*}
which is the kernel of a GiG distribution. The posterior distribution of $w_l$ is
\begin{align*}
p(w_l |g_l,\tau,\lambda_l) & \propto \exp\Big( -\frac{\lambda_l^2}{2} w_l \Big) \exp\Big( -\frac{(\mathbf{g}_l-\bar{\boldsymbol{\zeta}}_l)' (\mathbf{g}_l-\bar{\boldsymbol{\zeta}}_l)}{2\tau w_l} \Big) \\
 & = \exp\Big( -\frac{1}{2} \Big( \lambda_l^2 w_l + \frac{(\mathbf{g}_l-\bar{\boldsymbol{\zeta}}_l)' (\mathbf{g}_l-\bar{\boldsymbol{\zeta}}_l)}{\tau} \frac{1}{w_l} \Big) \Big),
\end{align*}
which is the kernel of a GiG distribution. The posterior distribution of $\lambda_l$ (integrating out $w_l$) is again a GiG obtained from
\begin{align*}
p(\lambda_l | \tau, \mathbf{g}_l) & \propto \lambda_l^{\bar{a}_l^\lambda-1} \exp(-\bar{b}_l^\lambda \lambda_l) \frac{\sqrt{\tau}}{2\lambda_l} \exp\Big( -\frac{\norm{\mathbf{g}_l}_1 \sqrt{\tau}}{\lambda_l} \Big) \\
 & \propto \lambda_l^{\bar{a}_l^\lambda-2} \exp\Big( -\frac{1}{2} \Big( 2\bar{b}_l^\lambda \lambda_l + \norm{\mathbf{g}_l}_1 \sqrt{\tau} \big/ \lambda_l \Big) \Big).
\end{align*}

\end{document}